\documentclass[preprint,12pt]{elsarticle}

\usepackage{amssymb}
\usepackage{amsmath}
\usepackage{graphicx} % Required for inserting images
\usepackage{natbib}
\usepackage{url}
\usepackage{subfig}
\usepackage{dirtytalk}
\usepackage[table]{xcolor}
\usepackage{amsmath,amssymb,amsfonts}%
\usepackage{amsthm}%
\usepackage{mathrsfs}%
\usepackage{booktabs}%
\usepackage{multicol}
\usepackage{multirow}
\usepackage{hyperref}
\usepackage{amsmath}
\usepackage{mathtools}
\usepackage{float}
\usepackage{soul}
\usepackage[linesnumbered,ruled,vlined]{algorithm2e}

\renewcommand{\vec}[1]{\mathbf{#1}}
\newcommand{\wi}{\vec{w}_{i}}
\newcommand{\mean}{\boldsymbol{\mu}}
\newcommand{\meank}{\boldsymbol{\mu}_{K}}
\newcommand{\meanki}{\boldsymbol{\mu}_{k[\vec{w}_{i}]}}

\newcommand{\sigu}{\boldsymbol{\Sigma}}
\newcommand{\siguk}{\boldsymbol{\Sigma}_{K}}
\newcommand{\sigukii}{\boldsymbol{\Sigma}_{k[\vec{w}_{i}, \vec{w}_{i}]}}

\definecolor{col1}{HTML}{4DAF4A}
\definecolor{col2}{HTML}{377EB8}
\definecolor{col3}{HTML}{E41A1C}
\definecolor{col4}{HTML}{984EA3}
\definecolor{col5}{HTML}{FF7F00}
\newcommand\crule[3][black]{\textcolor{#1}{\rule{#2}{#3}}}

\newcommand{\pkg}[1]{{\normalfont\fontseries{b}\selectfont #1}}
\let\proglang=\textsf \let\code=\texttt

\newcommand{\beginsupplement}{% complete program to generate appendix with tables
    \setcounter{table}{0}
    \renewcommand{\thetable}{A\arabic{table}}%
    \setcounter{figure}{0}
    \renewcommand{\thefigure}{A\arabic{figure}}%    
}

\journal{International Journal of Approximate Reasoning}

\begin{document}

\begin{frontmatter}
\title{Robust fuzzy clustering with cellwise outliers}

\author{Giorgia Zaccaria$^a$\footnote{Corresponding author: \href{giorgia.zaccaria@unimib.it}{giorgia.zaccaria@unimib.it}}, Lorenzo Benzakour$^b$, Luis A. García-Escudero$^c$, Francesca Greselin$^b$, Agustín Mayo-Íscar$^c$} %% Author name

\affiliation{organization={Department of Economics, Management and Statistics, University of Milano-Bicocca},%Department and Organization
            addressline={Via Bicocca degli Arcimboldi 8}, 
            city={Milan},
            postcode={20100}, 
            country={Italy}}
\affiliation{organization={Department of Statistics and Quantitative Methods, University of Milano-Bicocca},%Department and Organization
            addressline={Via Bicocca degli Arcimboldi 8}, 
            city={Milan},
            postcode={20100}, 
            country={Italy}}
\affiliation{organization={Department of Statistics and Operational Research, University of Valladolid},%Department and Organization
            addressline={Paseo de Belén 7}, 
            city={Valladolid},
            postcode={47011}, 
            country={Spain}}

%% Abstract
\begin{abstract}
In a data matrix, we may distinguish between cases, each represented by a row vector for a statistical unit, and cells, which correspond to single entries of the data matrix. Recent developments in Robust Statistics have introduced the cellwise contamination paradigm, which assumes contamination on cells rather than on entire cases. This approach becomes particularly relevant as the number of variables increases. Indeed, discarding or downweighting entire cases because of a few anomalous cells in them, as done by traditional (casewise) robust methods, can result in substantial information loss, since the non-contaminated (or reliable) cells can still be highly informative. This philosophy can also be considered in fuzzy clustering, by assuming that reliable cells within a case may still provide useful information for determining fuzzy memberships. A robust fuzzy clustering proposal is thus introduced in this work, combining the advantages of dealing with outlying cells and simultaneously controlling the degree of fuzziness of unit assignments. The cluster-specific relationships among variables, detected by the fuzzy clustering approach, are also key to better identifying outlying cells and correct them. The strengths of the proposed methodology are illustrated through a simulation study and two real-world applications. The effects of the model's tuning parameters are explored, and some guidance for users on how to set them suitably is provided.
\end{abstract}

%% Keywords
\end{frontmatter}

% \linenumbers

\section{Introduction}\label{sec: intro}
The goal of the unsupervised clustering is to discover subpopulations, called clusters, within the data which share common characteristics, and to estimate their statistical features (e.g., centers, scatter/covariance matrices, etc.). In many cases, clusters are not perfectly separated, and the information provided by partitioning units may offer a narrow perspective on the underlying structure of the data. To overcome this limitation of ``hard'' clustering algorithms such as, for instance, $k$-means \cite{M:1967, BH:1967}, fuzzy clustering approaches have been introduced \cite{B:1981}, in which units are not fully assigned to a single cluster but can have positive membership degrees to more than one cluster. Many of these fuzzy clustering approaches are reviewed in \cite{HKKR:1999,VDOP:2007,GFM:2020}.
One of the main advantages of fuzzy clustering methods is that they allow for varying degrees of fuzzification through a tuning parameter $m$, say the fuzzifier parameter, which can be set depending on the purpose of the analysis. This feature cannot be achieved by other clustering methodologies, such as finite mixture models \cite{MLP:2000}, which, although they quantify membership uncertainty and provide \say{soft} clustering through the estimation of posterior probabilities, do not provide control over the desired degree of fuzziness. 

That flexibility of the fuzzy clustering methods is especially important in domains such as, for instance, medical diagnosis and treatment planning for subgroups of patients. A concrete example comes from clinical studies on certain diseases, where some patients may present overlapping symptoms or ambiguous laboratory test results, making it difficult to assign them uniquely to a diagnostic category. In such cases, fuzzy clustering enables, on the one hand, the identification of a subset of patients for whom further testing may be particularly recommended, and on the other hand, the possibility of adjusting the size of this subset based on the purpose of the analysis or specific clinical priorities. However, as this example highlights, the interest in fuzzification typically concerns only a subset of units rather than all of them: for many units, the membership to a specific cluster may be sufficiently strong that a hard assignment is appropriate. This motivates the use of fuzzy clustering methods with the so-called \say{high contrast} property \cite{RTK:1995}, where hard and soft assignments coexist within the same framework. 

In real-world applications, data often contain measurement errors, anomalies, or, more generally, outliers, some of which may actually correspond to behaviors of genuine interest in the data. In Robust Statistics \cite{H:1981,MMYS:2019}, outliers have been traditionally defined as cases (or units) that do not follow the pattern of the majority of the data, and therefore referred to as casewise outliers. Specifically, given an $(n \times p)$ data matrix $\vec{X}$, where $p$ variables are measured on $n$ units, we may distinguish between \textit{cases}, represented by the rows $\vec{x}_i = (x_{i1}, \ldots, x_{ip})$, $i = 1, \ldots, n$, of $\vec{X}$, and \textit{cells}, which correspond to the single measurements  $x_{ij}$ of the $j$-th variable ($j = 1, \ldots, p$) for the $i$-th unit ($i = 1, \ldots, n$). With this notation, the common assumption in Robust Statistics is that an entire case $\vec{x}_i$ may be considered atypical.

\begin{figure}[t]
    \centering
    \includegraphics[width=0.9\linewidth]{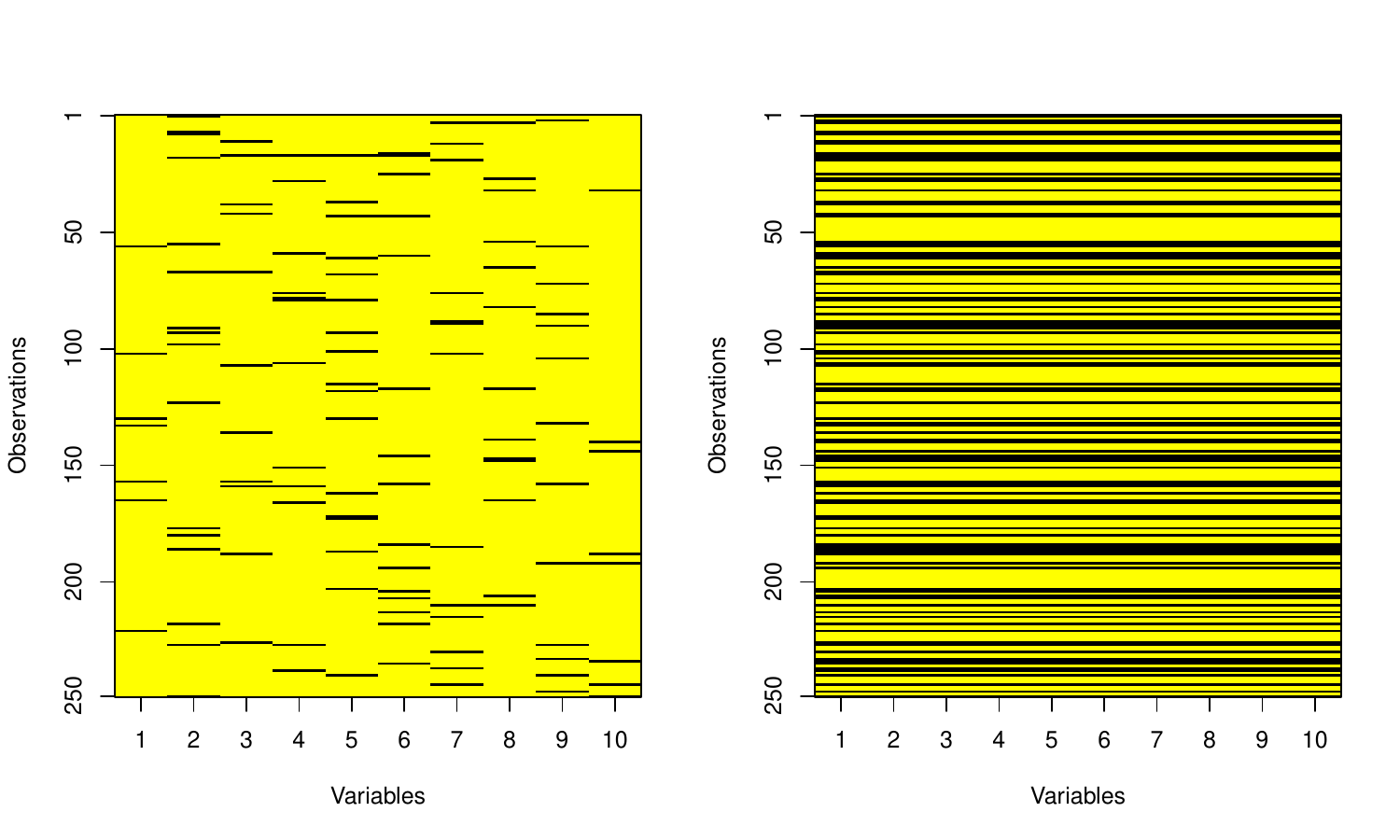}
    \caption{Data matrix with 250 units, $10$ variables, and $5\%$ of contaminated cells shown in black (left); corresponding rows in black that would be trimmed by a casewise robust method due to containing at least one outlying cell (right).}
    \label{fig: outlierparadigms}
\end{figure}

The presence of contaminating cases $\vec{x}_i$ -- even in very small numbers -- may cause methods traditionally applied to hard and fuzzy clustering to fail dramatically, resulting in the identification of clusters of very limited practical interest. This well-known phenomenon has motivated the development of numerous robust clustering techniques, from both hard and fuzzy perspectives (see \cite{DK:2002,GEGMMI:2010,BD:2012,GEGMMIH:2015} and references therein), essentially from a casewise point of view. Until the first decade of the 2000s, this was indeed the dominant approach for dealing with outliers. However, the increasing dimensionality of the data makes it reasonable to assume that units may exhibit only a few outlying cells $x_{ij}$, while the remaining (non-outlying) cells may still contain reliable information for data analysis. This has led to the development of robust approaches aimed at addressing cellwise contamination \cite{AVAYZ:2009,RR:2026}, in which only single entries $x_{ij}$ of a data matrix, rather than entire rows $\vec{x}_i$, are assumed to be contaminated by arbitrary values. Under this type of contamination, discarding, assigning to a noise component or down-weighting entire cases, as done by traditional casewise robust methods for fuzzy clustering, can result in substantial information loss, even when the overall proportion of atypical cells is very small. Additionally, note that, as the number of variables $p$ increases, cellwise contamination is likely to affect nearly all cases, often through a single outlying value in many of them, causing (even robust) fuzzy clustering to fail (see Figure \ref{fig: outlierparadigms}). Moreover, in the cellwise paradigm, it is possible to take advantage of the reliable information within a case to obtain a predicted value for each cell and compare it with the observed one. If the former deviates from the latter, the cells are flagged as contaminated, and corrected through imputation for the parameter estimation, rather than discarding or down-weighing them. Motivated by the cellwise robust approach, Raymaekers and Rousseeuw introduced cellMCD \cite{RR:2023} for estimating the location and scatter parameters in single-population problems under cellwise contamination, building on the casewise-robust Minimum Covariance Determinant (MCD) estimator \cite{R:1984, R:1985}. The cellMCD approach has been recently extended to mixture modeling in cellGMM \cite{ZACetal:2025} (see also \cite{PWF:2025}). However, none of these proposals is designed to addresses cellwise contamination in the fuzzy clustering framework.

The main contribution of this paper is a novel approach, called cellFCLUST, which combines three key features: (i) cellwise outlier detection and robust parameter estimation; (ii) fuzzy clustering; and (iii) the flexibility of tuning hard and soft assignments for units (the previously commented \say{high contrast} property \cite{RTK:1995}). CellFCLUST arises as a cellwise extension of the F-TCLUST casewise robust fuzzy clustering method introduced in  \cite{FGEMI:2013b}. F-TCLUST aims to produce a fuzzy partition of the units and to estimate the location and scatter parameters of the clusters using a trimmed classification maximum likelihood approach. The latter relies on the individual contribution to the objective function to identify a subset of cases that are retained as reliable for parameter estimation, while detecting and trimming (removing) the most anomalous cases. Previous related methods can be found in \cite{KKD:1996}, where fuzzy $c$-means \cite{D:1974} is robustified via trimming, and in \cite{D:1991}, in which outliers are not trimmed but instead modeled as belonging to a \say{noise} cluster. Further approaches for achieving robustness, such as possibilistic clustering \cite{KK:2002} or models with heavy-tailed components \cite{CV:2008}, are, to the best of our knowledge, just focused on the casewise robustness paradigm.

The novel proposal introduced in this work assumes a Gaussian distribution for the clusters and allows cluster covariance matrices to have distinct elliptical shapes, similar to F-TCLUST. The dependence relationships among variables may therefore vary across clusters and can be used for detecting and correcting outlying cells. Specifically, by setting a proportion of cells to flag for the variables, cellFCLUST identifies and imputes them using the information contained in the remaining reliable cells of each case, making the use of an algorithm inspired by the Expectation-Maximization (EM, \cite{DLR:1977}) one suitable for performing this correction. As a result of the imputation, \textit{all} units are assigned to clusters with a certain degree of fuzziness, unlike trimming approaches, which entirely discard cases, or noise clustering methods, where outliers are not grouped with regular data. Following this rationale, cellFCLUST can naturally handle missing information, which often occurs in real data sets. Missing entries usually affect cells $x_{ij}$ of a data matrix, but their positions $(i,j)$ are known in advance, unlike cellwise outlying values; both types of cells are referred to as \textit{unreliable} in this paper. Several fuzzy clustering methods have been extended to cope with incomplete data sets, such as fuzzy $c$-means with incomplete information \cite{HB:2001} (see \cite{JG:2023} for an overview); however, none of them can deal with cellwise contamination, in which the position of the outlying cells is completely unknown. Moreover, in cellFCLUST, hard and soft assignments coexist within the same framework, and the proportions of each type of membership can be tuned by adjusting relevant parameters. This represents a novel feature compared to cellGMM, where this property cannot be achieved.

The paper is organized as follows. A brief review of F-TCLUST and its key features is provided in Section \ref{sec: literature} to help the reader follow its extension from casewise to cellwise robust fuzzy clustering. Section \ref{sec: method} formally introduces cellFCLUST, including its parameter estimation and algorithm. A simulation study was conducted to evaluate the performance of cellFCLUST in comparison with alternative methods for fuzzy clustering with and without outlier detection; the results are reported in Section \ref{sec: simulations}. In Section \ref{sec: paramchoice}, we provide guidance on the tuning parameter selection through examples that illustrate their impact on cluster recovery and outlier detection. Two real data applications from different fields are presented in Section \ref{sec: applications}. Finally, Section \ref{sec: conclusions} concludes the paper with a discussion on the proposed methodology, as well as potential directions for future developments in the cellwise clustering literature. The notation used throughout the paper is summarized in \ref{sec: notation}, facilitating the reading of the methodological sections.

\section{Background: F-TCLUST for casewise robust fuzzy clustering}\label{sec: literature}
Fuzziness in clustering was first introduced through the fuzzy $k$-means (or fuzzy $c$-means \cite{D:1974}, here denoted as FKM), which assigns units to clusters based on their squared Euclidean distance from the cluster mean vectors, often called centroids. Specifically, let $\vec{X} = [\vec{x}_{1}, \ldots, \vec{x}_{n}]^\prime$ be an $(n \times p)$ data matrix. The FKM objective function to be minimized is
\begin{equation}\label{eqn: fkm}
    J_{\text{FKM}}(\vec{U}, \{\mean_{k}\}_{k = 1}^{K}) = \sum _{i = 1}^{n} \sum _{k = 1}^{K} u_{ik}^{m}\left\|\vec{x}_{i} - \mean_{k} \right\|^{2},
\end{equation}
where $m \geq 1$ is the fuzzifier tuning parameter (a larger $m$ increases the degree of fuzziness, allowing more overlap between clusters and making cluster memberships less crisp), $\vec{U}$ is an $(n \times K)$ membership matrix with $u_{ik} \in [0, 1]$ denoting the membership degree of the $i$-th unit to the $k$-th cluster and such that $\sum_{k = 1}^{K} u_{ik} = 1$, for $i = 1, \ldots, n$, and $\mean_{1}, \ldots, \mean_{K}$ are cluster centroids in $\mathbb{R}^{p}$.

One of the main disadvantages of FKM, as in classical hard $k$-means, is its preference for spherical clusters with the same covariance structure, which makes it poorly suited for detecting more general types of clusters. Several procedures have been proposed to relax the isotropic assumption (see \cite{GK:1979,TKR:1991,RTK:1996,GG:1989}, among others). These methods can be robustified through casewise trimming. In particular, in this section we focus on F-TCLUST \cite{FGEMI:2013b}, which integrates fuzzification into a classification maximum likelihood approach, accommodates different cluster shapes, and removes outlying cases from parameter estimation via impartial trimming. The term \say{impartial} means that the method itself detects the cases to trim, avoiding any user intervention in their identification; the user only needs to set the fraction $\alpha$ of casewise outliers. The objective function of F-TCLUST to be maximized is 
\begin{equation}\label{eqn: ftclust}
    J_{\text{F-TCLUST}}(\vec{U}, \{\pi_k,\mean_{k}, \sigu_{k}\}_{k = 1}^{K}) = \sum _{i = 1}^{n} \sum _{k = 1}^{K} u_{ik}^{m} \log(\pi_k\phi_{p}(\vec{x}_{i}; \mean_{k}, \sigu_{k})), 
\end{equation}
where $\phi_{p}(\vec{x}_{i}; \mean_{k}, \sigu_{k}) = (2\pi)^{- p/2} \lvert \sigu_{k}\rvert^{-1/2} \exp \big( -\frac{1}{2}(\vec{x}_{i} - \mean_{k})^{\prime} \sigu_{k}^{-1} (\vec{x}_{i} - \mean_{k}) \big)$ is the probability density function evaluated at $\vec{x}_i \in \mathbb{R}^p$ of a $p$-variate normal distribution with $p$-dimensional mean vectors $\mean_{k}$ and $(p \times p)$ positive definite covariance matrices $\sigu_k$ holding the dependence relationships among variables within clusters. The considered weights $\pi_k$ are positive numbers such that $\sum_{k = 1}^K \pi_k = 1$. The membership degrees $u_{ik} \in [0, 1]$ satisfy the following constraints
\begin{equation*}
    \sum_{k = 1}^{K} u_{ik} = 1 \text{ when } i \in \mathcal{I}  \text{ and } \sum_{k = 1}^{K} u_{ik} = 0  \text{ when } i \notin \mathcal{I},
\end{equation*}
where $\mathcal{I}$ is the subset of not trimmed cases, with $\#\mathcal{I}=\lceil n(1-\alpha) \rceil$, i.e., $\alpha$ corresponds to the trimming level.

Since we deal with a maximum likelihood-type problem in the clustering framework, unboundedness may potentially arise when maximizing (\ref{eqn: ftclust}). To overcome that issue, F-TCLUST is also subject to the following eigenvalue-ratio constraint
\begin{equation}\label{eqn: eigenratio}
    \dfrac{\max\limits_{k = 1, \ldots, K}\max\limits_{j = 1, \ldots, p} \lambda_{j}(\sigu_k)}{\min\limits_{k = 1, \ldots, K}\min\limits_{j = 1, \ldots, p} \lambda_{j}(\sigu_k)} \leq c  
\end{equation}
where $c \geq 1$ is a fixed constant, and $\lambda_{1}(\sigu_k), \ldots, \lambda_{p}(\sigu_k)$ are the $p$ eigenvalues of $\sigu_k$. Constraint (\ref{eqn: eigenratio}) ensures the maximization of $J_{\text{F-TCLUST}}$ in (\ref{eqn: ftclust}) as a well-posed problem, preventing its unboundedness (which may occur, for example, by taking $u_{i1} = 1$, $\pi_1 = 1$, $\mean_{1} = \vec{x}_i$, and letting $|\sigu_1|\downarrow 0$). As the constant $c$, which constrains the ratio between the largest and smallest eigenvalues of the covariance matrices within and across clusters, decreases, the cluster configurations become increasingly similar to each other and closer to sphericity, ultimately reaching spherical and equally dispersed clusters when $c = 1$. The eigenvalue-ratio constraint in (\ref{eqn: eigenratio}) is also an important tool to avoid detecting so-called \say{spurious} clusters with little statistical relevance, typically driven by a few nearly collinear cases with $|\sigu_k|\simeq 0$ (see \cite{GEGGIMI:2018} for details). The detection of spurious solutions constitutes an additional source of lack of robustness in (fuzzy) clustering.

F-TCLUST extends the hard clustering approach known as TCLUST \cite{GEGMMI:2008} to the robust fuzzy clustering framework. Since TCLUST is based on the classification trimmed likelihood, F-TCLUST reduces to it when $m = 1$, so that the associated $u_{ik}$ obtained by maximizing (\ref{eqn: ftclust}) take values in $\{0, 1\}$ -- not in the interval -- and sum to $1$ for the non-trimmed cases and to $0$ for the trimmed ones. A version of TCLUST for robust mixture modeling was introduced in \cite{GEGMY:2014}, where entire cases can be again trimmed. Note that both approaches -- TCLUST and its version for robust mixture fitting -- can be used to compute posterior probabilities according to the robustly estimated parameters, thus providing \say{soft assignments} for the non-trimmed cases. However, crucial differences from the membership degrees of F-TCLUST exist. First, F-TCLUST explicitly embeds fuzzy memberships into the objective function (\ref{eqn: ftclust}), treating them as parameters to be estimated jointly with the cluster weights, mean vectors and covariance matrices. Second, as noted in Section \ref{sec: intro}, the fuzzifier parameter $m$ allows the user to control the degree of fuzziness.

\section{CellFCLUST for cellwise robust fuzzy clustering}\label{sec: method}
In this section, we introduce a novel fuzzy clustering method that handles cellwise outliers. The proposal, called \textit{cellwise Fuzzy Clustering} (cellFCLUST), is formulated within a maximum likelihood framework and can be viewed as the F-TCLUST \say{counterpart} for dealing with outlying cells rather than entire outlying cases. The methodology is illustrated in Section \ref{subsec: model}, and a detailed description of the algorithm for its parameter estimation is provided in Section \ref{subsec: algo}.

\subsection{Methodology}\label{subsec: model}
The key feature of cellFCLUST is detecting unreliable cells and imputing them, rather than discarding them from the parameter estimation process. Identification is performed variable-by-variable by comparing the contribution of each unit to the objective function when its value is considered contaminated or not, while keeping the reliable values for the other variables. The unreliable cells, both contaminated or missing, are imputed by leveraging the reliable information per unit and cluster-specific relationships among variables. We denote by $\vec{W} = [\vec{w}_{1}, \ldots, \vec{w}_{n}]^\prime$ the $(n \times p)$ cellwise indicator matrix, where $w_{ij} = 1$ if the cell is reliable, and $w_{ij} = 0$ otherwise. Differently from the outlying cells, as mentioned, the $(i, j)$-positions of the missing values in a data matrix are known, and therefore the corresponding zeros into $\vec{W}$ can be set a priori. According to $\wi$, we partition $\vec{x}_{i}$ into $\vec{x}_{i[\wi]}$ and $\vec{x}_{i[\wi^{c}]}$, where $\wi^{c} = \vec{1}_{p} - \wi$ with $\vec{1}_{p}$ being the unitary vector of dimension $p$. These two sub-vectors represent the reliable and unreliable cells for the $i$-th unit, respectively.

The cellFCLUST objective function to maximize is
\begin{equation}
\begin{split}
J_{\text{cellFCLUST}} (\vec{W}, \vec{U}, \{\pi_k, \mean_{k}, \sigu_{k}\}_{k = 1}^{K}) &= \sum_{i = 1}^{n}\sum_{k = 1}^{K} u_{ik}^{m} \\
&\quad \times \log(\pi_k \phi_{p[\wi]}(\vec{x}_{i[\wi]}; \meanki, \sigukii)),
\end{split}
\label{eqn: obj}
\end{equation}
where both $\mean_{1}, \ldots, \meank$ and $\sigu_{1}, \ldots, \siguk$ are restricted in (\ref{eqn: obj}) to the sub-vectors and the sub-matrices, respectively, corresponding to the reliable cells for the $i$-th unit, and $p[\wi]$ is their number. As in F-TCLUST, $\vec{U}$ represents the $(n \times K)$ membership matrix, the weights $\pi_k$ are positive and such that $\sum_{k = 1}^K \pi_k = 1$, and the fuzzifier parameter $m$ must be greater than $1$ to obtain strictly fuzzy memberships, since $m = 1$ results in crisp $0$-$1$ assignments for every unit, even though $u_{ik} \in [0, 1]$, as shown in \cite{RTK:1995}. The maximization of (\ref{eqn: obj}) is subject to both the eigenvalue-ratio constraint in (\ref{eqn: eigenratio}) and the following one
\begin{equation}
    \sum_{i =1}^{n} w_{ij} = h, \; \text{for} \; j = 1, \ldots, p, \label{eqn: constrW} 
\end{equation}
which imposes that $h  = \left\lceil (1-\alpha) n \right\rceil$ cells per variable are reliable. Here, $\alpha$ represents the proportion of flagged cells and it should be at most $0.25$ to guarantee that pairs of variables overlap for some units, allowing their covariances to be safely computed. Note that F-TCLUST, instead of considering constraint (\ref{eqn: constrW}), assumed $w_{ij} = 0$ for all $j = 1, \ldots, p$ for the trimmed cases $\vec{x}_i$, without any imputation, and $w_{ij} = 1 \;$ for all $j = 1, \ldots, p$ for the untrimmed ones.
We could have considered the maximization of a version of (\ref{eqn: obj}) in which the weights $\pi_k$ are removed; the effect of this removal will be briefly discussed in Section \ref{sec: paramchoice}.

Moreover, it is worth noting that casewise methodologies, such as F-TCLUST, systematically require trimming a large fraction of units located in the tails of the distribution in order to eliminate even a small fraction of cellwise contamination. Consequently, this trimming causes covariance matrix estimators to be biased downward due to the removal of part of the clusters' variability. For this reason, covariance matrix estimators resulting from casewise trimming typically require some form of adjustment. This is, for instance, the case of MCD-based covariance matrix estimators, which need to be multiplied by a consistency factor greater than one \cite{CH:1999}. In contrast, cellFCLUST limits the amount of information discarded in the tails of the cluster components, making such a correction of the estimated covariance matrices much less critical.

\vspace{0.5cm}
\subsection{An EM-inspired algorithm for cellFCLUST}\label{subsec: algo}
We aim to solve the maximization problem of the objective function in (\ref{eqn: obj}) via an EM-inspired algorithm that extends the one for handling missing data in the model-based clustering framework \cite{GJ:1994}, with an additional step for cellwise outlier detection. Specifically, the algorithm is composed of four alternating steps. After the initialization in \textit{Step 0}, its key feature lies in \textit{Step 1}, where $n - h$ cells per variable considered as potentially contaminated are flagged, and the corresponding positions in $\vec{W}$ are set to zero. Conditional on the previous update of $\vec{W}$, the membership values in $\vec{U}$ are updated in \textit{Step 2} so as to decrease the objective function and directly satisfy the ``hard contrast'' property. Following the EM rationale and treating unreliable cells as missing entries, the parameters of the distribution for the missing part of the data, necessary for overall parameter estimation, are obtained in \textit{Step 3}, which also leads to the imputation of potentially contaminated cells rather than their removal. Finally, \textit{Step 4} uses that information to update the weights $\pi_k$, the mean vectors $\mean_{k}$, and the covariance matrices $\sigu_{k}$, $ k = 1, \ldots, K$ (the eigenvalue-ratio constraint in (\ref{eqn: eigenratio}) must also be enforced by truncating their eigenvalues). A detailed description of these steps is provided below, along with the pseudocode for the entire procedure given in Algorithm \ref{algorithm}.

\begin{algorithm}
\scriptsize
\SetAlgoLined
\caption{cellFCLUST} \label{algorithm}
\KwIn{$\vec{X}$, $K$, $\alpha$, $c$, $m$, $\epsilon$, \textit{maxiter}}

$t \leftarrow 0$; $J_0 \leftarrow -\infty$;

\SetKw{KwIniz}{Initialization:}
\SetKwRepeat{KwRepeat}{Repeat}{until}
\KwIniz{$\vec{W}^{(0)}$, $\boldsymbol{\theta}^{(0)} = \{\pi_{k}^{(0)}, \mean_{k}^{(0)}, \sigu_{k}^{(0)} \}_{k = 1}^{K}$ and $\vec{U}^{(0)}$ \hfill{\color{red}$\triangleright$ See \textit{Step 0}}}

\KwRepeat{$J_t - J_{t-1} < \epsilon$ or $t = $ maxiter}{

{\bf Update}  $\vec{W}$ (conditional on $\vec{U}$ and $\boldsymbol{\theta}$): \hfill {\color{red}$\triangleright$ See \textit{Step 1}}

\For{$j = 1,\ldots,p$}{
$\bullet$ Compute 
\begin{equation*}
\tag{\textcolor{red}{Eq. \ref{eqn: delta}}}
\Delta_{ij}  \leftarrow - \dfrac{1}{2} \sum_{k = 1}^{K} \big(u_{ik}^{(t)}\big)^{m}  \Bigg[\log(2\pi) + \log(C_{ij(k)}) + \dfrac{(x_{ij} - \hat{x}_{ij(k)})^2}{C_{ij(k)}} \Bigg], 
\end{equation*}
where $\hat{x}_{ij(k)} = \mean_{k[j]}^{(t)} + \sigu_{k[j, \wi]}^{(t)} (\sigu_{k[\wi, \wi]}^{(t)})^{-1}(\vec{x}_{i[\wi]} - \mean_{k[\wi]}^{(t)})$ and $C_{ij(k)} = \sigu_{k[j, j]}^{(t)} - \sigu_{k[j, \wi]}^{(t)} (\sigu_{k[\wi, \wi]}^{(t)})^{-1} \sigu_{k[\wi, j]}^{(t)}$ for $\wi=\wi^{(t)}$.

$\bullet$ Set
\begin{equation*}
w_{ij}^{(t+1)} \leftarrow \begin{cases}
    1 \quad \text{if $\Delta_{ij} \geq \Delta_{(n-h)j}$}\\
    0 \quad \text{if $\Delta_{ij} < \Delta_{(n-h)j}$}\\
\end{cases},\text{ for }i = 1, \ldots, n,
\end{equation*}
where $\Delta_{(1)j} \leq \Delta_{(2)j} \leq \ldots \Delta_{(n)j}$ are the non-decreasing $\Delta_{ij}$ values.
}

$\wi \leftarrow \wi^{(t+1)}$ for $i = 1, \ldots, n$

\noindent{\bf Update} $\vec{U}$ (conditional on $\vec{W}$ and $\boldsymbol{\theta}$):\hfill {\color{red}$\triangleright$ See \textit{Step 2}}
\begin{equation*}
\tag{\textcolor{red}{Eq. \ref{eqn: membership}}}
u_{ik}^{(t+1)} \leftarrow 
\begin{cases}
    I\Big\{f_{ik} = \underset{k^{\prime} = 1, \ldots, K}{\max}  f_{ik^{\prime}} \Big\} & \text{if} \, \underset{k^{\prime} = 1, \ldots, K}{\max}  f_{ik^{\prime}}  \geq 1 \\
     \Bigg( \sum_{k^{\prime} = 1}^{K} \Bigg( \dfrac{\log(f_{ik})}{\log(f_{ik^{\prime}})} \Bigg)^{\frac{1}{m-1}} \Bigg)^{-1} & \text{if} \, \underset{k^{\prime} = 1, \ldots, K}{\max}  f_{ik^{\prime}} < 1
\end{cases},
\end{equation*}
where $f_{ik} = \pi_{k}^{(t)} \phi_{p[\wi ]}\big(\vec{x}_{i[\wi]}; \mean_{k[\wi]}^{(t)} , \sigu_{k[\wi, \wi]}^{(t)}\big)$.

$u_{ik} \leftarrow u_{ik}^{(t+1)}$ for $i = 1, \ldots, n$, $k = 1, \ldots, K$
    
{\bf Update} $\boldsymbol{\theta}$ (conditional on $\vec{U}$ and $\vec{W}$):\hfill {\color{red}$\triangleright$ See \textit{Steps 3-4}}

\For{$k=1,\ldots,K$}{

$\widehat{\vec{x}}_{i[\wi^{c}](k)}  \leftarrow \mean_{k[\wi^{c} ]}^{(t)}  + \sigu_{k[\wi^{c}, \wi]}^{(t)} (\sigu_{k[\wi, \wi]}^{(t)} )^{-1} (\vec{x}_{i[\wi]} - \mean_{k[\wi]}^{(t)} )$ for $i=1,...,n$

$\Tilde{\vec{x}}_{i(k)} \leftarrow (\vec{x}_{i[\wi]}, \widehat{\vec{x}}_{i[\wi^{c}](k)})$ for $i = 1, \ldots, n$

$\bullet$ Compute
\begin{align*}
&\pi_{k}^{(t+1)} \leftarrow \dfrac{\sum_{i = 1}^{n} u_{ik}^{m}}{\sum_{i = 1}^{n} \sum_{k = 1}^{K} u_{ik}^{m}}, \tag{\textcolor{red}{Eq. \ref{eqn: csize_est}}}\\
&\mean_{k}^{(t+1)} \leftarrow \dfrac{\sum_{i = 1}^{n} u_{ik}^{m} \, \Tilde{\vec{x}}_{i(k)}}{\sum_{i = 1}^{n} u_{ik}^{m}}, \tag{\textcolor{red}{Eq. \ref{eqn: mu_est}}}\\
&\sigu_{k}^{(t+1)} \leftarrow \dfrac{\sum_{i = 1}^{n} u_{ik}^{m} \Big[(\Tilde{\vec{x}}_{i(k)} - \mean_{k}^{(t+1)}) (\Tilde{\vec{x}}_{i(k)} - \mean_{k}^{(t+1)})^{\prime} + \vec{C}_{i[\wi^{c},\wi^{c}](k)} \Big]}{\sum_{i = 1}^{n} u_{ik}^{m}}, \tag{\textcolor{red}{Eq. \ref{eqn: sigma_est}}} 
\end{align*}
where $\vec{C}_{i[\wi^{c},\wi^{c}](k)} = \sigu_{k[\wi^{c} , \wi^{c}]}^{(t)} - \sigu_{k[\wi^{c}, \wi]}^{(t)} (\sigu_{k[\wi, \wi]}^{(t)})^{-1} \sigu_{k[\wi, \wi^{c}]}^{(t)}$
(truncation of the eigenvalues of $\sigu_{k}^{(t+1)}$ depending on $c$ may be required).
}

$t \leftarrow t + 1$

\textbf{Compute objective function:} $J_t \leftarrow J_{\text{cellFCLUST}} (\vec{W}^{(t)}, \vec{U}^{(t)}, \boldsymbol{\theta}^{(t)})$
}
\end{algorithm}

\begin{enumerate}
    \item[\textit{Step 0.}] Initial solutions for the parameters are obtained using several applications of TCLUST method in \cite{GEGMMI:2008}. Specifically, TCLUST is applied to each variable and pairs of variables to initialize $\vec{W}$, and then on random subsets of variables to achieve feasible initializations for $\pi_{k}, \mean_{k}$, and $\sigu_{k}, k = 1, \ldots, K$ (see \cite{ZACetal:2025} for details). Accordingly, an initial solution for $\vec{U}$ is computed as described in \textit{Step 2}.
    
    \item[\textit{Step 1.}] Given the current parameters, the membership matrix $\vec{U}$, and the actual configuration of $\vec{W}$, we update the latter column-by-column. Let consider the objective function in Eq. (\ref{eqn: obj}) as the sum of individual contributions, i.e. $J_{\text{cellFCLUST}}(\vec{W}, \vec{U}, \{\pi_{k}, \mean_{k}, \sigu_{k}\}_{k = 1}^{K}) = \sum_{i = 1}^{n} J_{\text{cellFCLUST}}^{(i)}(\wi,$ $\vec{u}_{i}, \{\pi_{k}, \mean_{k}, \sigu_{k}\}_{k = 1}^{K})$. It can be seen that     
    \begin{align}\label{eqn: contri}
        &J_{\text{cellFCLUST}}^{(i)} (\wi, \vec{u}_{i}, \{\pi_{k}, \mean_{k}, \sigu_{k}\}_{k = 1}^{K}) = \sum_{k = 1}^{K} u_{ik}^{m} \log(\pi_{k}) -\dfrac{1}{2} \sum_{k = 1}^{K} u_{ik}^{m} \Big[p[\wi] \nonumber \\
        &\log(2\pi) + \log \lvert \sigu_{k[\wi, \wi]} \rvert + (\vec{x}_{i[\wi]} - \mean_{k[\wi]})^{\prime} \sigu_{k[\wi, \wi]}^{-1} (\vec{x}_{i[\wi]} - \mean_{k[\wi]}) \Big],
    \end{align}
    where the first addend does not depend on the elements of $\vec{W}$ and can thus be ignored. For each column $j$, we compare the individual contribution in (\ref{eqn: contri}) when we consider its $i$-th element reliable ($w_{ij} = 1$) or contaminated ($w_{ij} = 0$) as follows 
    \begin{align}
        \Delta_{ij} &= J_{\text{cellFCLUST}}^{(i)}(\wi, \vec{u}_{i}, \{\pi_{k}, \mean_{k}, \sigu_{k}\}_{k = 1}^{K} \lvert w_{ij} = 1) \nonumber \\
        &\quad - J_{\text{cellFCLUST}}^{(i)}(\wi, \vec{u}_{i}, \{\pi_{k}, \mean_{k}, \sigu_{k}\}_{k = 1}^{K} \lvert w_{ij} = 0) \nonumber\\
        &= - \dfrac{1}{2} \sum_{k = 1}^{K} u_{ik}^{m} \Bigg[\log(2\pi) + \log(C_{ij(k)}) + \dfrac{(x_{ij} - \hat{x}_{ij(k)})^2}{C_{ij(k)}} \Bigg], \label{eqn: delta}
    \end{align}
    where $\hat{x}_{ij(k)} = \mean_{k[j]} + \sigu_{k[j, \wi]} (\sigu_{k[\wi, \wi]})^{-1}(\vec{x}_{i[\wi]} - \mean_{k[\wi]})$ and $C_{ij(k)} = \sigu_{k[j, j]} - \sigu_{k[j, \wi]} (\sigu_{k[\wi, \wi]})^{-1} \sigu_{k[\wi, j]}$ are two scalars denoting the expectation and variance, respectively, of the cell associated with the $j$-th variable in the $i$-th unit, conditional on the other reliable cells for the $i$-th unit and given the parameters of the $k$-th cluster. The proof for the final expression of (\ref{eqn: delta}) is provided in the Supplementary Material.
    
    By sorting the $\Delta_{ij}$ values in an increasing order, i.e., $\Delta_{(1)j} \leq \Delta_{(2)j} \leq \cdots \leq \Delta_{(n)j}$, we flag as unreliable the cells with indexes $\{i: \Delta_{ij} < \Delta_{(n-h)j} \}$, where $h = \left\lceil (1-\alpha) n \right\rceil$, and set the corresponding elements of the $j$-th column of $\vec{W}$ to zero. Equivalently, the cells with $\{i: \Delta_{ij} \geq \Delta_{(n-h)j} \}$ are considered reliable, which is denoted by ones into the $j$-th column of $\vec{W}$. Therefore, the supposedly contaminated cells are those whose $\Delta_{ij}$ is negative or small.
    
    If missing values occur for the $j$-th variable, the corresponding cells of $\vec{W}$ are set to zero. Consequently, the $\Delta_{ij}$ values are computed only on the observed cells. This approach may result in varying proportions of zeros across variables, depending on the extent of missingness they contain. However, using $\alpha$ to account for both the contaminated and missing cells -- and therefore grounding $h$ on $n$, independently of the number of missing values -- could potentially lead to an incorrect identification of the truly contaminated cells. In general, we recommend excluding variables with a high proportion of missing data, particularly when moderate contamination is expected; specifically, the proportion of missing \textit{and} outlying values must not exceed $0.25$, as also mentioned in Section \ref{subsec: model}.

    \item[\textit{Step 2.}] Given the cellwise indicator matrix and the current parameters, the membership values are updated as
     \begin{equation}\label{eqn: membership}
     u_{ik} = \begin{cases}
     I\Big\{f_{ik} = \underset{k^{\prime} = 1, \ldots, K}{\max}  f_{ik^{\prime}} \Big\} & \text{if} \, \underset{k^{\prime} = 1, \ldots, K}{\max}  f_{ik^{\prime}}  \geq 1 \\
     \Bigg( \sum_{k^{\prime} = 1}^{K} \Bigg( \dfrac{\log(f_{ik})}{\log(f_{ik^{\prime}})} \Bigg)^{\frac{1}{m-1}} \Bigg)^{-1} & \text{if} \, \underset{k^{\prime} = 1, \ldots, K}{\max}  f_{ik^{\prime}} < 1
     \end{cases},
    \end{equation}
    where $$f_{ik} = \pi_{k} \phi_{p[\wi]}(\vec{x}_{i[\wi]}; \mean_{k[\wi]}, \sigu_{k[\wi, \wi]}),$$ and $I\{\cdot\}$ represents the indicator function. In the first case, the $i$-th unit is fully assigned to the $k$-th cluster, while in the second case, the assignment is fuzzy. This is a desirable property of a fuzzy clustering approach since not all units necessarily require a soft assignment. Some of them, especially those in the \say{core} of the clusters, can be unequivocally assigned and, thus, this update of $u_{ik}$ automatically results in the \say{high contrast} property in \cite{RTK:1995}.
    
    In order to justify the update of $u_{ik}$ in (\ref{eqn: membership}), note that the maximization of the objective function (\ref{eqn: obj}) is equivalent to the minimization of $\sum_{i = 1}^n \sum_{k = 1}^K u_{ik}^m D_{ik}$, with $D_{ik} = -\log(f_{ik})$. For fixed $\vec{W}$ and $\{\pi_{k}, \mean_{k}, \sigu_{k}$ $\}_{k = 1}^{K}$, the function to be minimized is separable with respect to the index $i$, and hence it suffices to determine how to minimize
    \begin{equation}\label{eqn: aux_membership}
    \sum_{k=1}^K u_{ik}^m D_{ik}
    \end{equation}
    with respect to $u_{i1}, \ldots, u_{iK}$, for a fixed $i$. If $\max_{k' = 1, \ldots, K} f_{ik'} < 1$, that is, whenever $D_{ik} > 0$ for $k = 1, \ldots, K$, then (\ref{eqn: aux_membership}) is a strictly convex function and a unique global minimizer exists. Standard Lagrange multiplier theory shows that the values $u_{ik}$ must be proportional to $D_{ik}^{-1/(m-1)}$, as traditionally occurs in FKM. When combined with the constraint $\sum_{k = 1}^K u_{ik} = 1$, updating $u_{ik}$ results in fuzzy memberships, that is the second case in (\ref{eqn: membership}). On the other hand, if $\max_{k' = 1,\cdots,K} f_{ik'} \geq 1$ and we define $k^\ast = \arg\max_{k' = 1, \ldots, K} f_{ik'}$, it trivially follows that
    \begin{equation*}
        \sum_{k = 1}^K u_{ik}^m \log(f_{ik}) \leq \sum_{k = 1}^K u_{ik}^m \log(f_{ik^\ast}) \leq \log(f_{ik^\ast}) \sum_{k = 1}^K u_{ik} = \log(f_{ik^\ast}),
    \end{equation*}
    and, consequently, the largest decrease in (\ref{eqn: aux_membership}) occurs when $u_{ik^\ast} = 1$ and $u_{ik} = 0$ for all $k \neq k^\ast$. Therefore, the complete update for the membership values in (\ref{eqn: membership}) is demonstrated.
    
    \item[\textit{Step 3.}] Given the cellwise indicator matrix and the current parameters, we compute the parameters of the distribution for the unreliable data $\vec{x}_{i[\wi^{c}]}$ conditional on the observed $ \vec{x}_{i[\wi]}$ as
    \begin{align}
    &\mean_{k[\wi^{c} \lvert \wi]} = \mean_{k[\wi^{c} ]} + \sigu_{k[\wi^{c}, \wi]}^{} (\sigu_{k[\wi, \wi]})^{-1} (\vec{x}_{i[\wi]} - \mean_{k[\wi]}) := \widehat{\vec{x}}_{i[\wi^{c}](k)}, \label{eqn: mean_miss} \\
    &\sigu_{k[\wi^{c}, \wi^{c} \lvert \wi]} = \sigu_{k[\wi^{c} , \wi^{c}]} - \sigu_{k[\wi^{c}, \wi]} (\sigu_{k[\wi, \wi]})^{-1} \sigu_{k[\wi, \wi^{c}]} := \vec{C}_{i[\wi^{c},\wi^{c}](k)}. \label{eqn: sigma_miss} 
    \end{align}
    Further details on the derivation of these parameters are provided in the Supplementary Material.
    \item[\textit{Step 4.}] Given the membership matrix and considering the completed data $\Tilde{\vec{x}}_{i(k)} = (\vec{x}_{i[\wi]}, \widehat{\vec{x}}_{i[\wi^{c}](k)})$, for $i = 1, \ldots, n$ and $k = 1, \ldots, K$, where both contaminated and missing values are imputed through \textit{Step 3}, we update the parameters as follows
    \begin{align}
        &\pi_{k} = \dfrac{\sum_{i = 1}^{n} u_{ik}^{m}}{\sum_{i = 1}^{n} \sum_{k = 1}^{K} u_{ik}^{m}}, \label{eqn: csize_est} \\
        &\mean_{k} = \dfrac{\sum_{i = 1}^{n} u_{ik}^{m} \, \Tilde{\vec{x}}_{i(k)}}{\sum_{i = 1}^{n} u_{ik}^{m}}, \label{eqn: mu_est} \\
        &\sigu_{k} = \dfrac{\sum_{i = 1}^{n} u_{ik}^{m} \Big[(\Tilde{\vec{x}}_{i(k)} - \mean_{k}) (\Tilde{\vec{x}}_{i(k)} - \mean_{k})^{\prime} + \vec{C}_{i[\wi^{c},\wi^{c}](k)} \Big]}{\sum_{i = 1}^{n} u_{ik}^{m}}. \label{eqn: sigma_est} 
    \end{align}
    In (\ref{eqn: sigma_est}), the term $\vec{C}_{i[\wi^{c},\wi^{c}](k)}$ only affects the covariance of the imputed values. Further details on the derivation of (\ref{eqn: sigma_est}), which corresponds to the traditional covariance matrix estimation in presence of missing values, are provided in the Supplementary Material. If the update of the covariance matrices in (\ref{eqn: sigma_est}) does not satisfy the eigenvalue-ratio constraint in (\ref{eqn: eigenratio}), we apply the efficient eigenvalue-truncation procedure proposed by Fritz et al. \cite{FGEMI:2013}. This ensures that the constraint in (\ref{eqn: eigenratio}) is met.
\end{enumerate}

At the end of \textit{Step 4}, the objective function in (\ref{eqn: obj}) is computed. If the focus is on clusters with $\sum_{i = 1}^{n} u_{i1}^{m} = \ldots = \sum_{i = 1}^{n} u_{iK}^{m}$, the objective function in (\ref{eqn: obj}) should be modified by removing the $\pi_k$ weights, which approximately corresponds to set $\pi_{k} = 1/K$ in \textit{Step 4} throughout all iterations (this is an exact correspondence for $m = 1$). \textit{Steps 1-4} are repeated until the increase (and, more precisely, non-decrease) in the objective function falls below a small tolerance value $\epsilon$ (e.g., $10^{-6}$ in our experiments), or until a maximum number of iterations ($500$ in the examples shown in this work) is reached. The cellFCLUST objective function is non-decreasing at every iteration (see the Supplementary Material for details on the algorithm’s monotonicity). To increase the chances of finding the global constrained maximum of the objective function, the algorithm should be run several times with different initializations, retaining the best solution. 

\section{Simulation study}\label{sec: simulations}
We conduct a simulation study to evaluate the performance of cellFCLUST relative to several alternative methodologies across three main goals: cluster recovery, parameter estimation, and outlier detection. The competitors we consider are fuzzy clustering methods, both robust and non-robust. It is worth noting that the robust approaches included in the simulation study are designed to handle casewise outliers. Indeed, to the best of our knowledge, prior to cellFCLUST, no fuzzy clustering methodology addressing cellwise contamination has been proposed in the literature, as highlighted in Section \ref{sec: intro}. 

The first competitor is F-TCLUST, which represents the casewise counterpart of our proposal and it is run via the \proglang{R} package \pkg{tclust} (\cite{FGEMI:2012}, \proglang{R} version 3.3). Two other classes of methods encompass those which are not specifically tailored for outliers: fuzzy $k$-means (FKM, \proglang{R} function \code{FKM}) and Gustafson, Kessel and Babuska-like fuzzy $k$-means (FKMGKB, \proglang{R} function \code{FKM.gkb}), which relaxes the spherical assumption for the clusters \cite{BvdVK:2002}. Their robust versions based on the identification of a noise cluster are implemented via the \proglang{R} functions \code{FKM.noise} and \code{FKM.gkb.noise}, respectively. All these methods are included in the R package \pkg{fclust} \cite{FGS:2019}. Additionally, we also consider the Unsupervised Fuzzy Trimmed C Prototypes (UFTCP, \cite{KKD:1996}), where fuzzy $c$-means is extended via a trimming approach.

Finally, we compare the proposal with cellGMM \cite{ZACetal:2025}, a mixture model designed to deal with cellwise contamination. As discussed in Section \ref{sec: intro}, although mixture models provide soft assignments, they do not offer the same flexibility as fuzzy clustering methods in capturing cluster overlap. Results for cellGMM are compared with those for cellFCLUST in the Supplementary Material. 

\subsection{Design of the simulation study}\label{subsec: design}
Two scenarios with varying numbers of units and clusters are considered, while the number of variables is fixed to $p = 10$. In \textit{Scenario 1}, $n = 250$, $G = 2$: $30\%$ of units are generated from the first cluster, with $\mean_{1} = \vec{0}$ and $\sigu_{1} = [\sigma_{jl} = (0.6^{\lvert j-l \rvert})/16, j, l = 1, \ldots, p]$; and $70\%$ from the second cluster, with $\mean_{2} = [\mu_{j} = (-1)^{j} \times 0.5, j = 1, \ldots, p]$ and $\sigu_{2} = [\sigma_{jl} =  ((-0.6)^{\lvert j-l \rvert})/16, j, l = 1, \ldots, p]$. In \textit{Scenario 2}, $n = 500$, $G = 4$, and the sizes of the clusters are more balanced: the first two clusters, each containing $20\%$ of the units, have the same mean vectors and covariance matrices as in Scenario 1; the other two clusters, each composed of $30\%$ of the units, have mean vectors $\mean_{3} = [\mu_{j} = (j \mod 3) - 1, j = 1, \ldots, p]$ and $\mean_{4} = [\mu_{j} = \big( (j+1 \mod 4) - 1 \big)/2, j = 1, \ldots, p]$, and covariance matrices $\sigu_{3} = [\sigma_{jl} = (0.7^{\lvert j-l \rvert})/16, j, l = 1, \ldots, p]$ and $\sigu_{4} = [\sigma_{jl} = \big((-0.7)^{\lvert j-l \rvert}\big)/16, j, l = 1, \ldots, p]$. The maximum overlap between pairs of clusters -- where the overlap is defined in \cite{MM:2010} as the sum of the two misclassification probabilities -- ranges from $0.01$ to $0.05$ in both scenarios, indicating moderate cluster separation, which represents a suitable configuration for fuzzy clustering methodologies. For each scenario, $100$ random samples are obtained and contaminated with $0\%$ (baseline), $1\%, 5\%$, and $10\%$ of outlying cells per variable (i.e., in each column of the data matrix) randomly drawn from a uniform distribution on the interval $[-30, 30]$. The contamination is also performed in such a way that no outlying unit lies within the $99$-th percentile ellipsoid of any cluster, comparing their Mahalanobis distance, computed using the parameters employed for data generation, with the $99$-th percentile of a chi-squared distribution with $p$ degrees of freedom. 

The fuzzifier parameter $m$ is set to $2$ for cellFCLUST and F-TCLUST, to $1.35$ for FKM, FKM.noise and UFTCP, and to $1.5$ for FKM.gkb and FKM.gkb.noise. To select these values, we compare the proportion of Weak Assignments (WA) -- defined as the proportion of units whose highest membership is below $0.90$ -- resulting from the estimation of each methodology under $\alpha = 0$ (baseline) with the theoretical ones. The latter are computed from the membership matrices in (\ref{eqn: membership}) using the generating parameters and $\alpha = 0$, yielding WA proportions of $0.05$ and $0.07$ in the two scenarios, respectively. It should be noted that we focus on $u_{ik}$ for the unit $i$ belonging to cluster $k$ rather than the fuzziness itself, i.e., $u_{ik^{\prime}}$ with $k^{\prime} \neq k$, since the latter depends on the number of clusters considered. Additionally, analyzing the highest membership value across clusters provides information on the \say{weight} a unit has in the cluster it is assigned to: if it is a representative unit of that cluster, its membership will be high -- potentially approaching one -- regardless of the number of clusters. In our experience, when a unit is weakly assigned to a cluster, the remaining membership is typically shared among only a few clusters, that is, it is not evenly split across all $K-1$ clusters. For the methods that require eigenvalue-ratio constraints given by $c$, i.e., cellFCLUST and F-TCLUST, this is set according to the ratio of the eigenvalues computed from the true (i.e., data-generating) covariance structure. For FKM.gkb and FKM.gkb.noise, we specify precise constraints: the volume parameters $\rho_g$ as obtained from the aforementioned covariance matrices, and $\gamma = 0.1$ to ensure numerical stability. These two parameters are therefore changed from their default values, which were set only to prevent numerical singularities. Instead, we apply fine-tuning to choose more appropriate values that improve the performance of these clustering methods. The flagging level $\alpha$ is fixed to the true generating contamination for cellFCLUST, while the trimming one is set to $0.25$ for F-TCLUST, as previously employed in \cite{ZACetal:2025}. However, due to the nature of cellwise contamination, with $p = 10$ and cellwise outliers representing $1\%$, $5\%$, and $10\%$ of the cells, the percentage of cases affected by at least one outlying cell can reach $10\%$, $40\%$, and $65\%$ of the total cases, respectively. For noise clustering methodologies, the number of outliers is automatically selected within the algorithm. Similarly, UFTCP chooses the appropriate outlier proportion from a set of possible levels ranging from 0 to 0.5, in increments of 0.05, ultimately retaining the value that maximizes a density criterion \cite{GG:1989} as the validity measure.

\subsection{Results}
The results of the simulation study are evaluated in terms of cluster recovery, parameter estimation, and outlier detection. In the following sections, we analyze the performance of the proposed and alternative methods focusing separately on these objectives.

\subsubsection{Cluster recovery}\label{subsubsec: cluster_recovery}
Cluster recovery is assessed through the Misclassification Rate (MR) and the Adjusted Rand Index (ARI, \cite{HA:1985}) by comparing the theoretical and estimated assignments of units to clusters, determined from the maximum values of $u_{ik}$ across $k = 1 \ldots K$ for each $i = 1, \ldots, n$. Lower MR values and higher ARI values indicate better recovery, with perfect agreement between theoretical and estimated partitions when MR is zero and ARI is one.

\begin{figure}[!htbp]
    \centering
    \includegraphics[width=\linewidth, height=0.8\textheight]{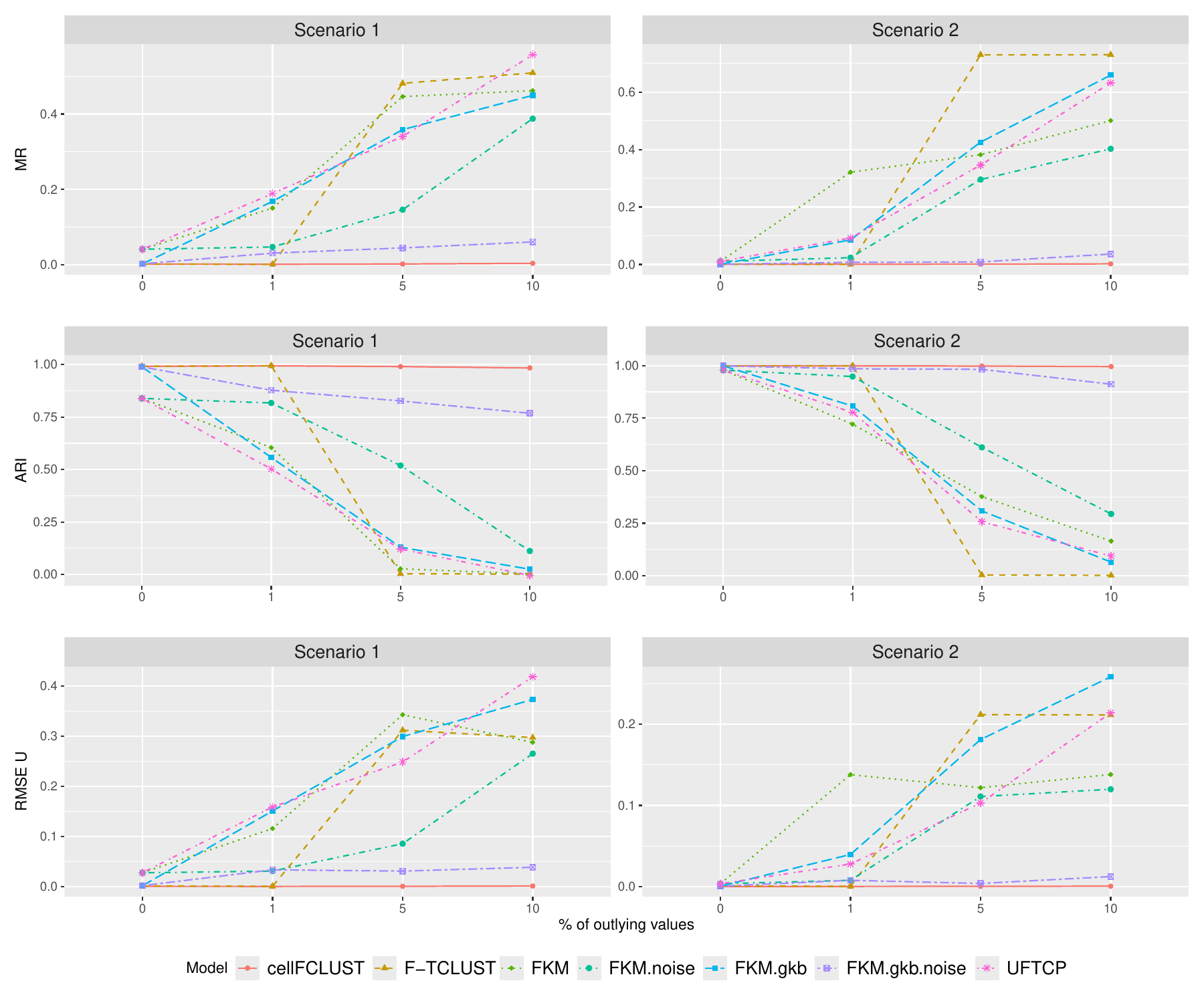}
    \caption{Results of the simulation study: Misclassification Rate (MR), Adjusted Rand Index (ARI), and Root Mean Squared Error (RMSE) of the membership matrix averaged over $100$ samples per scenario, percentage of contamination, and model. For F-TCLUST, FKM.noise, FKM.gkb.noise, and UFTCP the indices are computed only on the units not flagged as outliers.}
    \label{fig: simstudy_results_MR_U}
\end{figure}

As shown in Figure \ref{fig: simstudy_results_MR_U}, cellFCLUST has substantially lower MR and higher ARI (close to $1$) in both scenarios across contamination levels, since it is the only model designed to address cellwise contamination. The difference with the other methods increases as the proportion of outlying cells in the data grows, since even robust (casewise) methodologies can break down when contamination spreads across the cases. Among the competitors, FKM, FKM.gkb, and UFTCP show the highest MR (and lowest ARI) for cellwise contamination of $1\%$. However, at $5\%$ and $10\%$ contamination, only cellFCLUST and FKM.gkb.noise maintain good clustering performance. This result is mainly due to the large number of cases containing outlying cells as the contamination level increases, and their ability to accommodate non-spherical clusters. Moreover, unlike F-TCLUST, where $\alpha$ is fixed in advance, FKM.gkb.noise automatically selects the proportion of outlying cases, allowing it to stabilize at lower MR (and higher ARI) values than cellFCLUST's natural casewise counterpart. Although this flexibility, cellFCLUST consistently outperforms FKM.gkb.noise due to its higher efficiency in handling cellwise outliers, without discarding or downweighting  entire contaminated units during the estimation procedure. 

\subsubsection{Parameter estimation}\label{subsubsec: param_estimation}
Parameter estimation is evaluated via the Root Mean Squared Error (RMSE) of the estimated membership matrix $\vec{U}$ and cluster mean vectors $\{\mean_{k}\}_{i = 1}^{K}$, and the Kullback-Leibler (KL) discrepancy of the estimated cluster covariance matrices $\{\sigu_{k}\}_{i = 1}^{K}$, compared to the corresponding theoretical parameters used to generate the data (Section \ref{subsec: design}). To properly perform this comparison, the label switching problem is resolved by finding the cluster label ordering that minimizes RMSE between the theoretical and estimated cluster means over all possible permutations.

\begin{figure}[!ht]
    \centering
    \includegraphics[width=\linewidth, height=0.8\textheight]{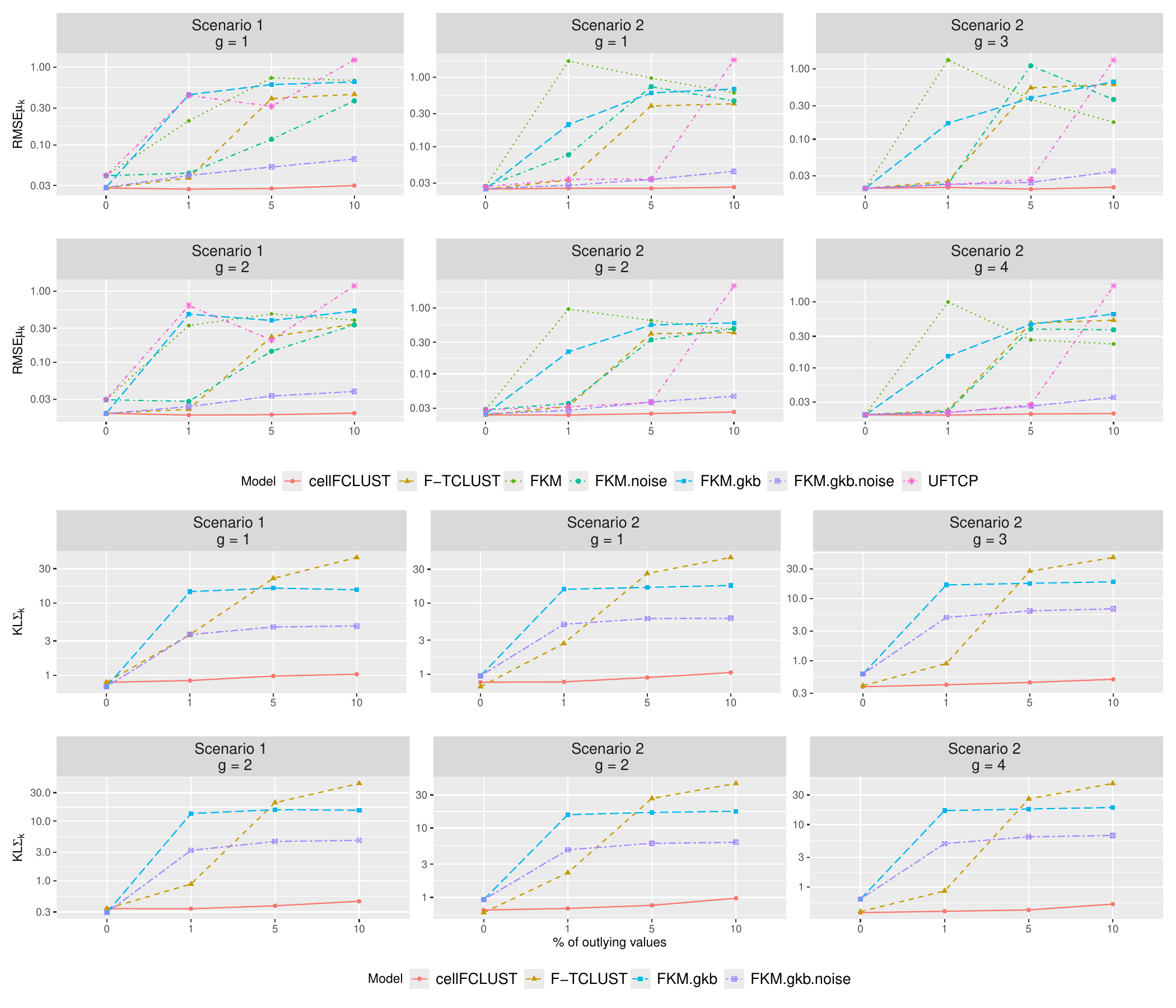}
    \caption{Results of the simulation study: Root Mean Squared Error (RMSE) of the cluster mean vectors and Kullback-Leibler (KL) discrepancy for the cluster covariance matrices averaged over $100$ samples per scenario, percentage of contamination, and model. The values are represented via log-transformation, while the y-axis ticks are labeled using the original scale.}
    \label{fig: simstudy_results_Parameters}
\end{figure}

The results for the membership matrix are shown in Figure \ref{fig: simstudy_results_MR_U}, while those for the cluster mean vectors and covariance matrices are reported in Figure \ref{fig: simstudy_results_Parameters}. RMSE of the membership matrices exhibits a pattern similar to MR and ARI, as does RMSE of the cluster mean vectors, which deteriorates at $1\%$ of contamination for the non-robust models. Additionally, the latter significantly degrades at $5\%$ and $10\%$ for F-TCLUST and FKM.noise, whereas the FKM.gkb.noise estimates remain acceptable but suboptimal with respect to cellFCLUST. In the estimation of the cluster covariance matrices, cellFCLUST outperforms the competitors from the first contamination level, as indicated by the KL discrepancy measure. On the other hand, F-TCLUST provides good estimates only when the number of contaminated cases in at least one cell is smaller than the number of trimmed cases, handling at most $1\%$ of cellwise contamination when $p = 10$. Unlike the other indices, the difference between cellFCLUST and FKM.gkb.noise becomes evident in the estimation of the covariance matrices. Indeed, for FKM.gkb-type models -- especially the noise version -- increasing the smallest eigenvalue of the covariance matrix $\sigu_{k}$ to ensure its determinant equals $\rho_{g}$ does not yield results comparable to those obtained with the efficient procedure proposed in \cite{FGEMI:2013}, which provides a closed-form solution for the truncated eigenvalues satisfying constraint (\ref{eqn: eigenratio}). This is particularly pronounced in Figure \ref{fig: simstudy_results_Parameters} at $1\%$ contamination, where the two models implementing the aforementioned procedure -- namely, cellFCLUST and F-TCLUST -- produce better estimates of the covariance matrices. It is worth noting that F-TCLUST is implemented with $\alpha = 0.25$, while the maximum percentage of cases with at least one contaminated cell is $10\%$ when the true $\alpha = 0.01$. At higher contamination levels, the KL discrepancy of FKM.gkb.noise is lower than that of F-TCLUST due to the larger proportion of outliers not flagged by the latter. Regarding cellFCLUST, the combined effect of the eigenvalue-ratio constraint and the cellwise outlier detection mechanism allows this methodology to achieve lower KL values than FKM.gkb.noise.

\subsubsection{Outlier detection}\label{subsubsec: outlier_detection}

\begin{table}[!ht]
\centering
\caption{Memberships and outlier detection: proportion of Weak Assignments (WA), True and False Positive Rate (TPR and FPR), and False Negative Rate (FNR) averaged over $100$ samples per scenario, percentage of contamination, and model.}\label{tab: simstudy_wa_out}
\resizebox{0.9\textwidth}{!}{
\begin{tabular}{cl|cccc|cccc}
  \hline
\multicolumn{2}{c}{} & \multicolumn{4}{c}{Scenario 1} & \multicolumn{4}{c}{Scenario 2} \\
$\%$ out & Method & WA & TPR & FPR & FNR & WA & TPR& FPR & FNR \\ 
  \hline
  \multirow{6}{*}{0} & cellFCLUST & 0.03 & - & - & - & 0.05 & - & - & - \\
& F-TCLUST & 0.07 & - & - & - & 0.12 & - & - & - \\ 
& FKM & 0.10 & - & - & - & 0.04 & - & - & - \\ 
& FKM.noise & 0.10 & - & - & - & 0.04 & - & - & - \\
& FKM.gkb & 0.07 & - & - & - & 0.05 & - & - & - \\ 
& FKM.gkb.noise & 0.09 & - & - & - & 0.08 & - & - & - \\ 
& UFTCP & 0.10 & - & - & - & 0.04 & - & - & - \\ 
  \hline
\multirow{6}{*}{1} & cellFCLUST & 0.03 & 1.00 & 0.00 & 0.00 & 0.05 & 1.00 & 0.00 & 0.00 \\ 
& F-TCLUST & 0.00 & 1.00 & 0.25 & 0.00 & 0.01 & 1.00 & 0.24 & 0.00 \\ 
& FKM & 0.18 & - & - & - & 0.13 & - & - & - \\ 
& FKM.noise & 0.11 & 0.92 & 0.06 & 0.08 & 0.06 & 0.79 & 0.07 & 0.21 \\ 
& FKM.gkb & 0.20 & - & - & - & 0.22 & - & - & - \\ 
& FKM.gkb.noise & 0.57 & 1.00 & 0.08 & 0.00 & 0.47 & 1.00 & 0.09 & 0.00 \\ 
& UFTCP & 0.17 & 0.93 & 0.09 & 0.07 & 0.14 & 1.00 & 0.09 & 0.00 \\
\hline
\multirow{6}{*}{5} & cellFCLUST & 0.04 & 0.99 & 0.00 & 0.01 & 0.06 & 0.99 & 0.00 & 0.01 \\ 
& F-TCLUST & 1.00 & 0.69 & 0.23 & 0.31 & 1.00 & 0.67 & 0.23 & 0.33 \\ 
& FKM & 0.39 & - & - & - & 0.43 & - & - & - \\ 
& FKM.noise & 0.28 & 0.69 & 0.23 & 0.31 & 0.26 & 0.60 & 0.20 & 0.40 \\ 
& FKM.gkb & 0.35 & - & - & - & 0.47 & - & - & - \\ 
& FKM.gkb.noise & 0.22 & 1.00 & 0.36 & 0.00 & 0.17 & 1.00 & 0.37 & 0.00 \\ 
& UFTCP & 0.45 & 0.99 & 0.37 & 0.01 & 0.45 & 1.00 & 0.40 & 0.00 \\ 
\hline
\multirow{6}{*}{10} & cellFCLUST & 0.05 & 0.99 & 0.00 & 0.01 & 0.07 & 0.99 & 0.00 & 0.01 \\ 
& F-TCLUST & 1.00 & 0.47 & 0.23 & 0.53 & 1.00 & 0.46 & 0.23 & 0.54 \\ 
& FKM & 0.66 & - & - & - & 0.70 & - & - & - \\
& FKM.noise & 0.53 & 0.60 & 0.31 & 0.40 & 0.56 & 0.50 & 0.24 & 0.50 \\ 
& FKM.gkb & 0.33 & - & - & - & 0.54 & - & - & - \\ 
& FKM.gkb.noise & 0.22 & 0.99 & 0.60 & 0.01 & 0.20 & 0.98 & 0.60 & 0.02 \\
& UFTCP & 0.60 & 0.81 & 0.45 & 0.19 & 0.65 & 0.80 & 0.45 & 0.20 \\ 
\hline
\end{tabular}
}
\end{table}

We compute the True Positive Rate (TPR), False Positive Rate (FPR), and False Negative Rate (FNR) as the proportions of cells correctly flagged, reliable cells incorrectly flagged as contaminated, and contaminated cells not recognized as such, respectively, to evaluate outlier detection. For the robust (casewise) competitors, all the cells of the outlying cases are considered contaminated. It is worth noting that a high value of FNR is more problematic than a high value of FPR for a robust model, as it implies that contaminated cells are not properly accounted for and their values affect the parameter estimates.

Looking at Table \ref{tab: simstudy_wa_out}, we can notice that TPR shows a near-perfect score for cellFCLUST and FKM.gkb.noise in both scenarios and across contamination levels, whereas F-TCLUST and FKM.noise start to deteriorate at $5\%$, and UFTCP at $10\%$. Additionally, cellFCLUST is the only model capable of achieving a zero FPR. It is worth recalling that, for the casewise robust models, all cells of the cases flagged as outlier are considered as contaminated, leading to a higher number of false positives. Consequently, the reliable information discarded by casewise robust models results in higher errors even when outlying values are detected. This effect is noticeable when comparing the overall performance of cellFCLUST and FKM.gkb.noise.

Similarly to TPR, the best performance for FNR is achieved by cellFCLUST and FKM.gkb.noise. As mentioned before, a high FNR is more detrimental than a high FPR, and this effect is also reflected in cluster recovery and parameter estimation. This can be seen, for instance, in F-TCLUST: at $1\%$ contamination, it exhibits a high FPR ($0.25$) and no false negatives, resulting in good cluster recovery (see Figure \ref{fig: simstudy_results_MR_U}). When contamination increases, FPR slightly decreases ($0.23$), while FNR increases substantially ($0.31$ at $5\%$ of contamination), leading to a marked deterioration in clustering performance, with a sharp increase in MR and a corresponding decrease in ARI.

\bigskip
\section{Effects of the cellFCLUST tuning parameters}\label{sec: paramchoice}
The proposed methodology depends on several tuning parameters: the number of clusters ($K$), the flagging level ($\alpha$), the constant for the eigenvalue-ratio constraint ($c$), and the fuzzifier ($m$). Additionally, we can also consider the scale factor ($S$), which arises when we modify $x_{ij}$ by $x_{ij}/S$ for $i = 1\ldots, n$ and $j = 1, \ldots p$, as a further parameter. The effect of the scale factor was already noted in \cite{FGEMI:2013b} and is inherent when considering general covariance matrices and the definition of the membership values in (\ref{eqn: membership}). However, $S$ can be viewed as an actionable tuning parameter as well.

We illustrate the role of all tuning parameters using one of the samples generated in Scenario 1 of the simulation study (Section \ref{sec: simulations}) with $5\%$ of contamination. It has to be highlighted that these parameters are interrelated, and a unified approach for their setting is needed. In this section, we provide tools for helping the user in selecting these parameters, which are particularly useful when no prior information is available, as is often the case in real-data applications.

\subsection{Number of clusters \texorpdfstring{$K$}{TEXT} and flagging level \texorpdfstring{$\alpha$}{TEXT}}
\begin{figure}[t]
    \centering
    \includegraphics[width=0.8\linewidth]{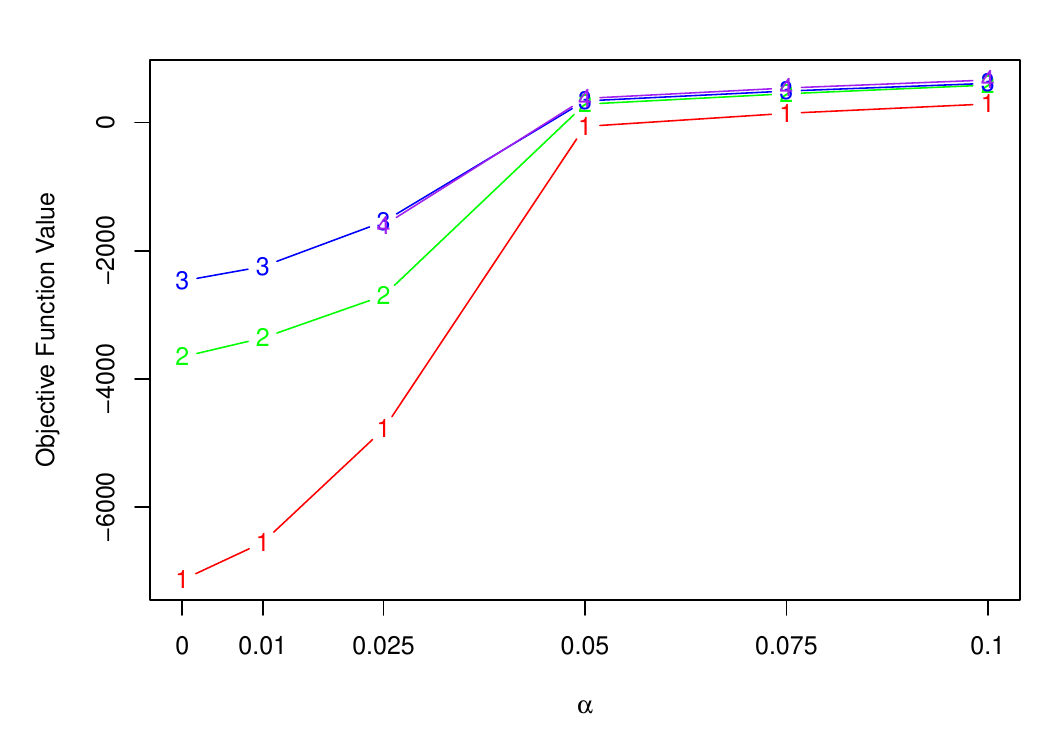}
    \caption{Synthetic data: objective function curves.}
    \label{fig: example_clt}
\end{figure}

The number of clusters in a data set cannot be chosen independently of the level of cells flagged as contaminated per variable. In this framework, we evaluate the behavior of the curves representing the objective function in (\ref{eqn: obj}) at convergence, by running cellFCLUST with different values of $K$ and $\alpha$, when $c = 14$ and $m = 2$. We vary $\alpha$ over the set $\{0, 0.01, 0.025, 0.05, 0.075, 0.10\}$, considering that the true contamination level is $0.05$. As shown in Figure \ref{fig: example_clt}, the objective functions differ significantly when $\alpha = 0, 0.01, 0.025$, since underestimating the contamination level requires additional clusters beyond those originally generated. It is worth noting that when $K = 4$ and $\alpha = 0$ or $0.01$, cellFCLUST finds solutions with an empty cluster, causing the algorithm to stop -- the corresponding objective function values are not reported in the figure. On the other hand, when $\alpha = 0.05$, which is the so generated contamination level in the data set, the objective functions for $K = 2, 3, 4$ are close to each other, indicating that two is a suitable choice for the number of clusters. Indeed, $K = 2$, which corresponds to the true value, captures the underlying structure of the data well, and increasing this number does not lead to a meaningful improvement in the objective function \cite{GEGMMI:2011}. 

\begin{figure}[!htbp]
\centering
\includegraphics[width=0.75\linewidth, height=0.38\textheight]{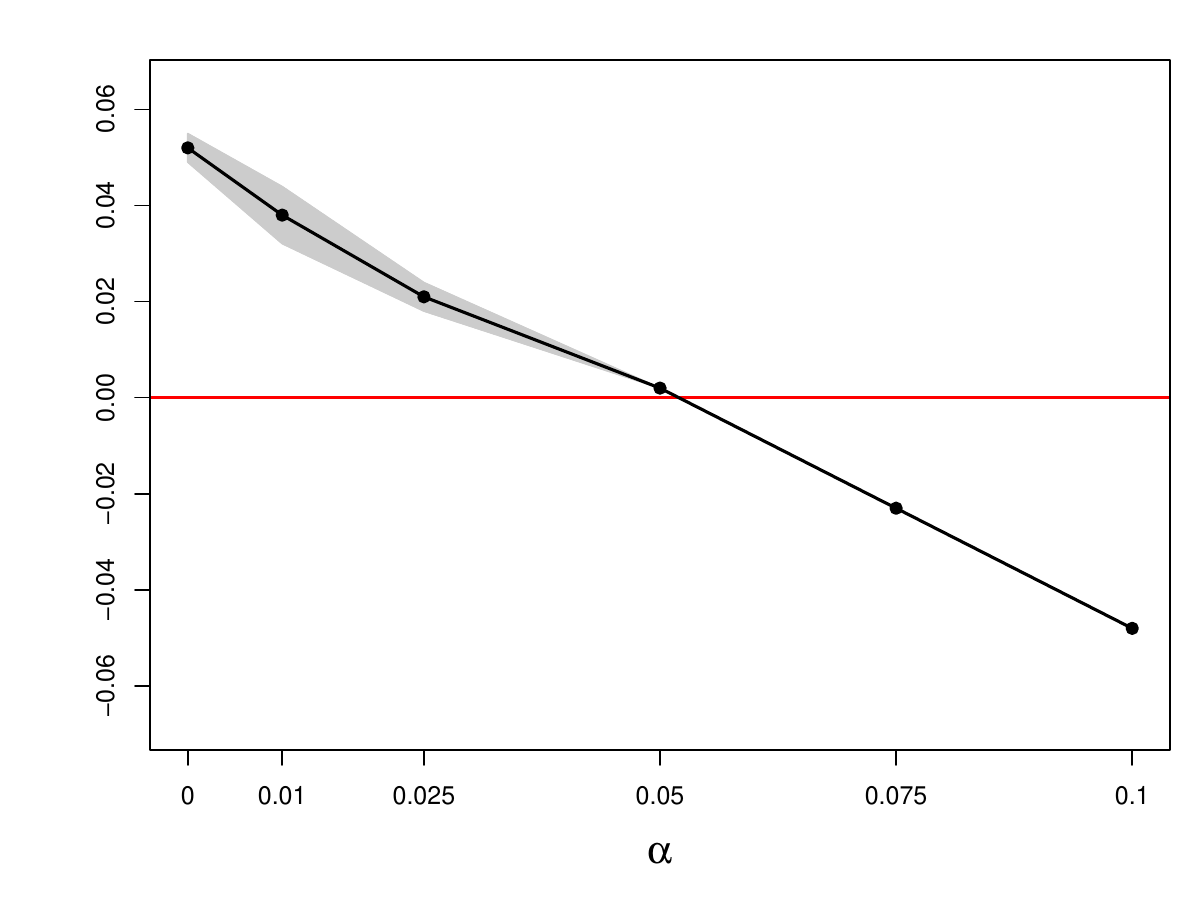}
\caption{Synthetic data: difference between the knee point of $\{\Delta_{ij}\}_{i = 1}^{n}$ and the $\alpha$ value. The curve shows the median differences across variables as $\alpha$ varies, with $K = 2$. The shaded gray area represents the variability.}
\label{fig: example_KneedlePlot}
\end{figure}

\begin{figure}[!htbp]
\centering
\subfloat[][$\alpha = 0.025$ \label{fig: example_Delta_2_5}] 
{\includegraphics[width=\linewidth, height=0.3\textheight]{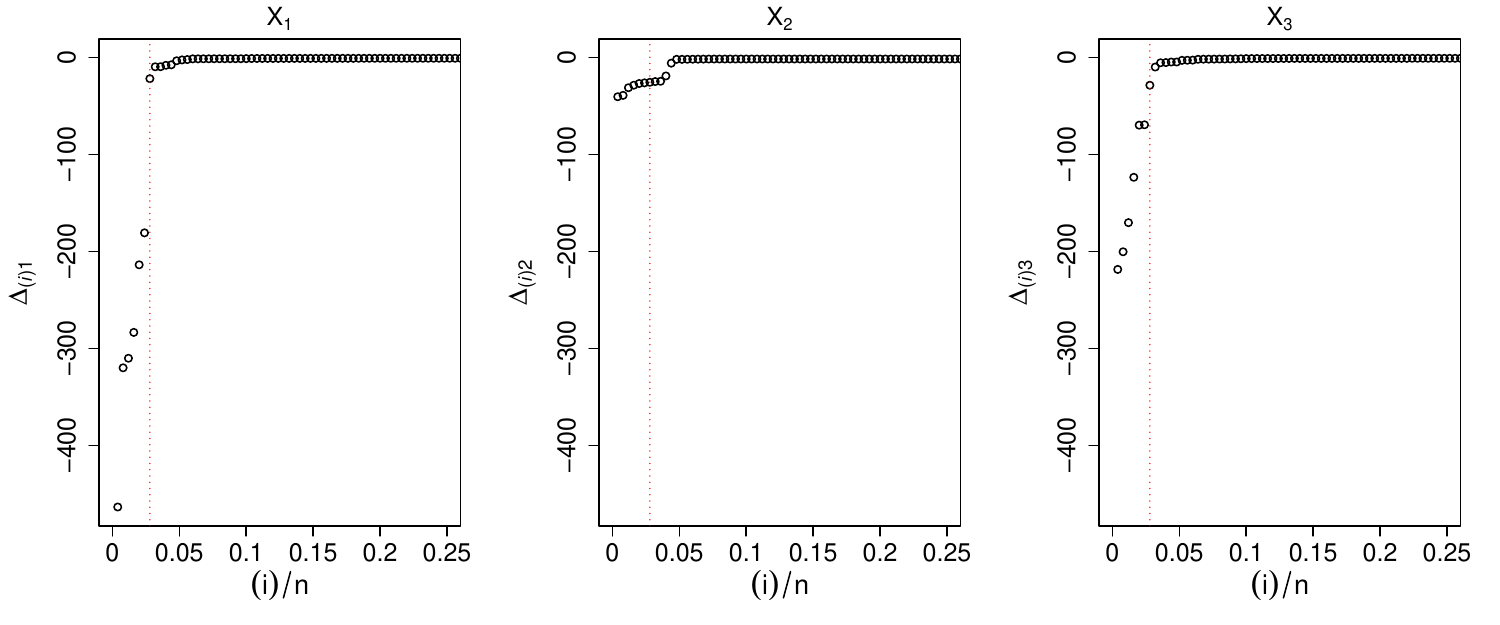}} \\
\subfloat[][$\alpha = 0.05$ \label{fig: example_Delta_5}] 
{\includegraphics[width=\linewidth, height=0.3\textheight]{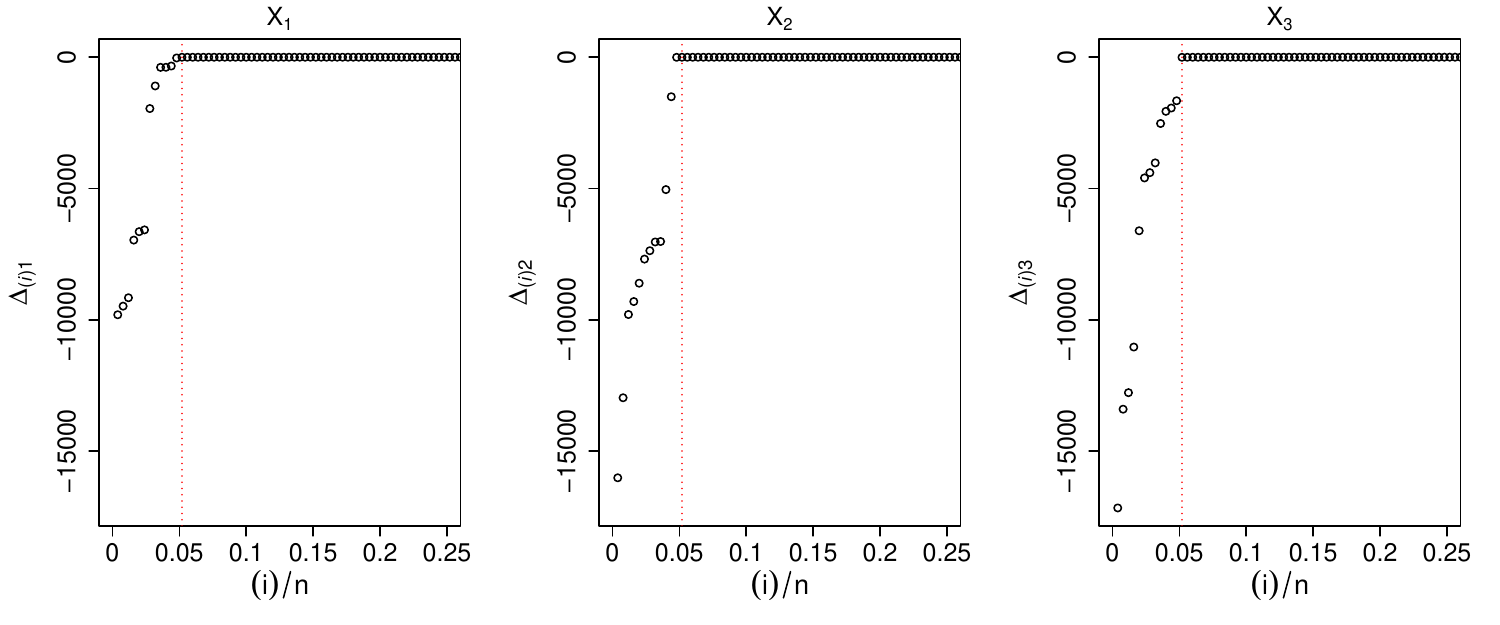}} 
\end{figure}
\addtocounter{figure}{+1}
\begin{figure}[t]\ContinuedFloat
\centering
\subfloat[][$\alpha = 0.075$ \label{fig: example_Delta_7_5}] 
{\includegraphics[width=\linewidth, height=0.3\textheight]{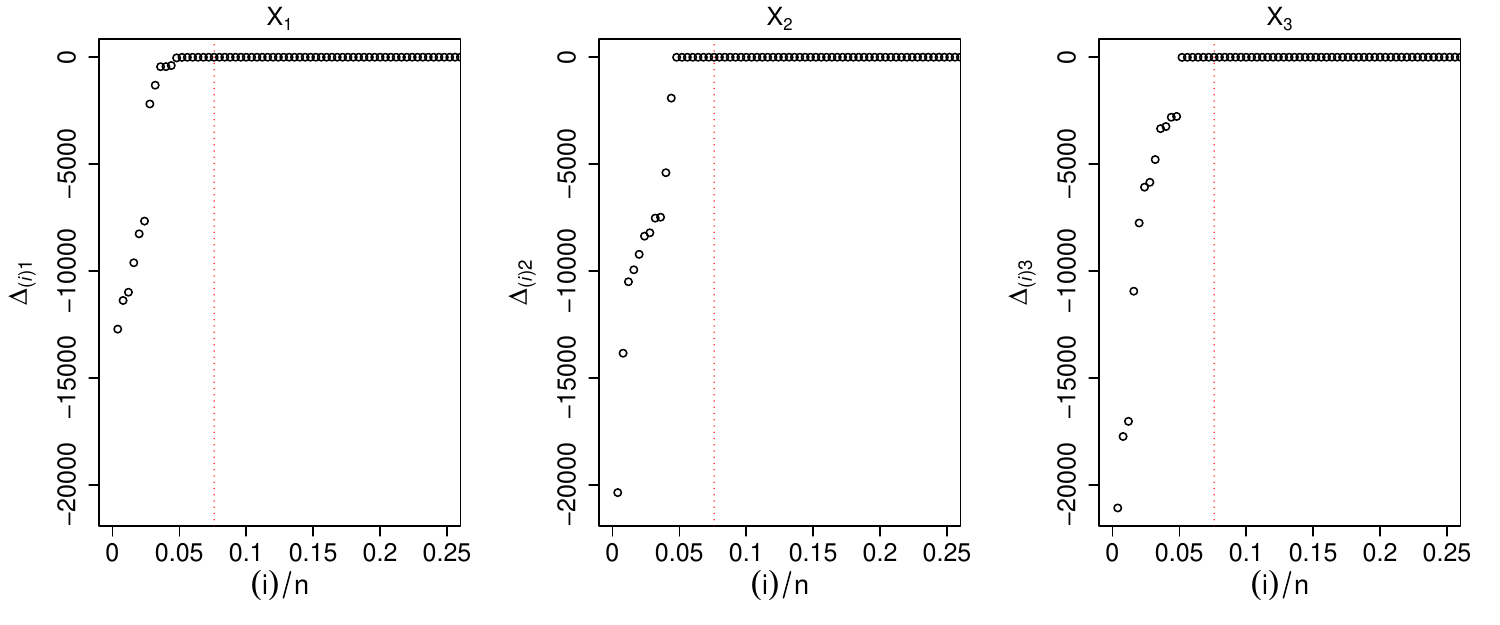}} 
\caption{Synthetic data: $\Delta$ plots where units are sorted according to their $\Delta$ values. Vertical, dashed, red line corresponds to the value of $\alpha$ used for the model implementation when $K = 2$.}
\label{fig: example_Delta}
\end{figure}

We further investigate how the cellFCLUST results change with the flagging level $\alpha$ to support the previous finding by analyzing the behavior of $\{\Delta_{ij}\}_{i = 1}^{n}$ for $j = 1, \ldots, p$. These are computed as defined in (\ref{eqn: delta}), using the parameters estimated at convergence for the selected $K$. For each variable, the ordered values $\Delta_{(i)j}$, where $\Delta_{(1)j} \leq \Delta_{(2)j} \leq \ldots \leq \Delta_{(n)j}$, can be plotted. In each plot -- which we discuss in detail later -- we identify the knee point, corresponding to the maximum distance between each $\Delta_{(i)j}$ and the straight line connecting the endpoints $\Delta_{(1)j}$ and $\Delta_{(n)j}$, as $\alpha$ varies. Therefore, we have $p$ knee points (one per variable) for each $\alpha$. Figure \ref{fig: example_KneedlePlot} shows the median difference between the $p = 10$ knee points and the corresponding $\alpha$, for different values of $\alpha$. The difference approaches zero when $\alpha = 0.05$, with variability (gray area, calculated using the Median Absolute Deviation) close to zero, confirming the selection of this value as the appropriate flagging level for these data. Furthermore, we can verify the behavior of $\Delta_{ij}$ for each variable, recalling that they correspond to the difference between the individual contribution to the objective function when a cell is considered reliable or unreliable. In Figure \ref{fig: example_Delta}, we plot points $\{\big((i)/n, \Delta_{(i)j}\big) \}_{i = 1}^{n}$ for the first three variables as examples, where units are ordered according to their $\Delta$ values (see the Supplementary Material for the plots of the remaining seven variables). When $\alpha = 0.025$, some units with a small value of $\Delta$ are not considered as contaminated, resulting in a lower flagging level than the theoretical contamination one (Figure \ref{fig: example_Delta_2_5}). The same occurs for smaller values of $\alpha$. On the other hand, too many cells than needed are flagged as unreliable when $\alpha = 0.075$, as shown in Figure \ref{fig: example_Delta_7_5}, where the values of $\Delta$ level off before the considered $\alpha$ -- the same happens for $\alpha = 0.10$. The appropriate choice turns out to be $0.05$ (Figure \ref{fig: example_Delta_5}), which corresponds to the true proportion of contaminated cells in the simulated sample. Indeed, at this level of flagged cells, $5\%$ of the $\Delta$ values per variable are significantly lower than the majority of the others. In general, we recommend initially adopting a conservative choice of $\alpha$, then gradually decreasing it while monitoring the behavior of the objective function curves. The additional tools provided for supporting the choice of $\alpha$, such as those in Figures \ref{fig: example_KneedlePlot} and \ref{fig: example_Delta}, are particularly useful when the objective function curves are smoother.

\subsection{Constant \texorpdfstring{$c$}{TEXT} for the eigenvalue-ratio constraint}
The constant for the eigenvalue-ratio constraint in (\ref{eqn: eigenratio}) allows for different cluster shapes. When $c = 1$, the clusters become spherical, and cellFCLUST with $\alpha = 0$ and $\pi_{1} = \ldots = \pi_{k}$ produces similar results to those of FKM. Indeed, a small value of $c$ reduces the cellFCLUST ability to detect elongated and/or differently dispersed clusters, as illustrated in \cite{FGEMI:2013b} and \cite{GEMI:2024}. However, in some applications, the user may be interested in more spherical types of clusters (see, for instance, \cite{HL:2013}).

\subsection{Fuzzifier parameter \texorpdfstring{$m$}{TEXT}, scale factor \texorpdfstring{$S$}{TEXT} and their interaction}
CellFCLUST shows different behavior depending on the fuzzifier parameter and the scaling of the data. The former affects the degree of fuzziness by letting the memberships become more similar as $m$ increases. Points that are more fuzzily assigned to the clusters are usually units located farther from the core of the clusters in the $p$-dimensional space. On the other hand, the scale factor $S$ influences the proportion of Hard Assignments (HA) obtained by cellFCLUST, which usually correspond to units within the core of the clusters carrying more weight in the estimation of their parameters. This effect was previously observed in \cite{GG:1989} and \cite{RTK:1996}. Given the property of high contrast, we thus treat the scale factor $S$ as a key tuning parameter that allows us to manage the mentioned level of hard assignment depending on the purpose of the data analysis. Furthermore, the scale factor and the fuzzifier interplay, resulting in a change in the clustering structure estimated by cellFCLUST when both $S$ and $m$ vary. Specifically, as the scale factor decreases and the fuzzifier increases, the assignments tend to become less crisp and more fuzzy. However, a complete fuzzification is not desirable from an interpretation point of view. It is worth noting that when $m = 1$, the results are scale-independent since cellFCLUST returns a completely hard assignment. Since the effect of the tuning parameters $m$ and $S$ in cellFCLUST is similar to its casewise counterpart F-TCLUST, the reader can refer to \cite{FGEMI:2013b} for further insights and numerical examples. 

\subsection{Restrictions on the weights \texorpdfstring{$\pi_{k}$}{TEXT}}
One of the advantages of cellFCLUST is its ability to accommodate different cluster sizes via the weights $\pi_k$ in the objective function $J_{\text{cellFCLUST}}$ (\ref{eqn: obj}), in a similar manner to F-TCLUST. It is worth noting that removing the weights $\pi_k$ in (\ref{eqn: obj}) would favor clusters with comparable values of $\sum_{i = 1}^{n} u_{ik}^{m}$, which are the quantities corresponding exactly to cluster sizes in a hard clustering approach ($m = 1$). We do not reproduce here examples illustrating the effect of constraining $\pi{_k}$, $k = 1, \ldots, K$, to be equal, which the reader can find in \cite{FGEMI:2013b}. Moreover, Fritz et al. in \cite{FGEMI:2013b} highlight that when $K$ is misspecified by the user with a value larger than necessary, some cluster weights $\pi_k$ can become very close to $0$, resulting in almost \say{empty} clusters, i.e., clusters with small $u_{ik}$ for all units. This occurrence can reveal the potential misspecification of $K$ to the user.

\subsection{Guidelines for the parameter setting}
The material presented above exemplify the role and effect of the cellFCLUST tuning parameters. All these parameters are closely related to each other, and their choice strongly depends on the application under analysis. If prior knowledge about the data set is available -- for instance, regarding the expected clustering structure or the desired degree of fuzzification -- this information can guide the choice of some (or, in rare cases, all) tuning parameters. When this does not occur, we suggest selecting appropriate values for the tuning parameters by taking into account the purpose of the analysis. Specifically, as a first step, a simple proposal is to choose the constant $c$ by considering the order of magnitude of the ratio between the maximum and the minimum eigenvalues computed on the entire data set. Secondly, the scale factor can be set by analyzing the proportion of HA as $S$ varies, selecting a reasonable subset of values for $S$ consistent with the data under study. Indeed, although $S$ and $m$ are interconnected, the former mostly affects HA, while the latter affects  the degree of fuzzification. In this regard, $m$ can be chosen by studying the proportion of WA. The user may prefer to select a value of $m$ that results in a proportion of WA within a specific range, depending on the goal of the analysis and the cluster configuration. Finally, the objective function curves presented in this section are a useful tool for selecting $K$ and $\alpha$, as previously described. The flagging level $\alpha$ can be deepened using the $\Delta$ plots. It is worth highlighting that the choice of $S$ and $m$ is connected to the selection of both the number of clusters and the level of flagged cells. 

\bigskip
\section{Real data analyses}\label{sec: applications}
In this section, we illustrate the potential of cellFCLUST through two real data applications. The first one focuses on identifying clusters of individuals with different levels of body fat-related risk (Section \ref{subsec: bodyfat}). In the second application, regions of the OECD countries are grouped based on eleven variables related to well-being (Section \ref{subsec: wellbeing}). 

\subsection{Body fat data}\label{subsec: bodyfat}
The data on body fat, available in the R package \texttt{UsingR}, contain physiological measurements of $250$ men -- units 172 and 182 have been discarded due to erroneous values. Among the $18$ variables, we consider only those directly observed. Additionally, we exclude the variable \textit{Age}, as it masks the clustering structure due to its cross effect on body phenotypes; the variables \textit{Weight} and \textit{Height}, as they define \textit{BMI} (i.e., Body Mass Index); and the variable \textit{Fat Free Weight}, since it is derived from estimated quantities not directly observed. Table \ref{tab: BF_var} lists the eleven variables used for the analysis. Due to their different scales, the variables are standardized using a robust procedure that replaces the mean with the median and the standard deviation with the median absolute deviation. The presence of outlying values can be observed both by examining the boxplot and the pair plot, which are reported in the Supplementary Material. CellFCLUST is particularly suitable for analyzing this data set for two main reasons. First, since the variables are highly correlated, there is more information available to impute the potentially outlying values. Second, although the elongated shape of the data suggests the presence of a single cluster, this is not reasonable in this context as it would merge individuals with very different physical characteristics. This motivates considering a small value of $c$, similarly to the reasoning in \cite{HL:2013} for the social stratification analysis. Assuming $c = 2$, we search for more spherical clusters that can sometimes be very close to each other, up to overlapping, making the fuzzy clustering approach extremely useful.

\begin{table}[t]
    \centering
    \caption{List of variables of the body fat data set.}
    \label{tab: BF_var}
    \resizebox{\linewidth}{!}{
    \begin{tabular}{c l c|c l c}
    \toprule
    ID & Name & Measur. unit & ID & Name & Measur. unit \\
    \midrule
    1 & Adiposity index (BMI) & kg/m$^2$ & 7 & Knee circumference (knee) & cm \\
    2 & Neck circumference (neck) & cm & 8 & Ankle circumference (ankle) & cm \\
    3 & Chest circumference (chest) & cm & 9 & Extended biceps circumference (bicep) & cm \\
    4 & Abdomen circumference (abdomen) & cm & 10 & Forearm circumference (forearm) & cm \\
    5 & Hip circumference (hip) & cm & 11 & Wrist circumference (wrist) & cm \\ 
    6 & Thigh circumference (thigh) & cm & & & \\
    \bottomrule
    \end{tabular}}
\end{table}

\begin{figure}[t]
\centering
\includegraphics[width=0.95\linewidth, height=0.6\linewidth]{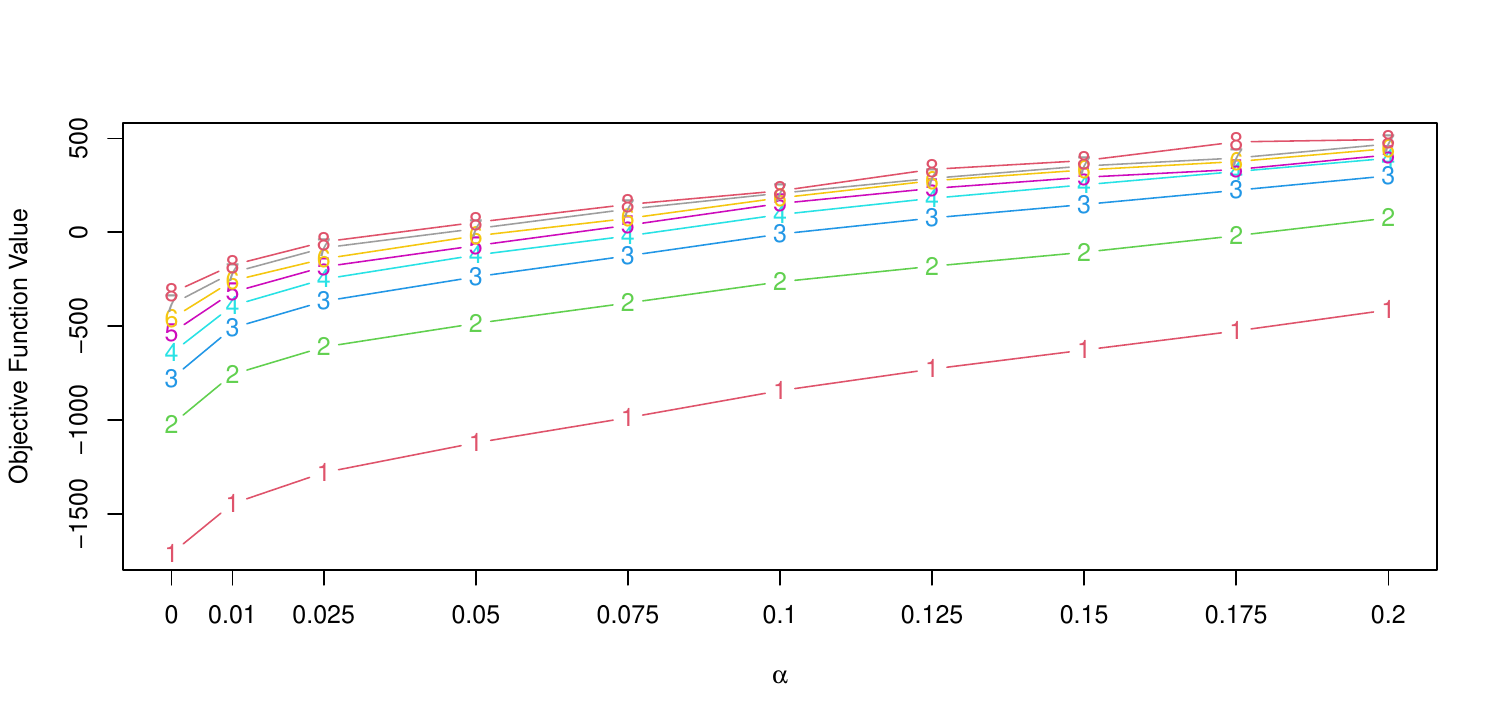} 
\caption{Body fat data: objective function curves.}
\label{fig: BF_CL}
\end{figure}

\begin{figure}[t]
\centering
\includegraphics[width=0.7\linewidth, height=0.5\linewidth]{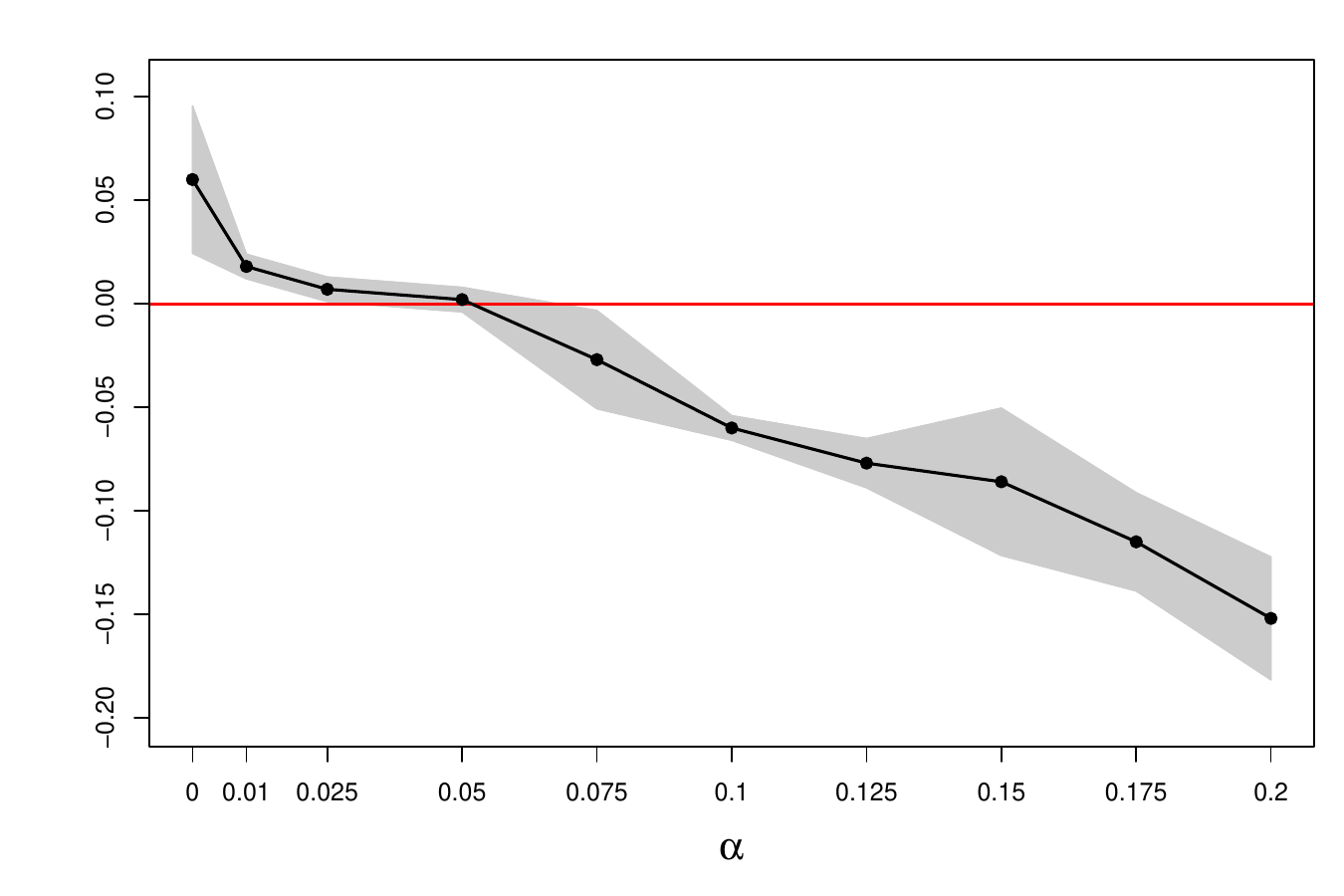} 
\caption{Body fat data: difference between the knee point of $\Delta_{ij}$ and the $\alpha$ value. The plot shows the median differences across variables as $\alpha$ varies, with $K = 4$. The shaded gray area represents the variability.}\label{fig: BF_KneedlePlot}
\end{figure}

As a preliminary analysis, we consider the proportions of HA and WA obtained by varying the scale factor. Fixing $m = 1.7$, which proves to be a suitable value for the degree of fuzzification in this data set (see the Supplementary Material), we choose $S = 2$, as it yields a proportion of HA and WA of 0.43 and 0.32, respectively. This results to be the best choice since a small value of $S$, i.e. $S = 1$, excessively fuzzifies all units, while higher values of $S$ correspond to an overly high proportion of hard assignments given the data configuration -- for $S = 3$ and $S= 4$, the proportions of HA are 0.89 and 0.98, while those of WA are 0.04 and 0.01, respectively. After selecting the fuzzifier parameter, the scale factor, and the constant for the eigenvalue ratio, we determine the values of $K$ and $\alpha$ based on the objective function curves reported in Figure \ref{fig: BF_CL}. Specifically, we vary $K \in \{1, \ldots, 8\}$ and $\alpha$ from the set that increases by $0.025$ increments from $0.025$ to $0.20$, including also $0$ and $0.01$. Looking at Figure \ref{fig: BF_CL}, we can choose $K = 4$, since it is the value from which the increase in the objective function does not justify an increase in the number of clusters. However, there is no clear indication of the optimal flagging level, which can be further investigated via the plots of the knee points. Following the same reasoning illustrated for the artificial data, we select $\alpha = 0.05$, as this is the level at which the median difference between the knee points of the $\Delta$ values and the corresponding flagging level is close to zero, and the variability is minimal (Figure \ref{fig: BF_KneedlePlot}). This choice can be corroborated via the inspection of the $\Delta$ plots provided in the Supplementary Material.

\begin{figure}[!htbp]
\centering
\includegraphics[width=0.7\linewidth, height=0.6\linewidth]{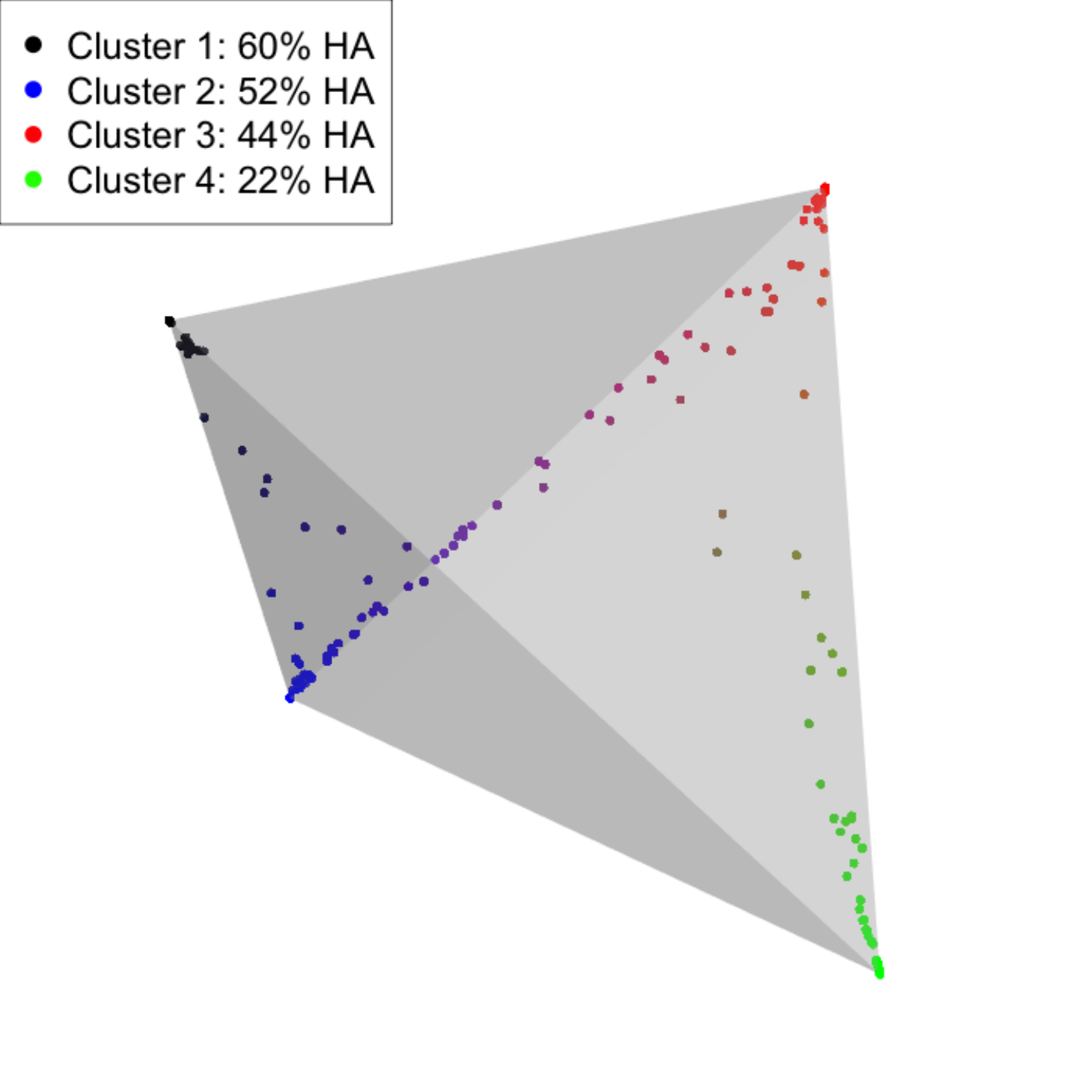}
\caption{Body fat data: tetrahedron plot showing cluster assignments (Cluster 1: black, Cluster 2: blue, Cluster 3: red, Cluster 4: green). Color intensity and point position reflect the degree of fuzziness (i.e., hard and soft memberships).}
\label{fig: BF_Fuzzy}
\end{figure}

The clustering results from cellFCLUST, using the tuning parameters set as previously detailed, are shown in Figure \ref{fig: BF_Fuzzy}. In the tetrahedral representation, points closer to the vertices have higher membership to the corresponding cluster, while those on the edges, or generally among vertices, are more fuzzified. The four clusters can be interpreted based on their configuration across variables (a graphical representation is provided in the Supplementary Material). \textit{Cluster 1} (black) consists of individuals with the lowest measurements for all variables, including \textit{BMI}, whose values, on the original scale, fall within the normal weight range, i.e. \textit{BMI} $< 25$. \textit{Cluster 2} (blue) includes men with generally higher measurements than Cluster 1 across the eleven variables, with \textit{BMI} values still mostly within the normal range. For \textit{Cluster 3} (red), some overlap with Cluster 2 is observed on specific variables (e.g., wrist), while \textit{BMI} predominantly falls within the overweight range, i.e. \textit{BMI} $\in [25, 30)$. Finally, \textit{Cluster 4} (green) groups individuals in the severe overweight to obese range -- the latter corresponds to \textit{BMI} $\geq 30$ -- with the highest values for all variables. As expected, Cluster 1 shows the highest proportion of HA (0.60 of the units in the cluster), and only $4$ units with WA. The fuzziness of these units is directed toward Cluster 2, with membership degrees to the latter ranging from $0.19$ to $0.37$. Cluster 2, which has a proportion of HA of 0.52, contains $31$ units with maximum membership below $0.9$; among them, $24$ have their second-highest membership in Cluster 3, and $7$ in Cluster 1. Cluster 3 includes $23$ weak assignments, with $13$ represented by men whose second-highest membership lies in Cluster 2. The remaining $10$ units would be secondarily assigned to Cluster 4, which, in turn, contains $17$ weakly assigned units whose second-highest membership is in Cluster 3. These individuals may represent a subgroup of men for whom further analyses could be conducted -- for instance, to assess whether they should follow specific dietary plans designed for individuals with obesity.

\begin{table}[t]
\centering
\caption{Proportion of outlying values detected by cellFCLUST in the body fat data set per variable and cluster. Total per row results in a fixed flagging level, which is $0.048$ in this case since $n - h = $ $250 - \left\lceil 0.95 \times 250 \right\rceil = 250 - 238 = 12$ units.}\label{tab: BF_out}
\begin{tabular}{lcccc}
\toprule
& Cluster 1 & Cluster 2 & Cluster 3 & Cluster 4 \\ 
\midrule
BMI & 0.000 & 0.012 & 0.020 & 0.016 \\ 
neck & 0.012 & 0.020 & 0.008 & 0.008 \\ 
chest & 0.004 & 0.012 & 0.020 & 0.012 \\ 
abdomen & 0.000 & 0.016 & 0.016 & 0.016 \\ 
hip & 0.004 & 0.012 & 0.012 & 0.020 \\ 
thigh & 0.004 & 0.004 & 0.012 & 0.028 \\ 
knee & 0.004 & 0.012 & 0.016 & 0.016 \\ 
ankle & 0.012 & 0.020 & 0.004 & 0.012 \\ 
bicep & 0.024 & 0.012 & 0.000 & 0.012 \\ 
forearm & 0.012 & 0.012 & 0.000 & 0.024 \\ 
wrist & 0.020 & 0.012 & 0.004 & 0.012 \\  
\bottomrule
\end{tabular}\\
\end{table}

Finally, we analyze the distribution of flagged cells per variable across the four clusters (Table \ref{tab: BF_out}). Specifically, it is worth noting that \textit{BMI} shows no unreliable cells in the first cluster, which corresponds to normal weight, while the cluster with the highest proportion of flagged cells -- considering a fixed total per variable -- is the third one. Similarly, Cluster 1 includes all reliable units for the variable \textit{abdomen}, while the other clusters share the same proportion of unreliable cells. Higher flagging levels are usually observed in Cluster 4, i.e., the group containing men with obesity. This cluster can also encompasses extremely obese individuals, whom the model is able to discover and assign to the correct cluster. It is important to highlight that the detection of outlying values by cellFCLUST does not occur marginally, but takes into account all reliable cells for each unit.

\subsection{Well-being: OECD regional data}\label{subsec: wellbeing}
The second empirical analysis focuses on well-being indicators published by the Organization for Economic Co-operation and Development (OECD) on a regional basis (\href{https://www.oecdregionalwellbeing.org/}{https://www.oecdregionalwellbeing.org/}). The dataset comprises $447$ statistical regions from $38$ OECD countries (Table \ref{fig: OECDregions}) and $11$ indicators, each ranging in $[0, 10]$. These are \textit{Education, Jobs, Income, Safety, Health, Environment, Civic Engagement, Accessibility to services, Housing, Community, Life satisfaction}. Although the scores have already been winsorized to mitigate the effect of marginal outlying values, the presence of non-marginal outliers justifies the need for a cellwise robust approach. Missing information is present in $1.6\%$ of the cells. For this data set a natural cluster structure can be expected, as dissimilarities between regions of different countries, or even continents, could be significant. Finally, the fuzzy approach can uncover hidden links between apparently distant regions.

\begin{table}[t]
\centering
\caption{Number of regions by OECD country.}\label{fig: OECDregions}
\resizebox{\linewidth}{!}{
\begin{tabular}{l r  l r  l r}
\toprule
Country & Regions & Country & Regions & Country & Regions\\ \midrule
Australia & 8 & Greece & 13 & New Zealand & 14\\
Austria & 9 & Hungary & 8 & Norway & 6\\
Belgium & 3 & Iceland & 2 & Poland & 17\\
Canada & 13 & Ireland & 3 & Portugal & 7\\
Chile & 16 & Israel & 6 & Slovak Republic & 4\\
Colombia & 33 & Italy & 21 & Slovenia & 2\\
Costa Rica & 6 & Japan & 10 & Spain & 17\\
Czech Republic & 8 & Korea & 7 & Sweden & 8\\
Denmark & 5 & Latvia & 6 & Switzerland & 7\\
Estonia & 5 & Lithuania & 10 & Türkiye & 26\\
Finland & 5 & Luxembourg & 1 & United Kingdom & 12\\
France & 18 & Mexico & 32 & United States & 51\\
Germany & 16 & Netherlands & 12 &  & \\ \bottomrule
\end{tabular}}
\end{table}

\begin{figure}[t]
    \centering
    \subfloat[][$S = 1$ \label{fig: OECD_S1}]{\includegraphics[width=.5\linewidth]{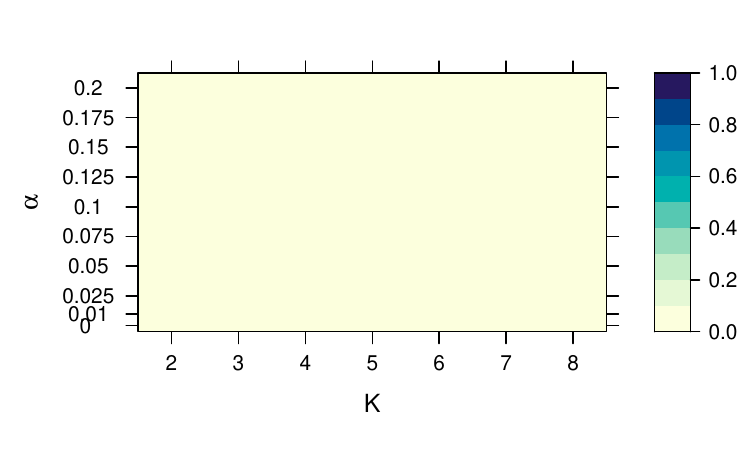}}
    \subfloat[][$S = 2$ \label{fig: OECD_S2}]{\includegraphics[width=.5\linewidth]{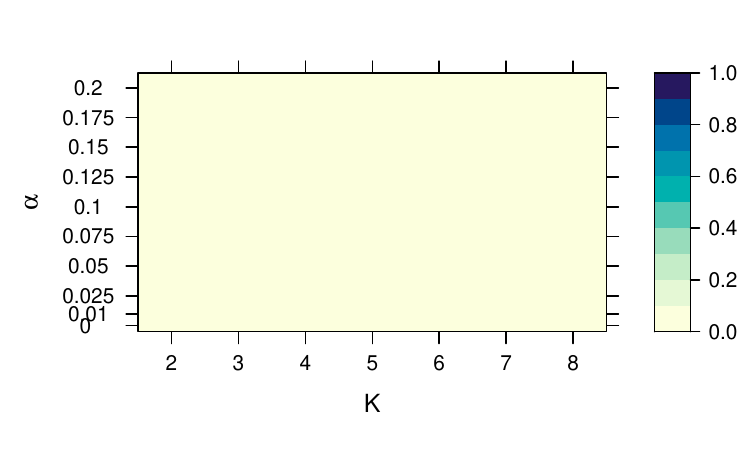}}\\
    \subfloat[][$S = 5$ \label{fig: OECD_S5}]{\includegraphics[width=.5\linewidth]{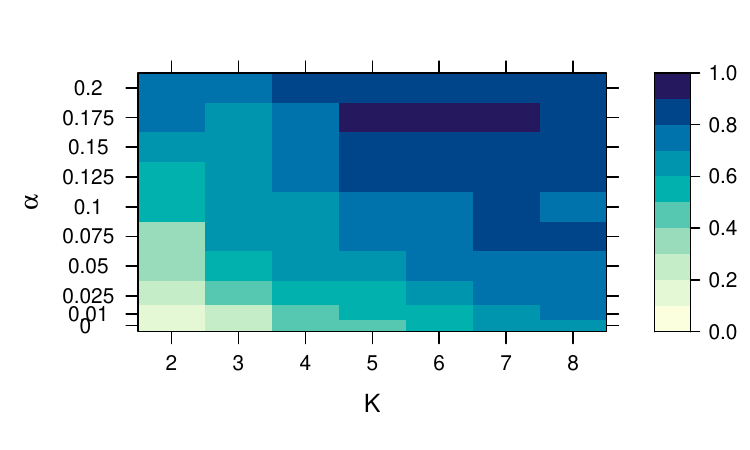}}
    \subfloat[][$S = 10$ \label{fig: OECD_S10}]{\includegraphics[width=.5\linewidth]{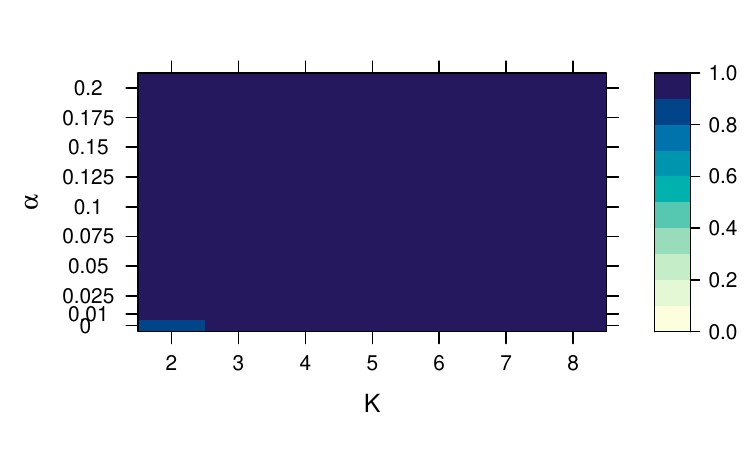}}
    \caption{OECD data: proportion of hard assignments depending on $K$ and $\alpha$ for four different levels of $S$.}
\end{figure}

\begin{figure}[t]
    \centering
    \includegraphics[width=0.9\linewidth,height=0.5\linewidth]{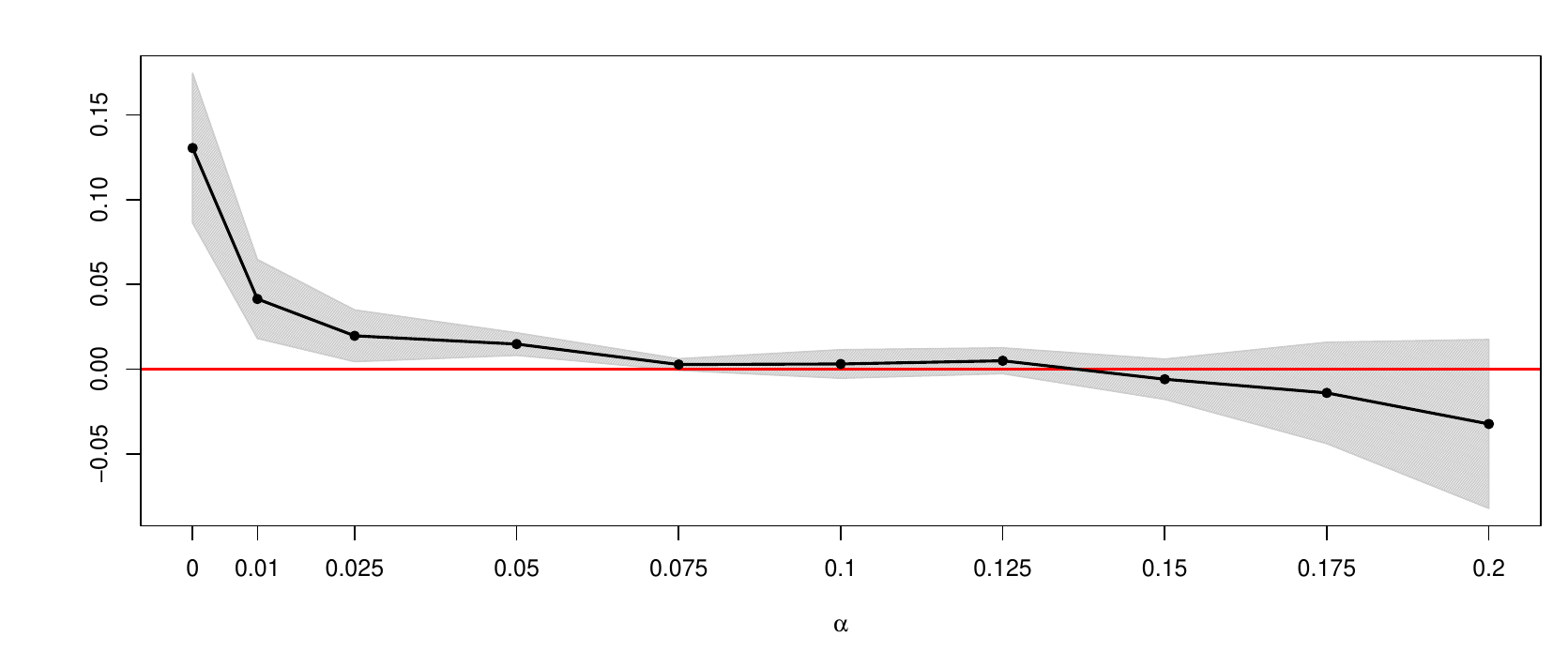}
    \caption{OECD data: difference between the knee point of $\Delta_{ij}$ and the $\alpha$ value. The plot shows the median differences across variables as $\alpha$ varies, with $K = 5$. The shaded gray area represents the variability.}\label{fig: oecd_med}
\end{figure}

Unlike the previous application, we choose a higher value of $c$, allowing for elongated clusters. This choice is motivated by the reasonable assumption of correlation between variables (e.g., \textit{Education} and \textit{Income}) within the same cluster, as well as different degrees of variability between clusters. Specifically, we impose $c = 50$, which corresponds to the order of magnitude of the ratio between the eigenvalues of the covariance matrix computed on the entire data set. Since fuzzy models are scale-dependent and a constraint on the eigenvalues is imposed, careful consideration should be given to the preprocessing of the variables. Given the nature of the indicators, we consider a selected number of scaling factors $S = \{1, 2, 5, 10\}$, which allow to adjust the proportion of hard assignments while maintaining high comprehensibility of the scaled scores. We select $S = 5$ as the proportion of HA approaches $0.74$  -- it ranges from $0.14$ to $0.91$ depending on the choices of $K$ and $\alpha$, as shown in Figure \ref{fig: OECD_S5}. The fuzzifier parameter $m$ is set to $1.8$ as it provides desirable levels of WA, with $29$ weakly assigned regions out of $447$ ($18$ for $m = 1.6$ and $55$ for $m = 2$, as shown in the Supplementary Material). Once $c$, $S$, and $m$ are defined, the objective function curves can be computed. As reported in the Supplementary Material, these curves suggest $K=5$ as a sensible choice. To support this choice, we include the plot of the knee points in Figure \ref{fig: oecd_med}.

\begin{figure}[!htbp]
      \includegraphics[width=.5\linewidth]{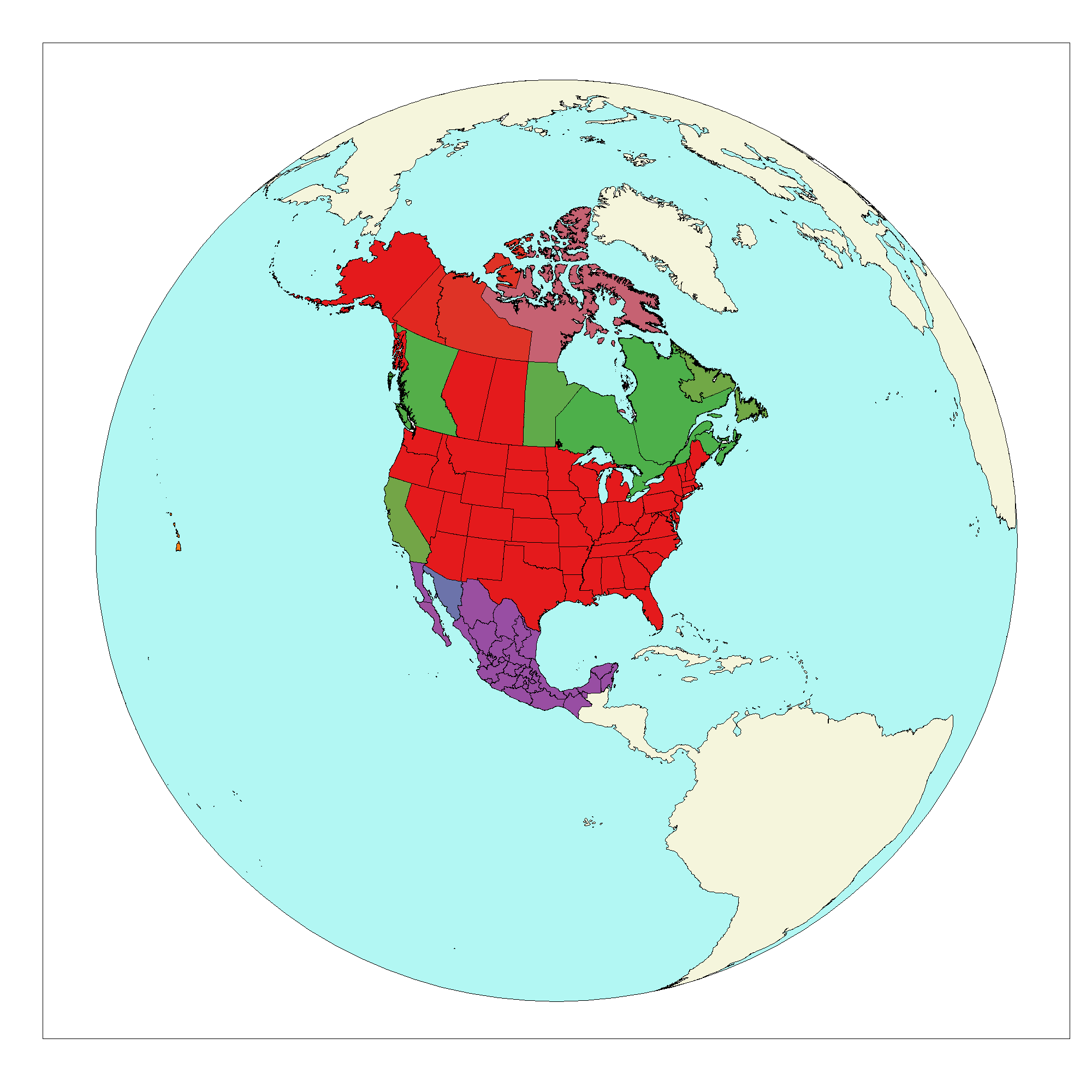}
      \includegraphics[width=.5\linewidth]{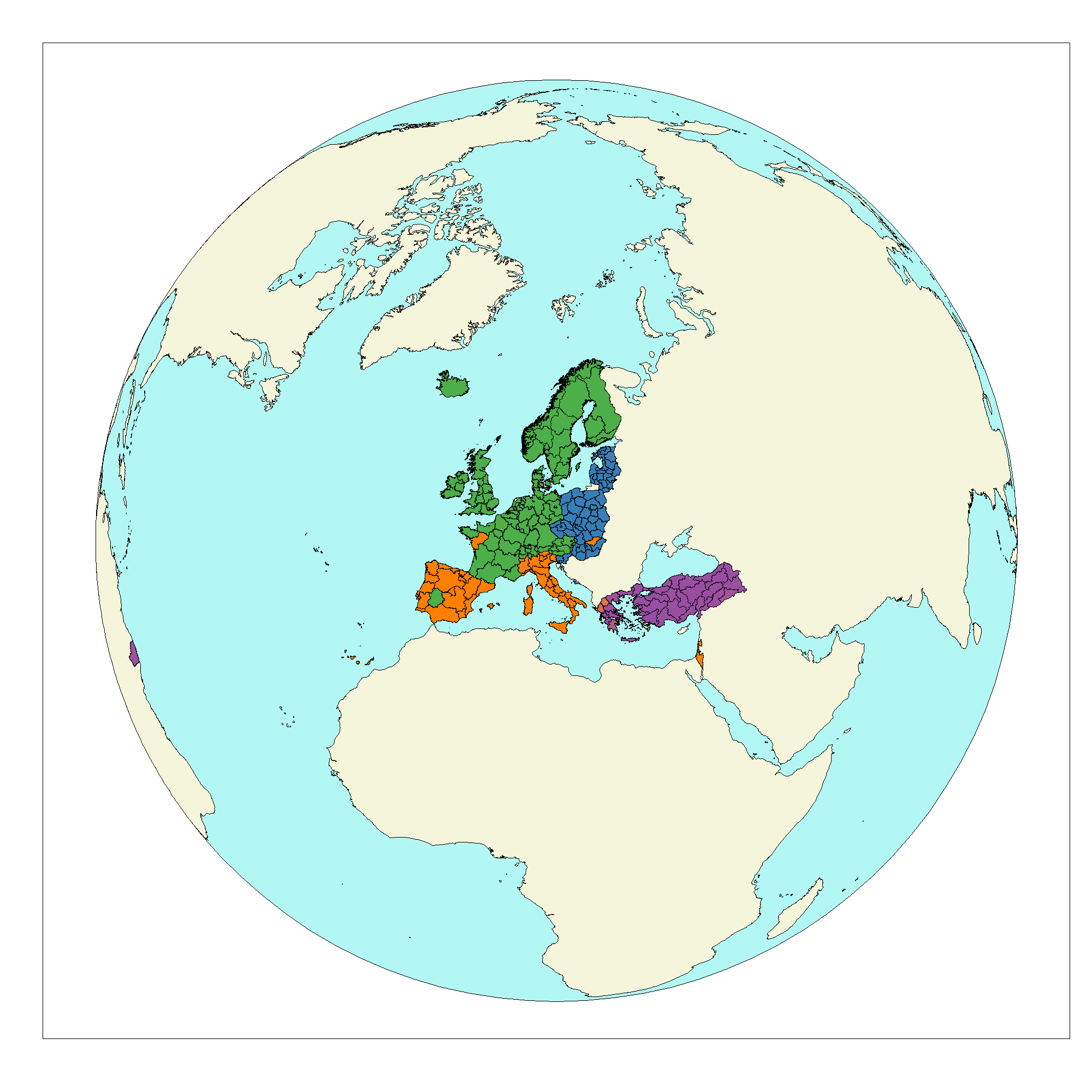}
      \includegraphics[width=.5\linewidth]{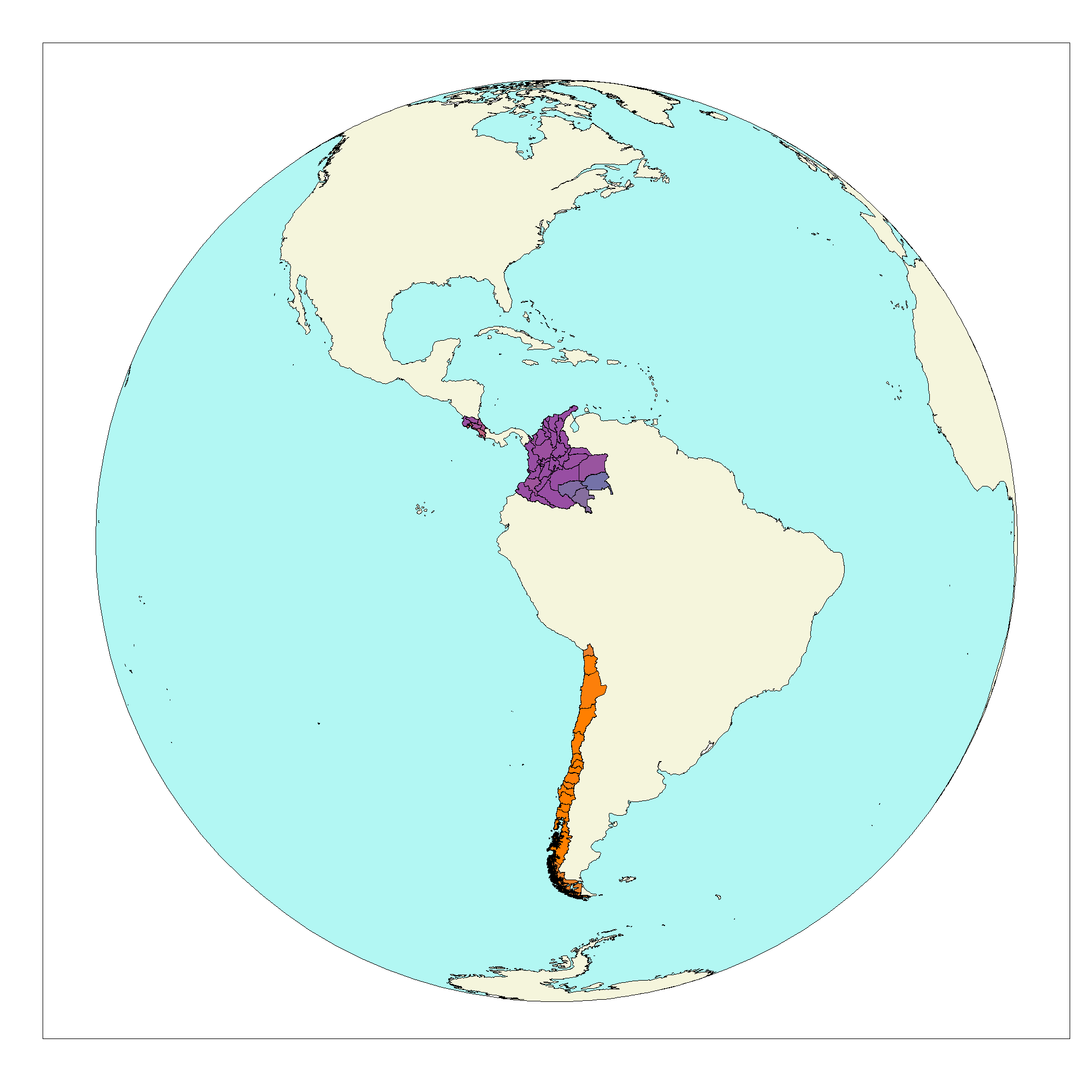}
      \includegraphics[width=.5\linewidth]{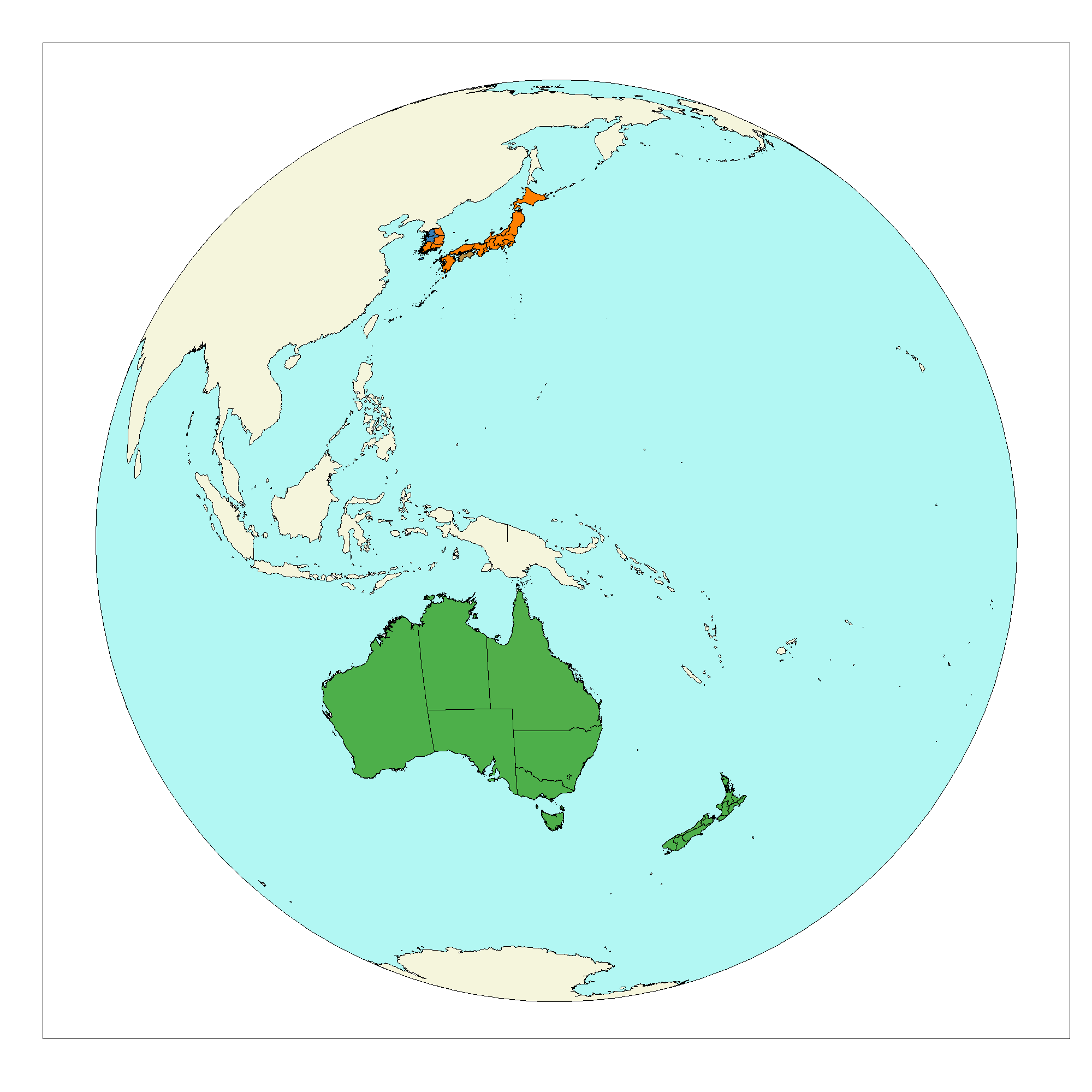}
      \caption{OECD data: clustering results. The color gradient indicates fuzzy assignments.\\ Cluster 1: \crule[col1]{0.3cm}{0.3cm} -- Cluster 2: \crule[col2]{0.3cm}{0.3cm} -- Cluster 3: \crule[col3]{0.3cm}{0.3cm} -- Cluster 4: \crule[col4]{0.3cm}{0.3cm} -- Cluster 5: \crule[col5]{0.3cm}{0.3cm}}
      \label{fig: oecd_map}
\end{figure}

As we can see in Figure \ref{fig: oecd_map}, clusters have a clear spatial characterization, which can be summarized as follows: Northern Europe and Oceania (Cluster 1), Eastern Europe (Cluster 2), United States of America (Cluster 3), Latin America and the Aegean Sea (Cluster 4), and Southern Europe, Asia and Chile (Cluster 5). Regions within the same country are often clustered together, with limited exceptions. Examples are Canadian regions, which are divided between United States of America and Northern Europe and Oceania, mainly driven by differences in \textit{Income} and \textit{Health}. Korean regions are also split between Eastern Europe and Southern Europe, Asia and Chile, as differences in \textit{Civic Engagement} and \textit{Life Satisfaction} discriminate the assignment. Another interesting case is that of France, where metropolitan regions are assigned to Northern Europe and Oceania, except for Pays de la Loire and Corsica, which are assigned to Southern Europe, Asia and Chile. However, overseas French regions such as Guadeloupe and French Guiana -- which are assigned to Latin America and the Aegean Sea -- do not fall into the same cluster as the mainland France. This may be due to their generally lower values of the indicators compared to those of Northern Europe and Oceania. Moreover, relevant fuzzy assignments can be found in many Greek regions, such as North Aegean, Peloponnese, South Aegean, Western Macedonia, Epirus, Ionian Islands, which belong to both Southern Europe, Asia and Chile, and Latin America and the Aegean Sea, with varying degrees. This is due to similarities within the cluster of Southern Europe, Asia and Chile offset by lower scores in \textit{Jobs}, \textit{Civic Engagement}, \textit{Community} and \textit{Life Satisfaction}, which are closer to those of Latin America and the Aegean Sea. Other fuzzy regions to highlight are the Colombian San Andrés, Amazonas, Guainía, and Guaviare which, although assigned to Eastern Europe, have a high membership value for Latin America and the Aegean Sea, and the Costa Rica Central, Central Pacific, Brunca and Huetar Caribbean, partly assigned to Southern Europe, Asia and Chile, due to similarities with the latter. A complete overview of the weakly assigned regions and their fuzzification is provided in the Supplementary Material.

\begin{figure}[!htbp]
  \centering
  \subfloat[][Northern Europe and Oceania\label{fig:OECD_out1}]{
    \includegraphics[width=0.48\linewidth]{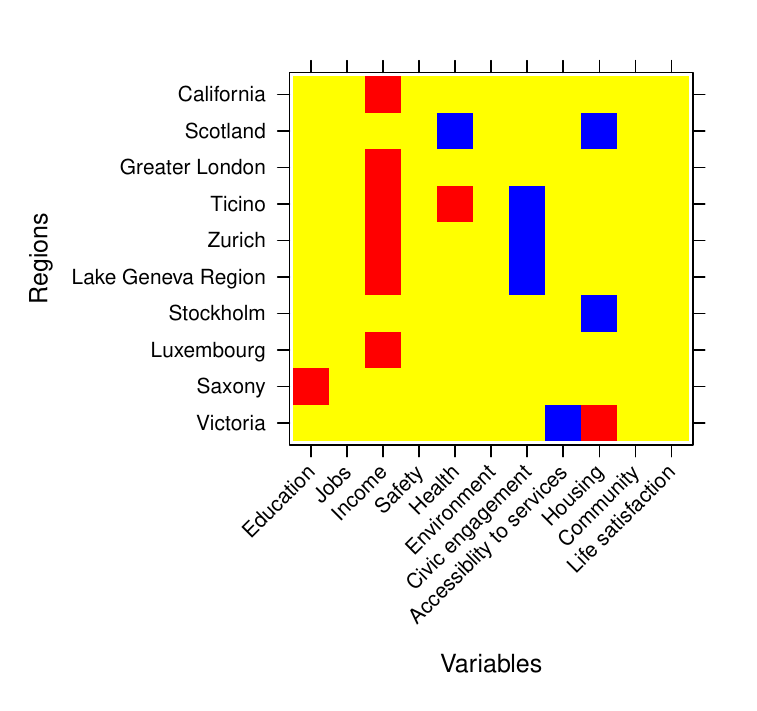}}
  \subfloat[][Eastern Europe\label{fig:OECD_out2}]{
    \includegraphics[width=0.48\linewidth]{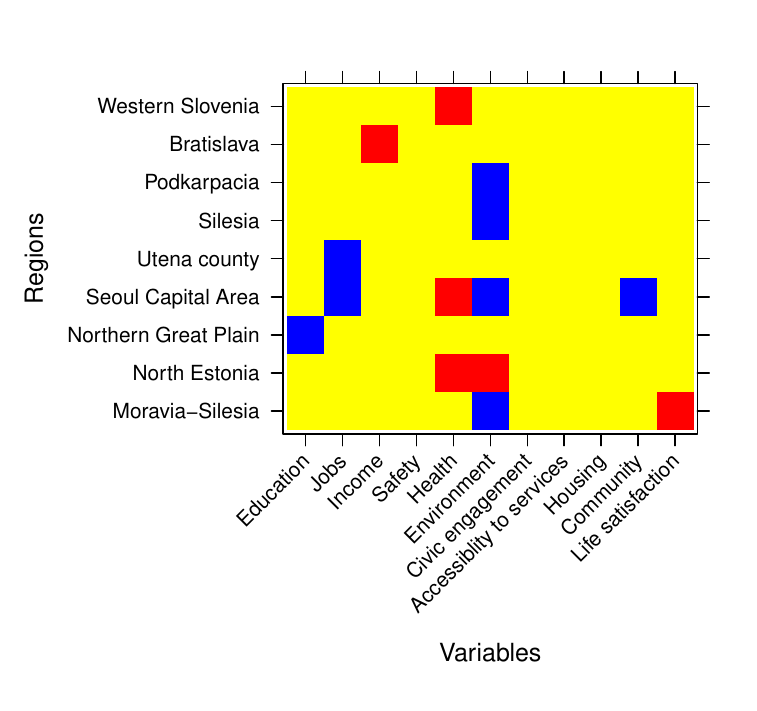}}\\
  \subfloat[][United States of America\label{fig:OECD_out3}]{
    \includegraphics[width=0.48\linewidth]{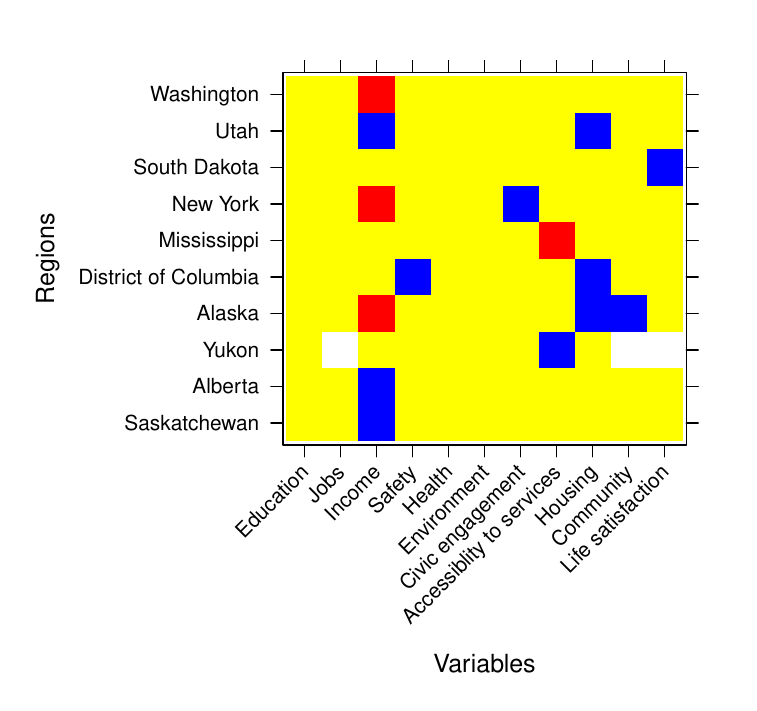}}
  \subfloat[][Latin America and the Aegean Sea\label{fig:OECD_out4}]{
    \includegraphics[width=0.48\linewidth]{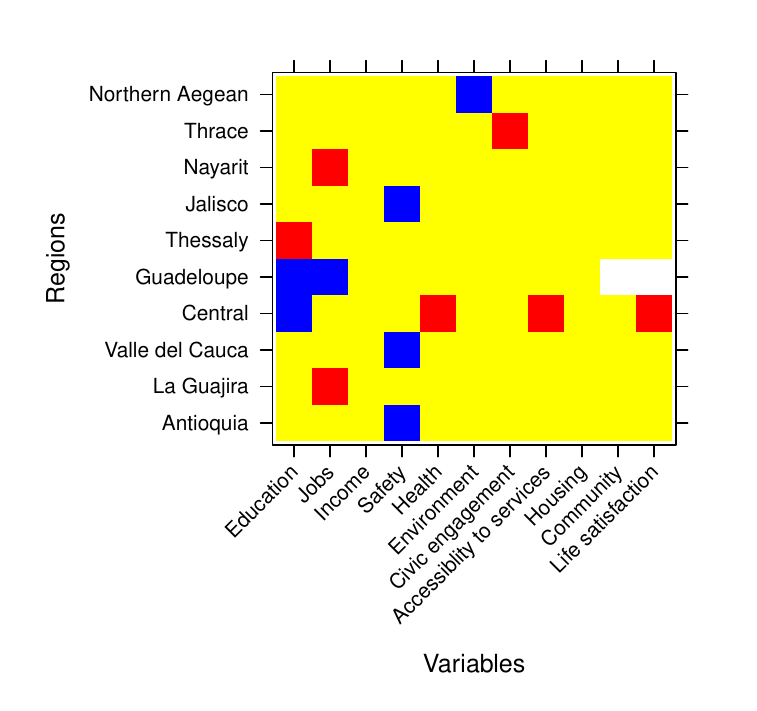}}
\end{figure}
\addtocounter{figure}{+1}
\begin{figure}[!htbp]\ContinuedFloat
  \centering
  \subfloat[][Southern Europe, Asia and Chile\label{fig:OECD_out5}]{
    \includegraphics[width=0.48\linewidth]{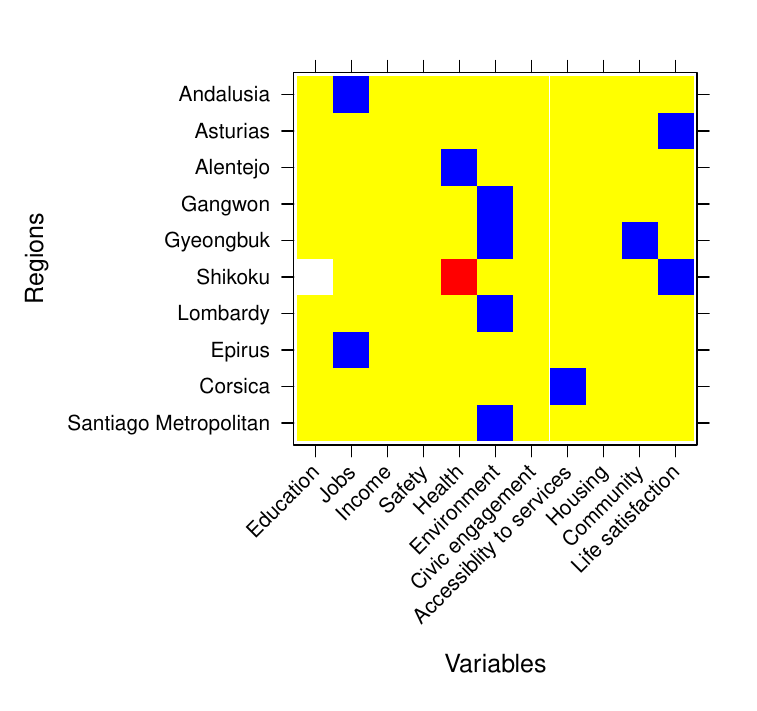}}
  \caption{OECD data: outlying cells of selected regions (yellow: reliable cells; blue: flagged cells imputed with a higher value than the original one; red: flagged cells imputed with a lower value than the original one; white: missing values).}\label{fig: OECD_W_selected}
\end{figure}

In terms of unreliable cells, we see in Figure \ref{fig: OECD_W_selected} that California is considered outlying with respect to \textit{Income}, as the difference between salaries in this region and those in Northern Europe and Oceania -- the cluster which California belongs -- is significant. Another example is Lombardy, which has a much lower score for \textit{Environment}, being a region with higher levels of industrialization, compared to other Italian regions or other regions of Southern European countries, causing significant pollution. Silesia represents another outlier in \textit{Environment}, due to the presence of a substantial coal mining industry in the region. With respect to \textit{Housing}, the Stockholm region presents an outlying value, as it hosts one of the most populated cities in all of Northern Europe and Oceania. New York is outlying both in \textit{Income}, due to notably high salaries, and \textit{Civic Engagement}, representing an example of a non-marginal outlier with lower and higher values than expected, respectively, given the other indicators. Jalisco, as other regions in its cluster, has a low score of \textit{Safety}, caused by increased exposure to criminal activities compared to other regions. It is also possible to observe countries where all the cells for a specific variable have been flagged as outlying. Particularly, Swiss regions are all considered outlying in \textit{Income}, which is unusually high compared to their peers, and \textit{Civic Engagement}, as voter turnout in the country is significantly lower than in most European nations. The same occurs for South Korea in \textit{Environment}, as the levels of pollution in the peninsula are consistently high. This result from cellFCLUST is interesting, as it helps uncover differences within clusters, where subgroups (e.g., all, or almost all, regions of a country) may differ from other units only in a small subset of variables. Depending on the purpose of the analysis, the user may increase the number of clusters and reduce the flagging level by merging these regions into additional clusters and avoiding the flagging of their values, if these are of interest. However, this comes at the cost of reduced model parsimony, which may not always be desirable. A complete overview of the flagged cells in each cluster is available in the Supplementary Material.

\section{Conclusions}\label{sec: conclusions}
A new methodology, called cellFCLUST, has been introduced for fuzzy clustering and cellwise outlier detection. Following the recent paradigm of cellwise contamination, which considers single cells of a data matrix to be potentially contaminated, we have developed an algorithm intended for flagging these cellwise outliers, which are then treated as missing values, and imputed rather than discarded, before estimating the model parameters. If not properly identified, outlying values and fuzzy assignments can interplay in ways that change the clustering structure and bias the parameter estimates. Notably, cellFCLUST encompasses both hard and soft robust clustering methodologies: crisp assignment decisions are made for cases located in the central regions or cores of the clusters, while fuzzy assignments are retained for ambiguous units.

Through a simulation study and two real-world applications, we have shown the effectiveness and usefulness of the proposed model. The former illustrate the performance of cellFCLUST in cluster recovery, parameter estimation and outlier detection compared to other fuzzy clustering methodologies. The results demonstrate the advantages of the proposal when cellwise contamination occurs. Regarding the real data analyses, the first one focuses on analyzing risk levels among individuals in the overweight to obese range, leveraging fuzzification to identify units whose measurements warrant closer inspection; the second application concerns indicators used to study well-being across regions of the OECD countries. In this framework, cellFCLUST enables the identification of common patterns among countries in terms of well-being, serving two main purposes: on one hand, to reveal different behaviors among regions within the same country -- for example, California is grouped with Northern European countries based on its measurement; and on the other hand, to highlight indicators on which some countries (and their regions) display anomalous values relative to the cluster they belong to. This is the case, for instance, of Switzerland, whose income is higher even than that of Northern European countries.  

A crucial role in the implementation of cellFCLUST is played by its tuning parameters, as the clustering and outlier detection results can be sensitive to their choice. We have illustrated their effects on artificial data, providing guidance to users for their selection. It is worth noting that all these parameters -- namely, the number of clusters $K$, the flagging level $\alpha$, the constant $c$ for the eigenvalue-ratio constraint, the fuzzifier parameter $m$, and the scale factor $S$ -- are interconnected and require interrelated analyses for their appropriate setting. The dependence of cellFCLUST on these parameters offers users a flexible methodology that accommodates on the purpose of the analysis and prevents issues in the parameter estimation. Simplifying these connected tasks remains an open issue, leaving room for improvement and future developments aimed at identifying a procedure for the simultaneous selection of all tuning parameters in cellFCLUST. Future work could also explore extending the robust fuzzy clustering approach with a factorial structure for the cluster covariance matrices \cite{GEGMI:2018} to the cellwise contamination framework.

\section*{Declaration of competing interest}
The authors declare that they have no known competing financial interests or personal relationships that could have appeared to influence the work reported in this paper.

\section*{Code availability}
The \texttt{R} code for the implementation of cellFCLUST and the simulation study in Section \ref{sec: simulations} is available at \url{https://github.com/giorgiazaccaria/cellFCLUST}.

\section*{Data availability}
The real data used in Section \ref{sec: applications} are publicly available, as detailed in the corresponding subsections.

\section*{Funding}
The research of Giorgia Zaccaria and Francesca Greselin was supported by Milano-Bicocca University Fund for Scientific Research, 2023-ATE-0448. Francesca Greselin's research was also supported by PRIN2022 - 2022LANNK\\C. The research of Luis A. García-Escudero and Agustín Mayo-Iscar was partially supported by the Spanish Ministerio de Ciencia, Innovaci\'{o}n y Universidades, grant PID2024-162240NB-I00, and Junta Castilla y Le\'{o}n grant VA064G24.

\appendix
\section{Notation}\label{sec: notation}
\beginsupplement
For the convenience of the reader, the notation used in this paper is listed in Table \ref{tab: notation}.
\small{
\begin{table}[!htbp]
\caption{Notation of the paper.}\label{tab: notation}
\begin{tabular}[h]{p{2.9cm}|p{9.5cm}}

\hline
Notation & Description \\

\hline
$n, p, K$ & Number of units, variables, clusters, respectively (scalars). \\
\hline
$m$ & Fuzzifier tuning parameter (scalar). \\ 
\hline
$\alpha$ & Proportion of cells per variable flagged as contaminated in a data matrix, referred to as \textit{flagging level} (scalar). \\
\hline
$\vec{X}$ & $(n \times p)$ data matrix, where $\vec{x}_{i}$ denotes its $i$-th row (unit) containing $p$ variable measurements, and $x_{ij}$ represents the cell corresponding to the $j$-th measurement for the $i$-th unit. \\
\hline
$\vec{W}$ & $(n \times p)$ cellwise indicator matrix, where zeros denote contaminated or missing cells. \\
\hline
$\vec{U}$ & $(n \times K)$ membership matrix of units to cluster. \\
\hline
$\pi_{k}$ & Weight of cluster $k$, $k = 1, \ldots, K$ (scalar). \\
\hline
$\mean_{k}$ & $p$-dimensional vector of variable means for cluster $k$, $k = 1, \ldots, K$. \\
\hline
$\mean_{k[j]}, \mean_{k[\wi]}$ & Sub-vectors of $\mean_{k}$ corresponding to the $j$-th variable and the variables for which $w_{ij} = 1, j \in \{1, \ldots, p\}$, respectively.\\
\hline
$\sigu_{k}$ & $(p\times p)$ covariance matrix for cluster $k$, $k = 1, \ldots, K$. \\
\hline
$\sigu_{k[j, j]}$ & $(j, j)$-element of $\sigu_{k}$ (scalar), corresponding to the variance of the $j$-th variable. \\
\hline
$\sigu_{k[j, \wi]}, \sigu_{k[\wi, \wi]}$ & Row vector and sub-matrix of $\sigu_{k}$, respectively, corresponding to the covariances between the $j$-th variable and the variables for which $w_{ij} = 1$, and to the covariances between the latter. \\
\hline
$c$ & Constant for the eigenvalue-ratio constraint on covariance matrices (scalar).\\
\hline
\end{tabular}
\end{table}
}

\bibliographystyle{elsarticle-num}
\bibliography{references}
\end{document}

\begin{frontmatter}
\title{Supplementary Material to ``Robust fuzzy clustering with cellwise outliers''}
\author{Giorgia Zaccaria$^a$, Lorenzo Benzakour$^b$, Luis A. García-Escudero$^c$, Francesca Greselin$^b$, Agustín Mayo-Íscar$^c$} %% Author name

\affiliation{organization={Department of Economics, Management and Statistics, University of Milano-Bicocca},%Department and Organization
            addressline={Via Bicocca degli Arcimboldi 8}, 
            city={Milan},
            postcode={20100}, 
            country={Italy}}
\affiliation{organization={Department of Statistics and Quantitative Methods, University of Milano-Bicocca},%Department and Organization
            addressline={Via Bicocca degli Arcimboldi 8}, 
            city={Milan},
            postcode={20100}, 
            country={Italy}}
\affiliation{organization={Department of Statistics and Operational Research, University of Valladolid},%Department and Organization
            addressline={Paseo de Belén 7}, 
            city={Valladolid},
            postcode={47011}, 
            country={Spain}}
\begin{abstract}
This document includes the supplementary material to the main article \say{Robust fuzzy clustering with cellwise outliers}. Specifically, it contains details on the proposed methodology and additional results for the simulation study, the tuning parameter selection, the analysis of the body fat data set and the OECD regional data set on well-being.
\end{abstract}
\end{frontmatter}

\section{Details on the cellFCLUST algorithm}
In this section, we provide additional details on the cellFCLUST algorithm presented in Section 3.2 of the main article. Specifically, we summarize the Gaussian distributions properties used throughout the algorithm (Section \ref{subsec: gaussianprop}); we derive the expression for the $\Delta_{ij}$ values (Section \ref{subsec_sup: delta}); we review the parameter estimates obtained when missing information occurs in the data, recalling that unreliable cells -- consisting of both contaminated and missing values -- are treated as missing (Section \ref{subsec_sup: missing}); and we prove the monotonicity of the cellFCLUST objective function (Section \ref{subsec_sup: convergence}). 

\subsection{Gaussian distribution properties}\label{subsec: gaussianprop}
Let $\vec{x} \in \mathbb{R}^{p}$ be a random vector with $\vec{x} \sim \mathcal{N}(\mean, \sigu)$. If we partition $\vec{x}$ into two non-empty blocks, and correspondingly the mean vector $\mean \in \mathbb{R}^{p}$ and the positive definite covariance matrix $\sigu \in \mathbb{R}^{p \times p}$, as follows 
\begin{equation*}
    \vec{x} =   \begin{bmatrix}
                \vec{x}_{1} \\
                \vec{x}_{2}
                \end{bmatrix}  
    \quad       
    \mean =   \begin{bmatrix}
                \mean_{1} \\
                \mean_{2}
                \end{bmatrix}  
    \quad   
    \sigu =   \begin{bmatrix}
                \sigu_{11} & \sigu_{12} \\
                \sigu_{21} & \sigu_{22}
                \end{bmatrix},         
\end{equation*}
then 
$\vec{x}_{1} \sim \mathcal{N}(\mean_{1}, \sigu_{11})$ and $\vec{x}_{2} \lvert \vec{x}_{1} \sim \mathcal{N}(\mean_{2 \lvert 1}, \sigu_{2 \lvert 1})$, where
\begin{align}
    &\mean_{2 \lvert 1} = \mean_{2} + \sigu_{21}(\sigu_{11})^{-1}(\vec{x}_{1} - \mean_{1}) := \widehat{\vec{x}}_{2} \label{eqn_sup: cond_mean}\\
    &\sigu_{2 \lvert 1} = \sigu_{22} - \sigu_{21} (\sigu_{11})^{-1} \sigu_{12}:= \vec{C}_{2} \label{eqn_sup: cond_cov}.
\end{align}
Note that $\sigu_{21} = (\sigu_{12})^{\prime}$ since $\sigu$ is symmetric by definition, and that $\sigu_{11}$ is positive definite because it is a leading principal minor of $\sigu$ (Sylvester's criterion; see \cite{G:1991}). Additionally, the determinant of $\sigu$, partitioned as described above, can be expressed as follows
\begin{equation}\label{eqn_sup: determinant}
    \lvert \sigu \rvert = \lvert \sigu_{11} \rvert \lvert \vec{C}_{2} \rvert,
\end{equation}
where $\vec{C}_{2}$ is defined in (\ref{eqn_sup: cond_cov}).

\subsection{Derivation of the $\Delta$ values}\label{subsec_sup: delta}
The $\Delta_{ij}$ values represent the difference between the individual contributions to the objective function when the $j$-th variable for the $i$-th unit is considered reliable ($w_{ij} = 1$) or contaminated ($w_{ij} = 0$), while the other values $w_{ij^{\prime}}, j^{\prime} \neq j$, remain unchanged. It is worth noting that when a cell is missing, $w_{ij}$ is set to $0$ and is not updated through iterations, i.e., for the $i$-th unit, $\Delta_{ij}$ is not computed for the $j$-th variable. Therefore, the $\Delta_{ij}$ values are computed only on the observed cells.

We recall that the individual contribution to the cellFCLUST objective function is given by
 \begin{align}
        &J_{\text{cellFCLUST}}^{(i)} (\wi, \vec{u}_{i}, \{\pi_{k}, \mean_{k}, \sigu_{k}\}_{k = 1}^{K})  = \sum_{k = 1}^{K} u_{ik}^{m} \log(\pi_{k}) -\dfrac{1}{2} \sum_{k = 1}^{K} u_{ik}^{m} \Big[p[\wi] \nonumber \\
        &\log(2\pi) + \log \lvert \sigu_{k[\wi, \wi]} \rvert + (\vec{x}_{i[\wi]} - \mean_{k[\wi]})^{\prime} \sigu_{k[\wi, \wi]}^{-1} (\vec{x}_{i[\wi]} - \mean_{k[\wi]}) \Big]. \label{eqn_sup: contri}
    \end{align}
When comparing Eq. (\ref{eqn_sup: contri}) under $w_{ij} = 1$ and $w_{ij} = 0$, the difference in the vector $\wi$ is confined to a single cell, i.e., that in position $(i, j)$, while the others remain unchanged. Denoting by $\text{MD}^{2}(\wi, \vec{x}_{i}, \mean_{k}, \sigu_{k}) =  (\vec{x}_{i[\wi]} - \mean_{k[\wi]})^{\prime} \sigu_{k[\wi, \wi]}^{-1} (\vec{x}_{i[\wi]} - \mean_{k[\wi]})$ the partial squared Mahalanobis distance, we obtain 
\begin{align*}
        \Delta_{ij} &= J_{\text{cellFCLUST}}^{(i)}(\wi, \vec{u}_{i}, \{\pi_{k}, \mean_{k}, \sigu_{k}\}_{k = 1}^{K} \lvert w_{ij} = 1) \\
        &\quad - J_{\text{cellFCLUST}}^{(i)}(\wi, \vec{u}_{i}, \{\pi_{k}, \mean_{k}, \sigu_{k}\}_{k = 1}^{K} \lvert w_{ij} = 0) \\
        &= \sum_{k = 1}^{K} u_{ik}^{m} \log(\pi_{k}) -\sum_{k = 1}^{K} u_{ik}^{m} \log(\pi_{k}) -\dfrac{1}{2} \sum_{k = 1}^{K} u_{ik}^{m}  \Big[ \\
        &\quad \; p[\wi \lvert w_{ij} = 1] \log(2\pi) - p[\wi \lvert w_{ij} = 0] \log(2\pi) \\
        &\quad + \log \lvert \sigu_{k[\wi, \wi \lvert w_{ij} = 1]} \rvert - \log \lvert \sigu_{k[\wi, \wi \lvert w_{ij} = 0]} \rvert \\
        &\quad + \text{MD}^{2}(\wi, \vec{x}_{i}, \mean_{k}, \sigu_{k} \lvert w_{ij} = 1) - \text{MD}^{2}(\wi, \vec{x}_{i}, \mean_{k}, \sigu_{k} \lvert w_{ij} = 0) \Big] \\
        &= - \dfrac{1}{2} \sum_{k = 1}^{K} u_{ik}^{m} \Bigg[\log(2\pi) + \log(C_{ij(k)}) + \dfrac{(x_{ij} - \hat{x}_{ij(k)})^2}{C_{ij(k)}} \Bigg], 
    \end{align*}
where $[\cdot \vert w_{ij} = 1]$ and $[\cdot \vert w_{ij} = 0]$ indicate that, in the vector $\wi$, the cell $(i, j)$ is considered reliable or contaminated, respectively, while the other cells $w_{ij^{\prime}}, j^{\prime} \neq j$, remain unchanged; $\hat{x}_{ij(k)}$ and ${C_{ij(k)}}$ are given by Eqs. (\ref{eqn_sup: cond_mean}) and (\ref{eqn_sup: cond_cov}), respectively, for $\vec{x}_{1} = \vec{x}_{i[\wi]}$ and $\vec{x}_{2} = x_{ij}$. To derive the third equality, we use Proposition 5 in \cite{RR:2023} and Eq. (\ref{eqn_sup: determinant}), i.e., $$\log \lvert \sigu_{k[\wi, \wi \lvert w_{ij} = 1]} \rvert = %\log(\lvert \sigu_{k[\wi, \wi \lvert w_{ij} = 0]} \rvert \lvert C_{ij(k)} \rvert)
\log \lvert \sigu_{k[\wi, \wi \lvert w_{ij} = 0]} \rvert + \log (C_{ij(k)})$$
where $\log (C_{ij(k)}) = \log \lvert C_{ij(k)} \rvert$ as $C_{ij(k)}$ is a scalar. 

\subsection{Imputation and parameter estimation with missing values}\label{subsec_sup: missing}
The conditional mean and covariance matrix computed in \textit{Step 3} of the cellFCLUST algorithm can be easily derived from Eqs. (\ref{eqn_sup: cond_mean}) and (\ref{eqn_sup: cond_cov}), respectively, for $\vec{x}_{1} = \vec{x}_{i[\wi]}$ and $\vec{x}_{2} = \vec{x}_{i[\wi^{c}]}$, given the current parameter $\mean_{k}$ and $\sigu_{k}$.

The update of $\mean_{k}$ and $\sigu_{k}$ are obtained as in \cite{DLR:1977, GJ:1994}. Specifically, plugging $\widehat{\vec{x}}_{i[\wi^{c}](k)} = \mean_{k[\wi^{c} \lvert \wi]}$ into $\vec{x}_{i} = (\vec{x}_{i[\wi]}, \vec{x}_{i[\wi^{c}]})$, we get  $\Tilde{\vec{x}}_{i(k)} = (\vec{x}_{i[\wi]}, \widehat{\vec{x}}_{i[\wi^{c}](k)})$, and we can consequently update $\mean_{k}$. For $\sigu_{k}$, it has to be noted that
\begin{align}
    &(\Tilde{\vec{x}}_{i(k)} - \mean_{k}) (\Tilde{\vec{x}}_{i(k)} - \mean_{k})^{\prime} + \vec{C}_{i[\wi^{c},\wi^{c}](k)} = \nonumber\\
    &= \begin{bmatrix}
        (\vec{x}_{i[\wi]} - \mean_{k[\wi]}) (\vec{x}_{i[\wi]} - \mean_{k[\wi]})^{\prime} & (\vec{x}_{i[\wi]} - \mean_{k[\wi]}) (\widehat{\vec{x}}_{i[\wi^{c}](k)} - \mean_{k[\wi^{c}]})^{\prime} \\
        (\widehat{\vec{x}}_{i[\wi^{c}](k)} - \mean_{k[\wi^{c}]}) (\vec{x}_{i[\wi]} - \mean_{k[\wi]})^{\prime} & (\widehat{\vec{x}}_{i[\wi^{c}](k)} - \mean_{k[\wi^{c}]}) (\widehat{\vec{x}}_{i[\wi^{c}](k)} - \mean_{k[\wi^{c}]})^\prime  \\
        & +\vec{C}_{i[\wi^{c},\wi^{c}](k)}
    \end{bmatrix}, \label{eqn_sup: covest}
\end{align}
where $\vec{C}_{i[\wi^{c},\wi^{c}](k)} = \sigu_{k[\wi^{c}, \wi^{c} \lvert \wi]}$. It is worth noting that, if we consider two iterations of the algorithm, say $t$ and $t+1$, $\widehat{\vec{x}}_{i[\wi^{c}](k)}$ is based on $\mean_{k}$ and $\sigu_{k}$ computed at iteration $t$, while $\mean_{k[\wi]}$ and $\mean_{k[\wi^{c}]}$ in (\ref{eqn_sup: covest}) are the sub-vectors of $\mean_{k}$ computed at iteration $t + 1$.

\subsection{Monotonicity of the cellFCLUST algorithm}\label{subsec_sup: convergence}
This subsection is devoted to showing that \textit{Steps 1}, \textit{2}, and \textit{3-4} of the proposed algorithm (Section 3.2 of the main article) yield non-decreasing values of the cellFCLUST objective function. As a result, each pass through these four steps produces a monotonic non-decreasing change in the objective function, and repeated applications of these steps lead to reach a local maximum of the objective function for any given initialization. To enhance the probability of reaching the global maximum, the algorithm should be run with multiple initializations, and the best solution among them should be retained as the output of the algorithm.

\textit{Step 1} -- starting from the current configuration of $\vec{W}$, with the membership values in $\vec{U}$, and the parameter estimates $\pi_{k}, \mean_{k}$ and $\sigu_{k}, k = 1, \ldots, K$, fixed at their values from the previous iterations -- uses the $\Delta_{ij}$ values to sequentially update (column by column) the $p$ columns of $\vec{W}$. Each of these updates modifies the objective function in a non-decreasing manner, so that the final outcome of Step 1 is an improvement of the objective function. See the arguments presented in the previous sections of this Supplementary Material for the rationale behind these updates.

On the other hand, with $\vec{W}$ and the parameters now assumed fixed, the best possible update of the values of $\vec{U}$, without decreasing the objective function, is obtained as described in \textit{Step 2} of the algorithm. This is properly justified in the main article, when presenting Step 2.

Finally, if we consider $\vec{W}$ and $\vec{U}$ as fixed, \textit{Steps 3-4} increase the objective function when updating the parameters in the proposed manner. To justify this, note that once $w_{ij}$ and $u_{ik}$ are fixed, the objective function corresponds to a sum of $K$ terms, each resembling the likelihood typically encountered in missing-data problems assuming a multivariate normal population \cite{DLR:1977}, but where units have weights $u_{ik}^m$ for $i = 1, \dots, n$, instead of equal weights $1/n$. Consequently, the proposed Steps 3-4 correspond to the application of the EM steps introduced in \cite{DLR:1977}, the only difference being the consideration of weights $u_{ik}^m$, which leads to the parameter updates described.

\section{Simulation study: comparison between cellFCLUST and cellGMM}
In Section 4 of the main article, we evaluate the performance of cellFCLUST in comparison with other \textit{fuzzy clustering} methods, both robust and non-robust. These fuzzy clustering methodologies are specifically tailored to handle cluster overlap through soft assignments. In the clustering literature, finite mixture models can also provide soft assignments of units via posterior probabilities, but do not allow the flexibility of fuzzy clustering to tune the levels of weak (and hard) assignments, which can be relevant in various applications.

In this section, we compare cellFCLUST with cellGMM \cite{ZACetal:2025}, a Gaussian mixture model for cellwise outlier detection. Regarding parameter recovery and outlier detection, it is worth noting that cellGMM results are comparable to those of cellFCLUST, as expected, as are the Misclassification Rate (MR) and Adjusted Rand Index (ARI), which are computed after assigning each unit to a cluster based on membership values (cellFCLUST) or posterior probabilities (cellGMM). On the other hand, differences in the estimation of $\vec{U}$ and the proportion of Weak Assignments (WA) can be highlighted. Specifically, cellFCLUST can adjust for cluster overlap and explicitly control the fuzziness of cluster assignments through the parameter $m$, whereas cellGMM does not. This difference is reflected into the estimation of the membership matrix $\vec{U}$ and the proportion of WA -- the latter defined as the fraction of units whose highest membership is below $0.90$ -- allowing them to belong to multiple clusters with non-negligible assignments. As shown in Table \ref{tab_sup: simstudy_cellmethods}, cellFCLUST better reconstructs the theoretical membership matrix $\vec{U}$ and captures fuzziness more effectively than cellGMM. This is also evident by looking at the WA values across scenarios and percentages of contamination, which are extremely small for cellGMM, whereas for cellFCLUST they approach the true values ($0.05$ for Scenario 1 and $0.07$ for Scenario 2). These results demonstrate the advantages of fuzzy clustering over model-based clustering in that particular sense -- which holds for both robust and non-robust approaches -- even though the broad applicability of the latter is unquestionable. Finally, simulations varying $m$ could offer further insight, but they are beyond the scope of this paper.

\begin{table}[!ht]
\centering
\caption{Results of the simulation study: Root Mean Squared Error (RMSE) of the membership matrix and proportion of Weak Assignments (WA) averaged over $100$ samples per scenario, percentage of contamination, and model. Entries are scaled by a factor of $10^{-3}$.}\label{tab_sup: simstudy_cellmethods}
\resizebox{0.8\textwidth}{!}{
\begin{tabular}{cl|cc|cc}
  \hline
\multicolumn{2}{c}{} & \multicolumn{2}{c}{Scenario 1} & \multicolumn{2}{c}{Scenario 2} \\
$\%$ out & Method & RMSE $\vec{U}$ & WA & RMSE $\vec{U}$ & WA \\ 
\hline
\multirow{2}{*}{0} & cellFCLUST & 0.83 & 33.64 & 0.26 & 50.72 \\
& cellGMM & 2.04 & ~2.12 & 1.05 & ~0.62 \\ 
\hline
\multirow{2}{*}{1} & cellFCLUST & 0.77 & 31.68 & 0.27 & 52.28 \\
& cellGMM & 2.11 & ~2.00 & 1.10 & ~0.54 \\ 
\hline
\multirow{2}{*}{5} & cellFCLUST & 0.97 & 38.84 & 0.34 & 57.02 \\
& cellGMM & 2.55 & ~2.56 & 1.24 & ~0.70\\
\hline
\multirow{2}{*}{10} & cellFCLUST & 1.54 & 46.28 & 0.54 & 73.86 \\
& cellGMM & 3.16 & ~3.72 & 1.71 & ~1.52 \\
\hline
\end{tabular}}
\end{table}

\newpage
\section{Additional results on the effect of the tuning parameters}
We provide in Figure \ref{fig_sup: example_Delta} the $\Delta$ plots for the remaining seven variables of the data set used in Section 5 of the main article. 

\begin{figure}[!htbp]
\centering
\subfloat[][$\alpha = 0.025$] 
{\includegraphics[width=\linewidth, height=0.75\textheight]{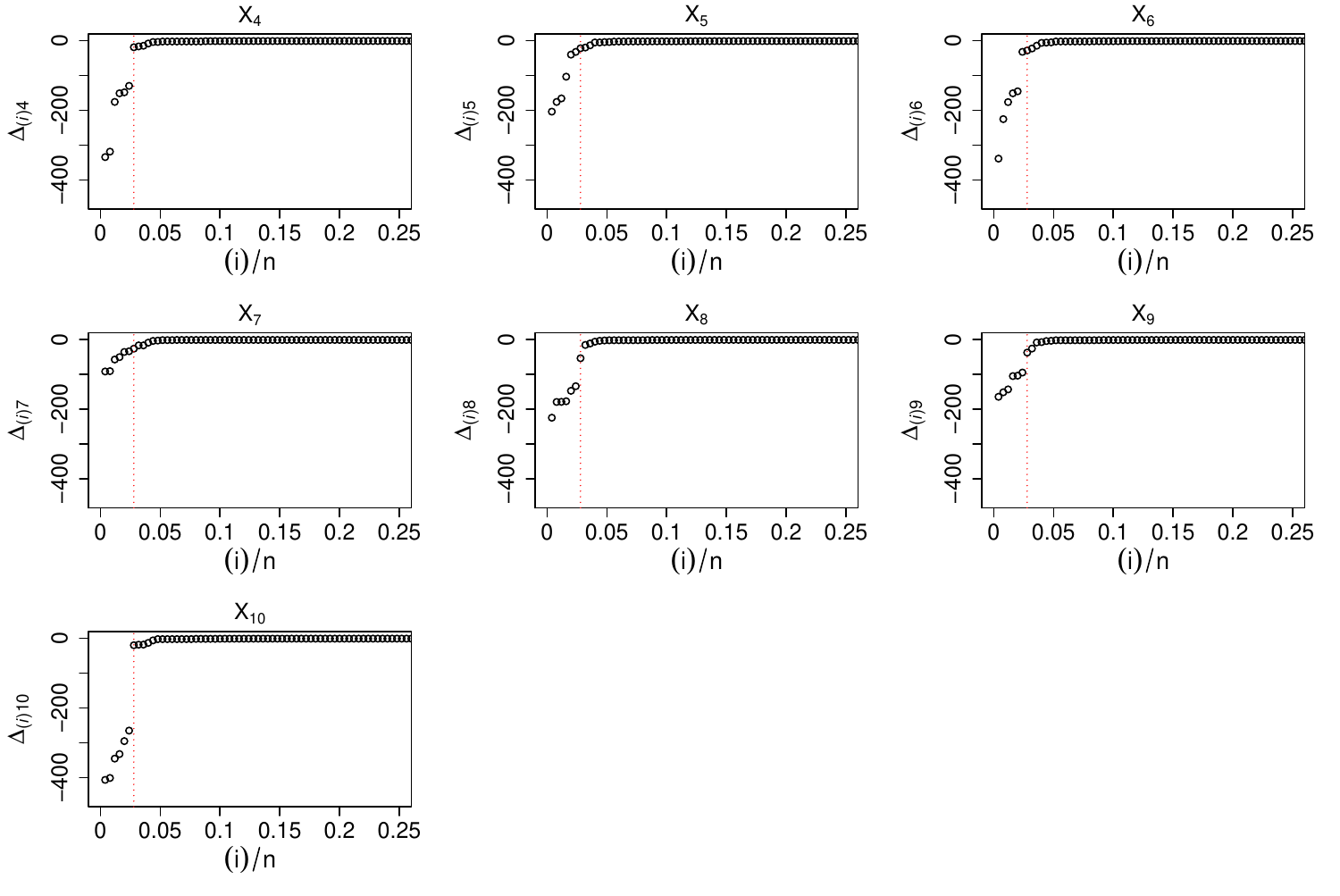}} 
\end{figure}
\addtocounter{figure}{+1}
\begin{figure}\ContinuedFloat
\subfloat[][$\alpha = 0.05$] 
{\includegraphics[width=\linewidth, height=0.75\textheight]{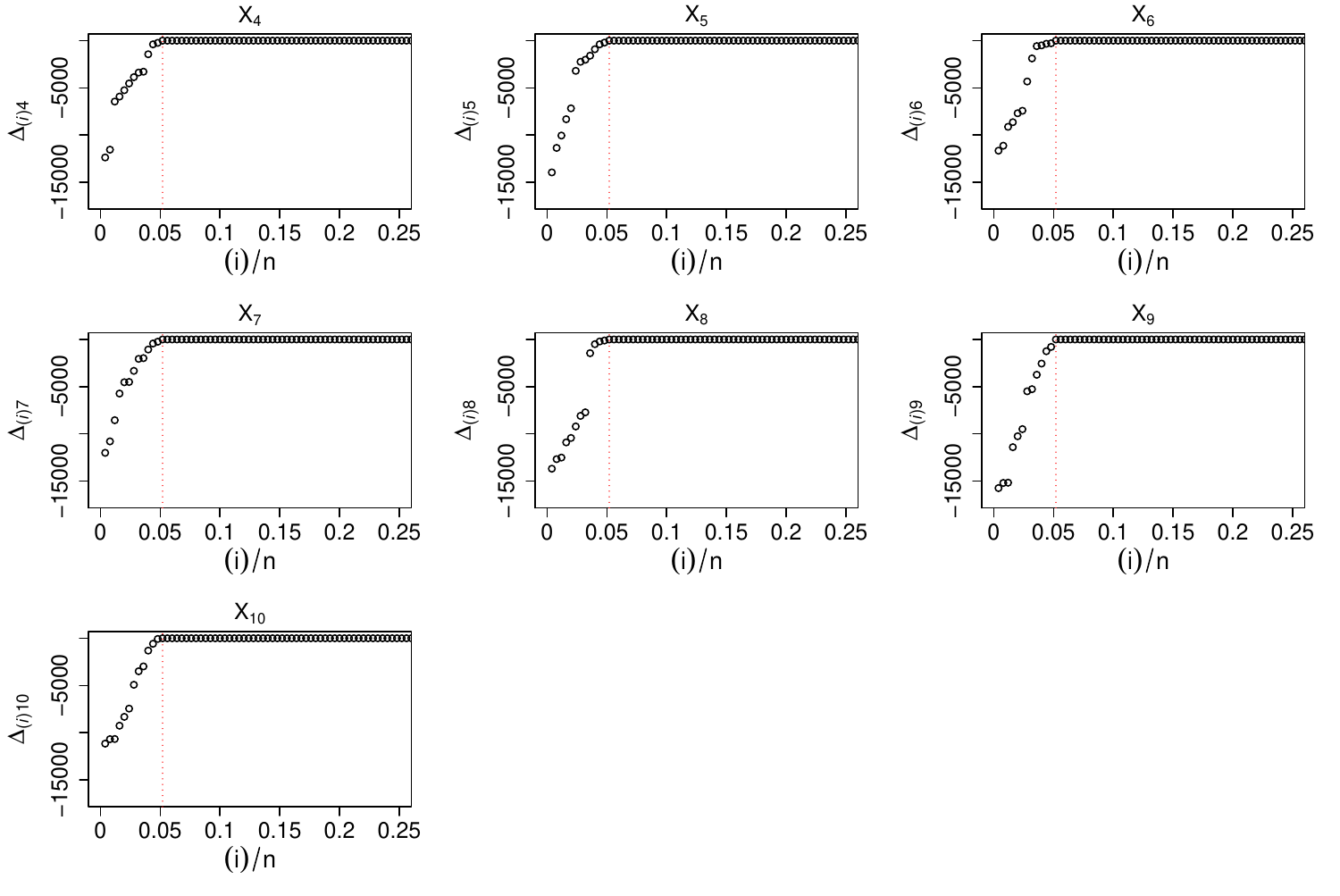}} 
\end{figure}
\addtocounter{figure}{+1}
\begin{figure}\ContinuedFloat
\centering
\subfloat[][$\alpha = 0.075$] 
{\includegraphics[width=\linewidth, height=0.75\textheight]{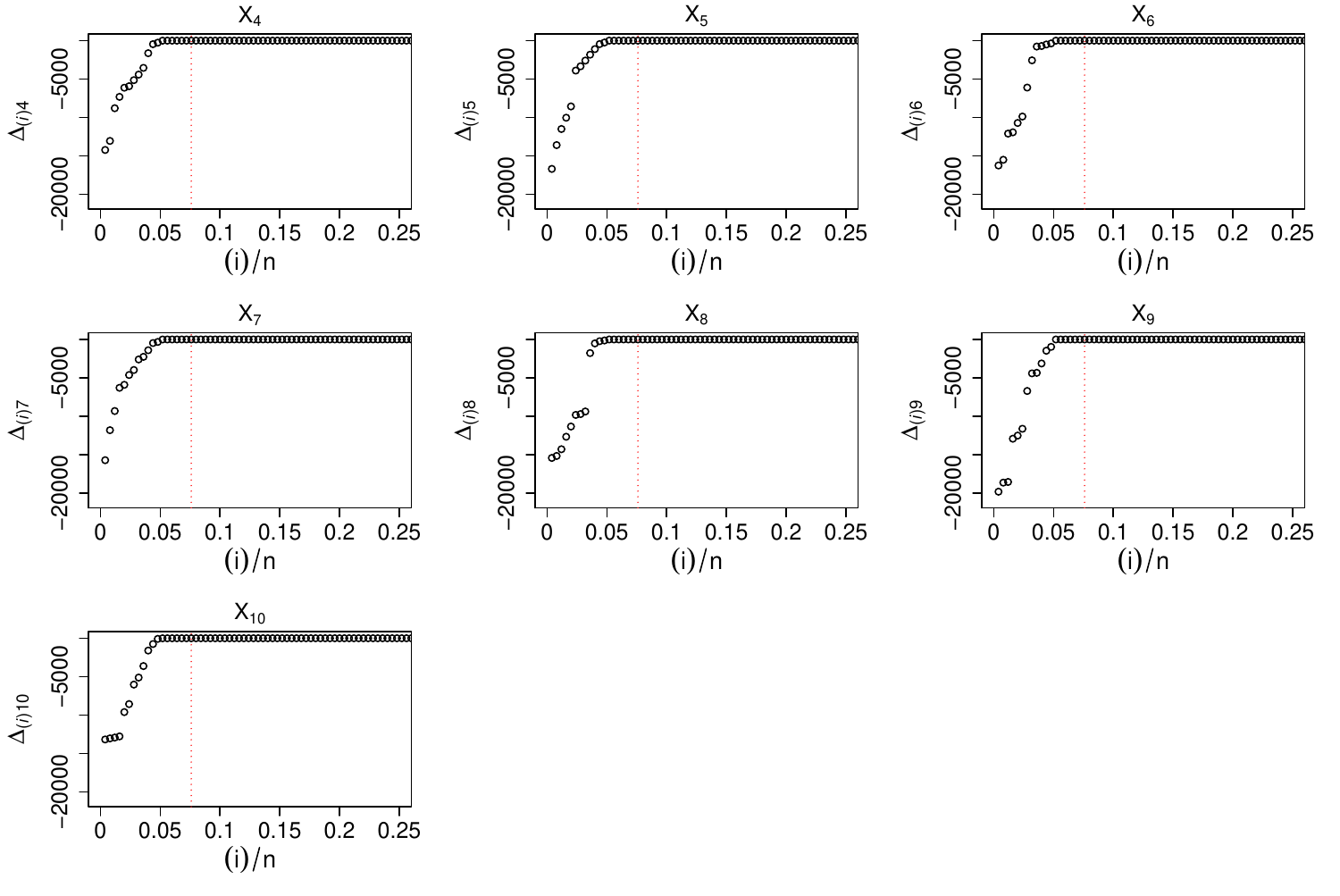}} 
\caption{Synthetic data: $\Delta$ plots where units are sorted according to their $\Delta$ values. Vertical, dashed, red line corresponds to the value of $\alpha$ used for the model implementation when $K = 2$.}
\label{fig_sup: example_Delta}
\end{figure}

\section{Additional results for the real data analyses}
\subsection{Body fat data set}
\begin{figure}[!htbp]
\centering
\subfloat[][Boxplot \label{fig_sup: BF_box}] 
{\includegraphics[width=0.5\linewidth, height=0.48\linewidth]{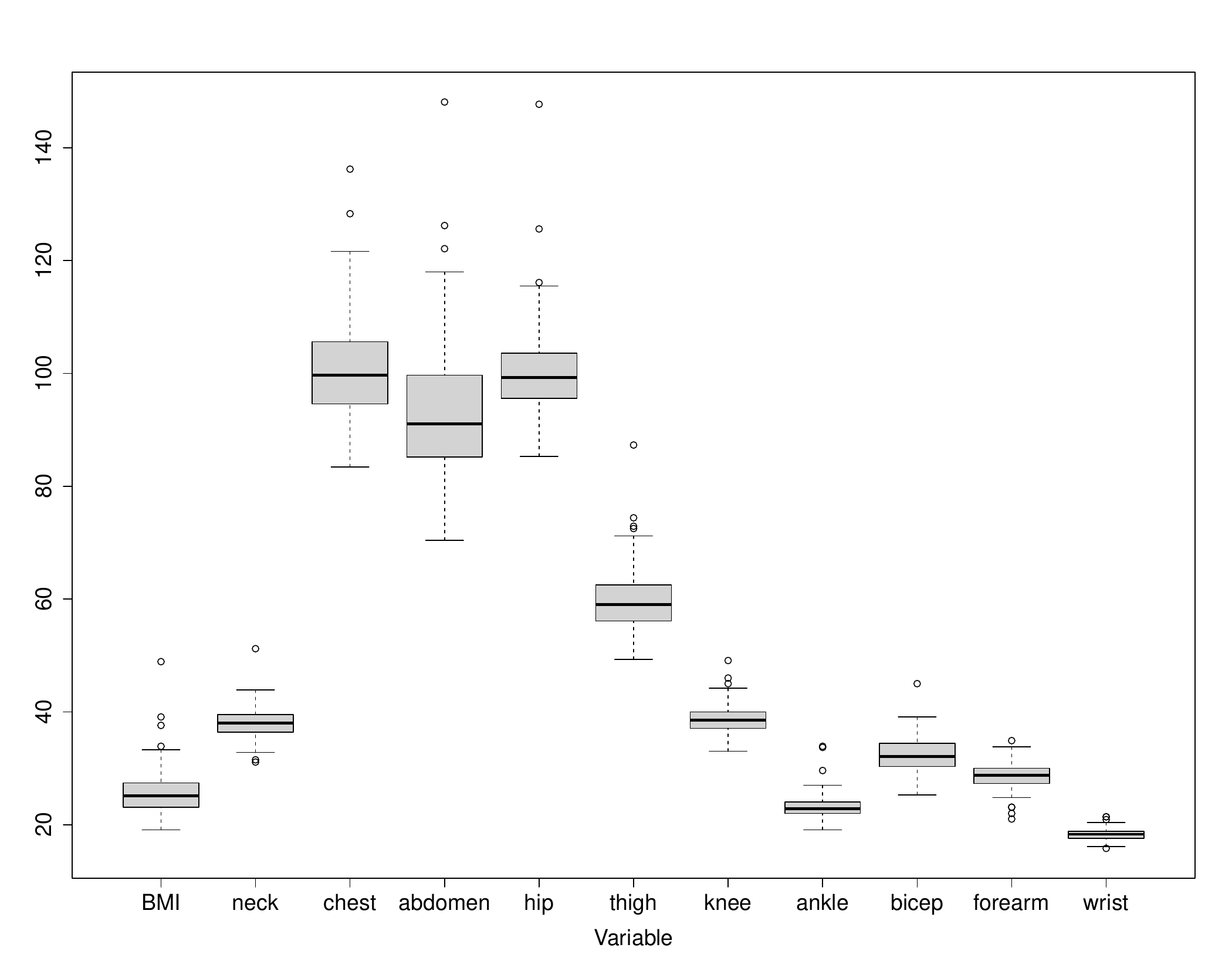}} 
\subfloat[][Pair plot \label{fig_sup: BF_pair}] 
{\includegraphics[width=0.5\linewidth, height=0.45\linewidth]{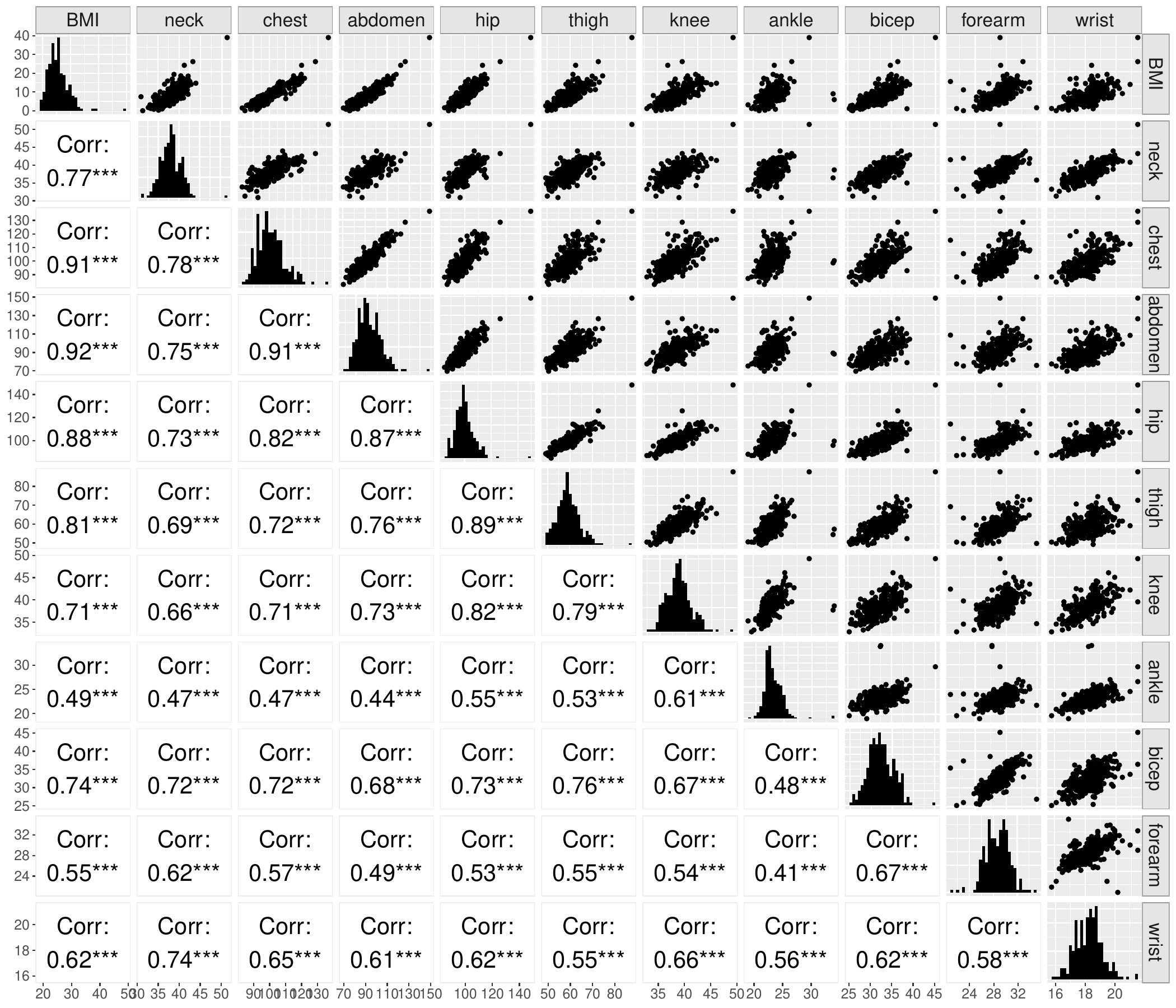}} 
\caption{Body fat data: analysis of the eleven variables selected.}
\label{fig_sup: BF}
\end{figure}

We provide the boxplot (Figure \ref{fig_sup: BF_box}) and the pair plot (Figure \ref{fig_sup: BF_pair}) for the body fat data set examined in Section 6.1 of the main article. These plots suggest the presence of both marginal and non-marginal outliers.

\begin{figure}[!htbp]
    \centering
    \subfloat[][$m = 1.3$]{\includegraphics[width=.5\linewidth]{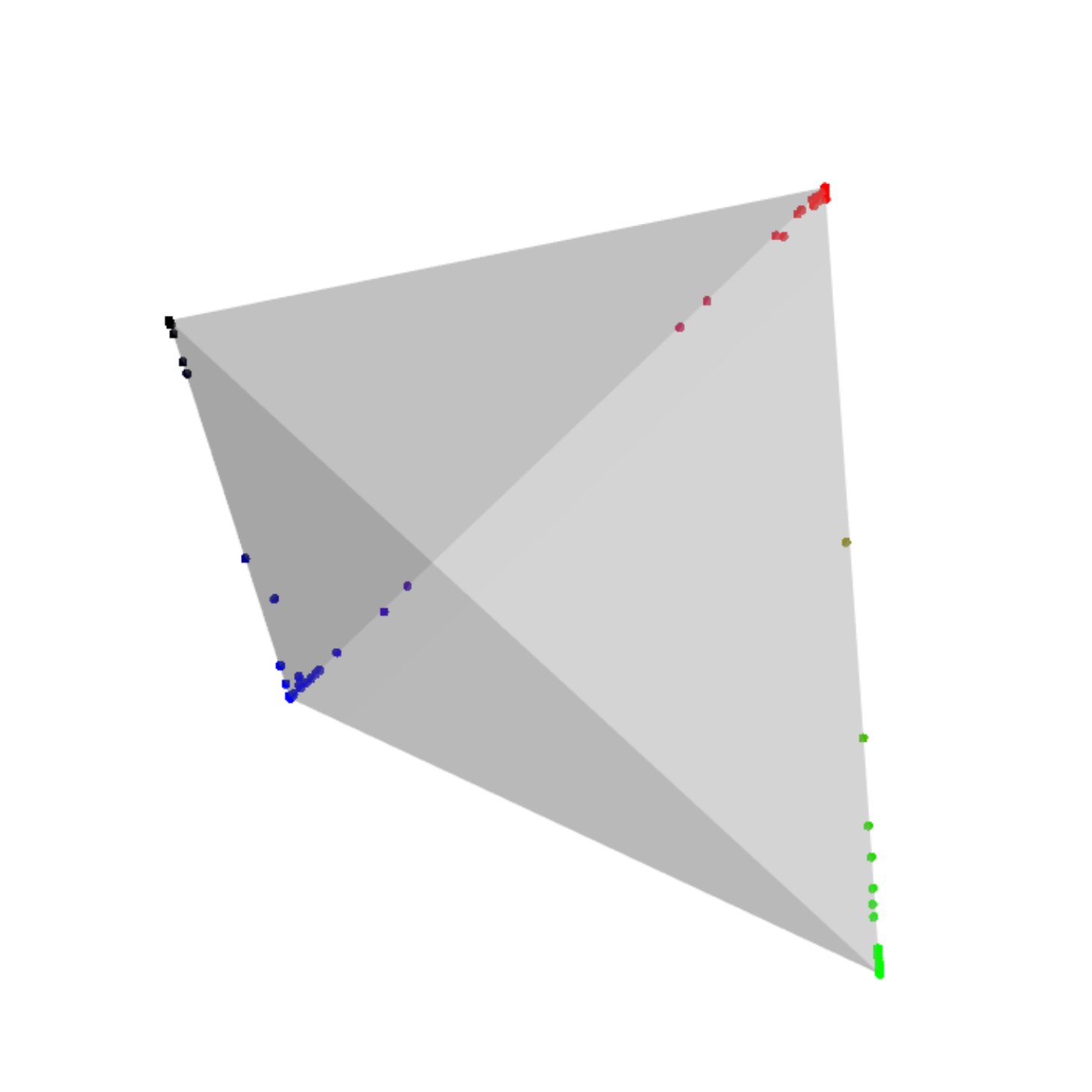}}
    \subfloat[][$m = 1.5$]{\includegraphics[width=.5\linewidth]{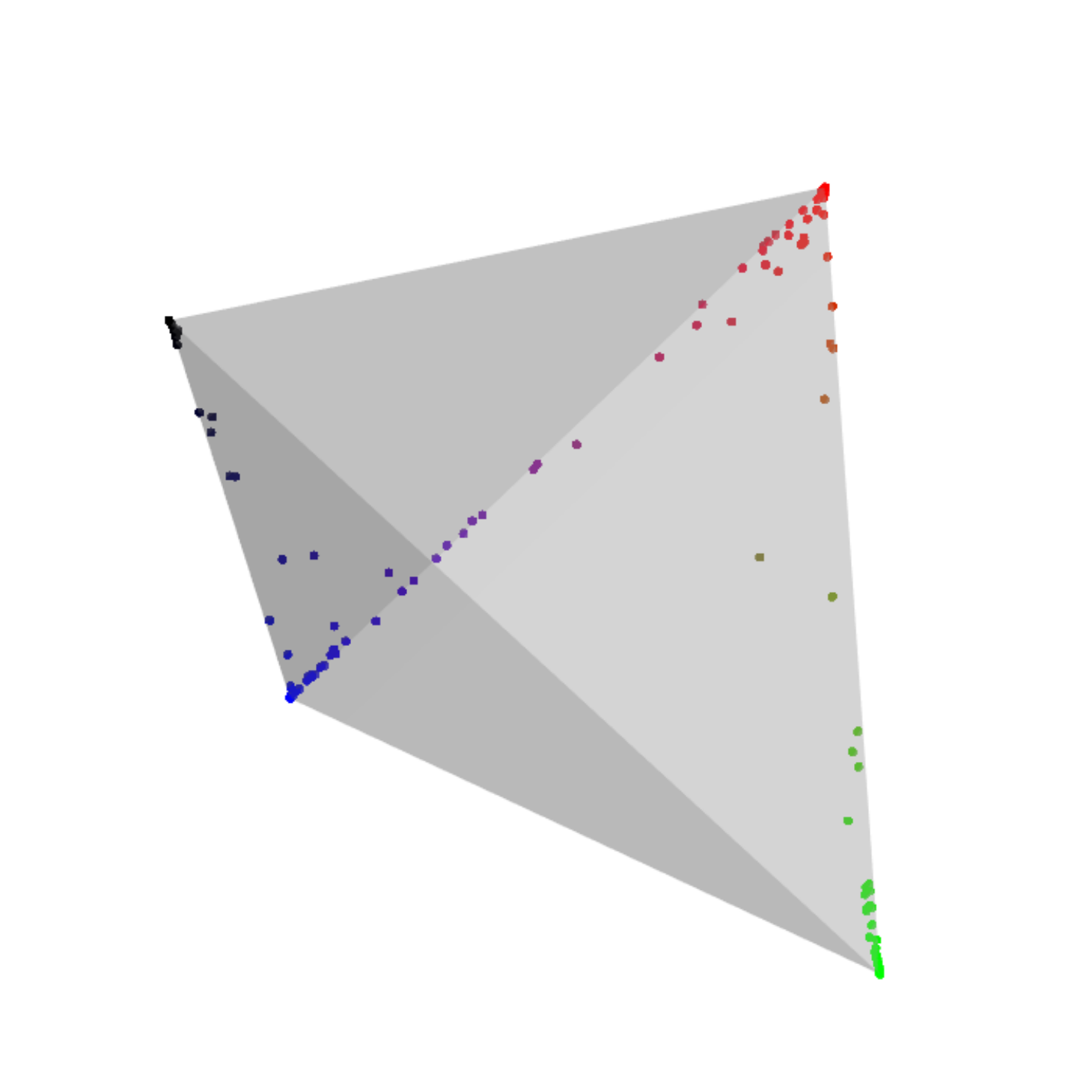}} \\
    \subfloat[][$m = 1.7$]{\includegraphics[width=.5\linewidth]{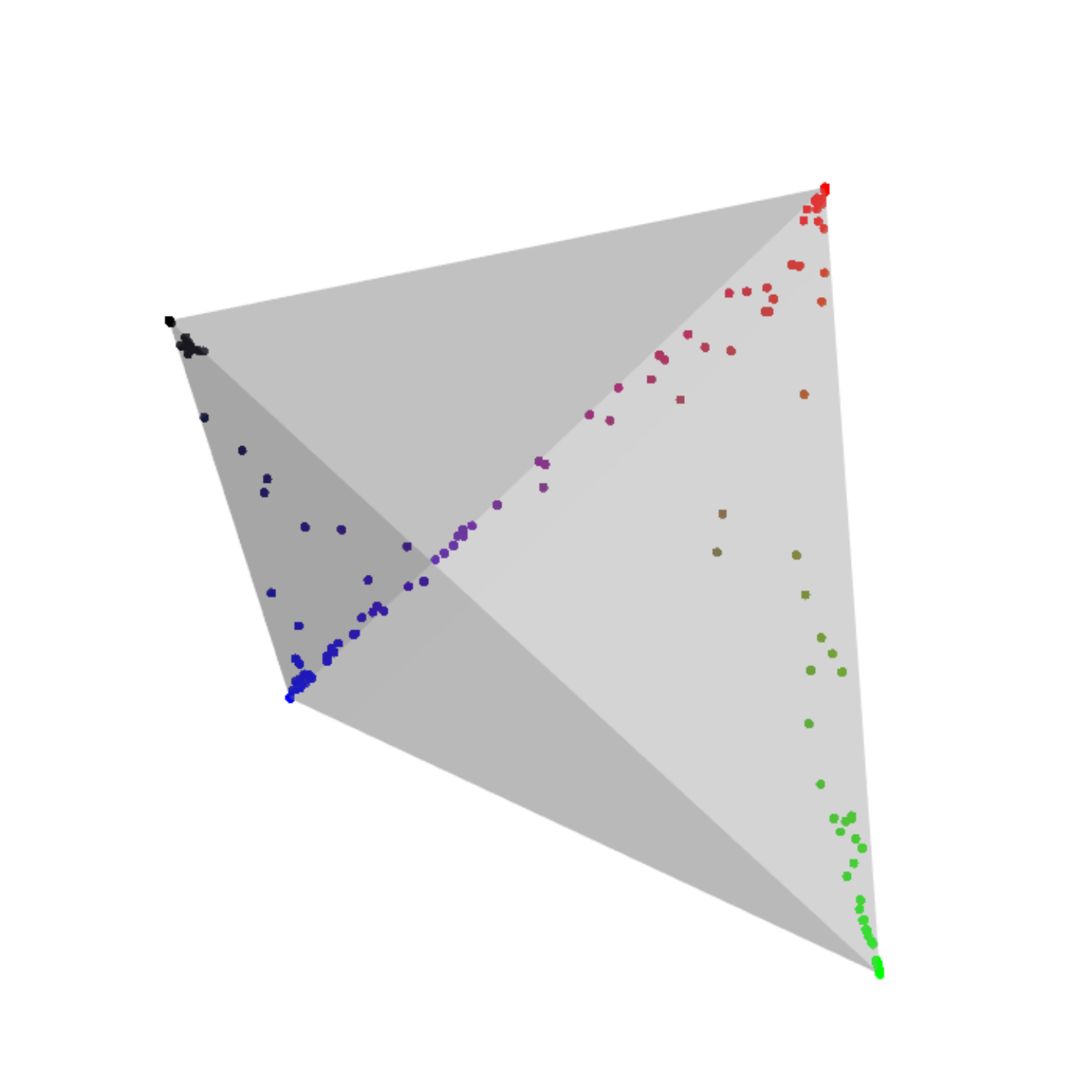}}
    \subfloat[][$m = 1.9$]{\includegraphics[width=.5\linewidth]{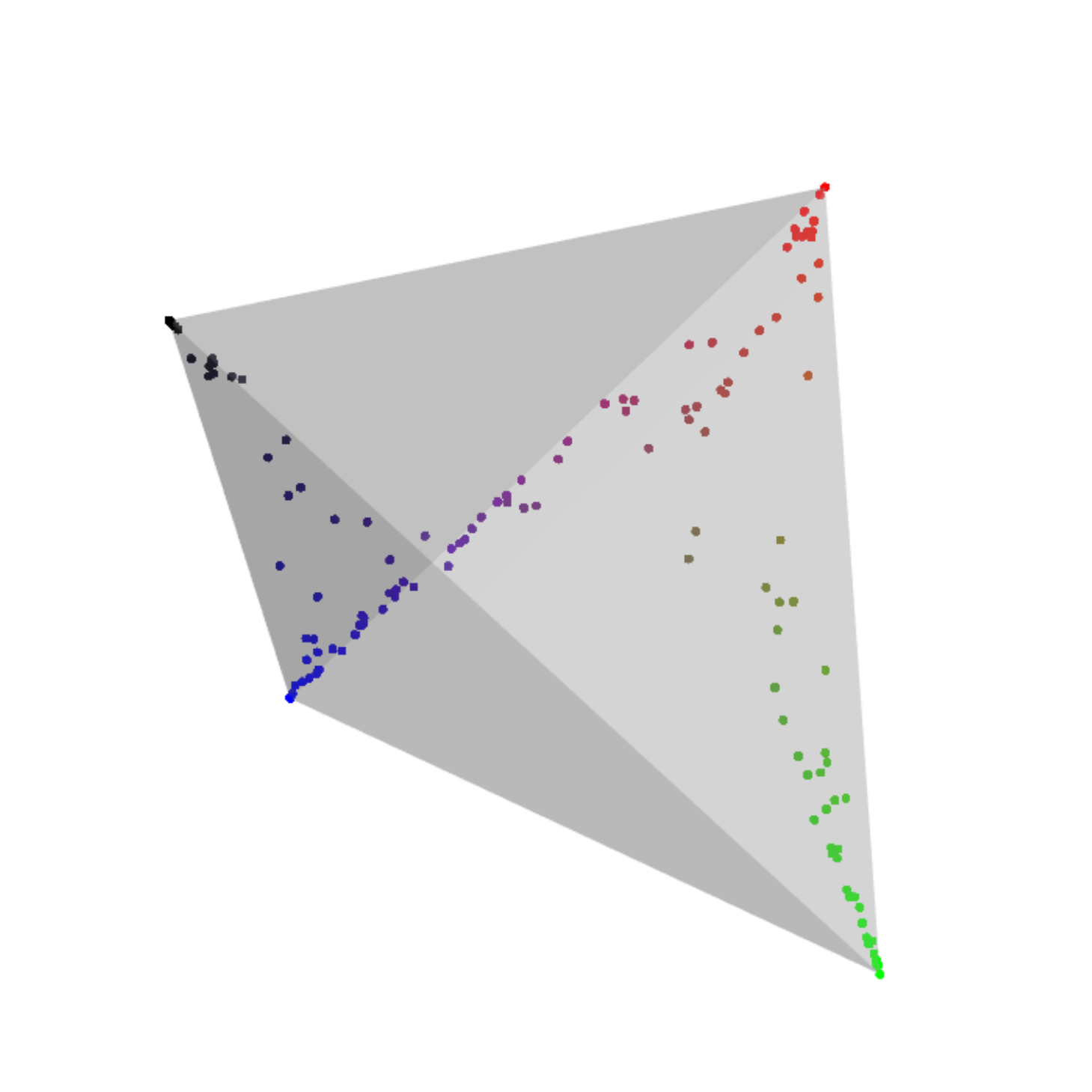}}
    \caption{Body fat data: tetrahedron plot showing cluster assignments (Cluster 1: black, Cluster 2: blue, Cluster 3: red, Cluster 4: green) when $m$ varies. Color intensity and point position reflect the degree of fuzziness (i.e., hard and soft memberships).}\label{fig_sup: BF_m}
\end{figure}

The preliminary analysis on this data set reported in Section 6.1 allows us to choose the fuzzifier parameter $m$. Specifically, we select $m$ by examining the fuzzification obtained by cellFCLUST. Recalling that we select $c = 2$ to avoid obtaining only one elongated cluster, we expect a sensible proportion of weakly assigned units (i.e., $u_{ik} < 0.9$, when observation $i$ belongs to cluster $k$) due to the closeness of the clusters. Figure \ref{fig_sup: BF_m} shows the hard and weak assignments to the clusters as $m$ varies. When $m = 1.3$, very few units are weakly assigned; the same holds for $m = 1.5$. More sensible values for the fuzzifier parameter appear to be $m = 1.7$ and $m = 1.9$. Between these two, we choose the former since the difference lies mostly in the fuzzification between Cluster 2 (blue) and Cluster 3 (red), while the main interest is likely between the third and fourth clusters, where the latter mainly contains obese men (green cluster). It is worth highlighting that the choices of the tuning parameters are strongly related to each other. Therefore, in-depth analyses of their settings are necessary for the user when no prior information is available about the data -- such as that used for setting $c$.

\begin{figure}[t]
    \centering
    \includegraphics[width=\linewidth, height=0.8\textwidth]{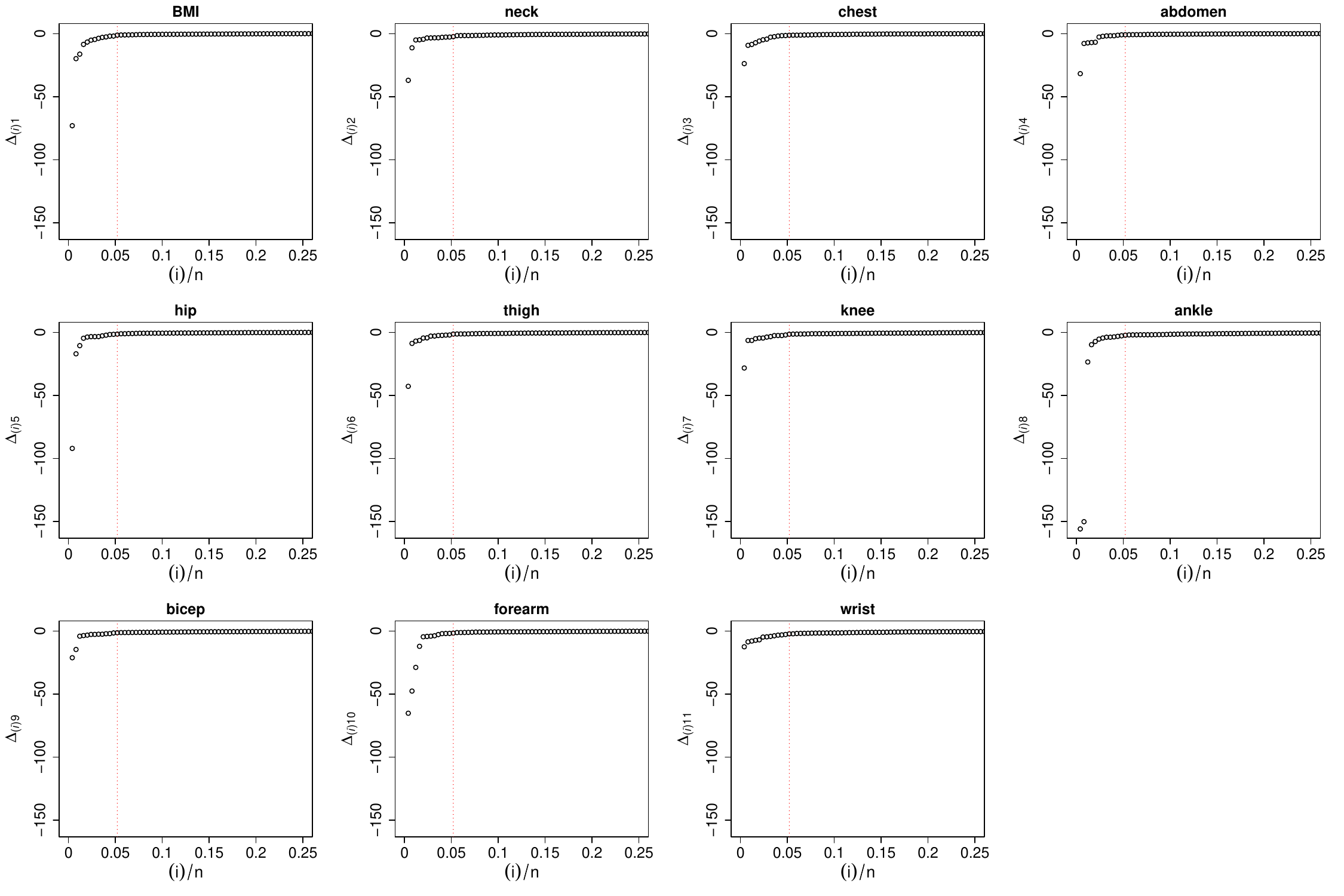}
    \caption{Body fat data: $\Delta$ plot for each variable, where units are sorted according to their $\Delta$ values. Vertical, dashed, red line corresponds to $\alpha = 0.05$ when $K = 4$.}
    \label{fig_sup: BF_Delta}
\end{figure}

We also provide here the plot of the $\Delta$ values per variable (Figure \ref{fig_sup: BF_Delta}) to corroborate the choice of $\alpha = 0.05$, as discussed in the main article. As we can see in the figure, the chosen flagging level turns out to be suitable.

\begin{figure}[!htbp]
\centering
\includegraphics[width=0.95\linewidth, height=0.7\linewidth]{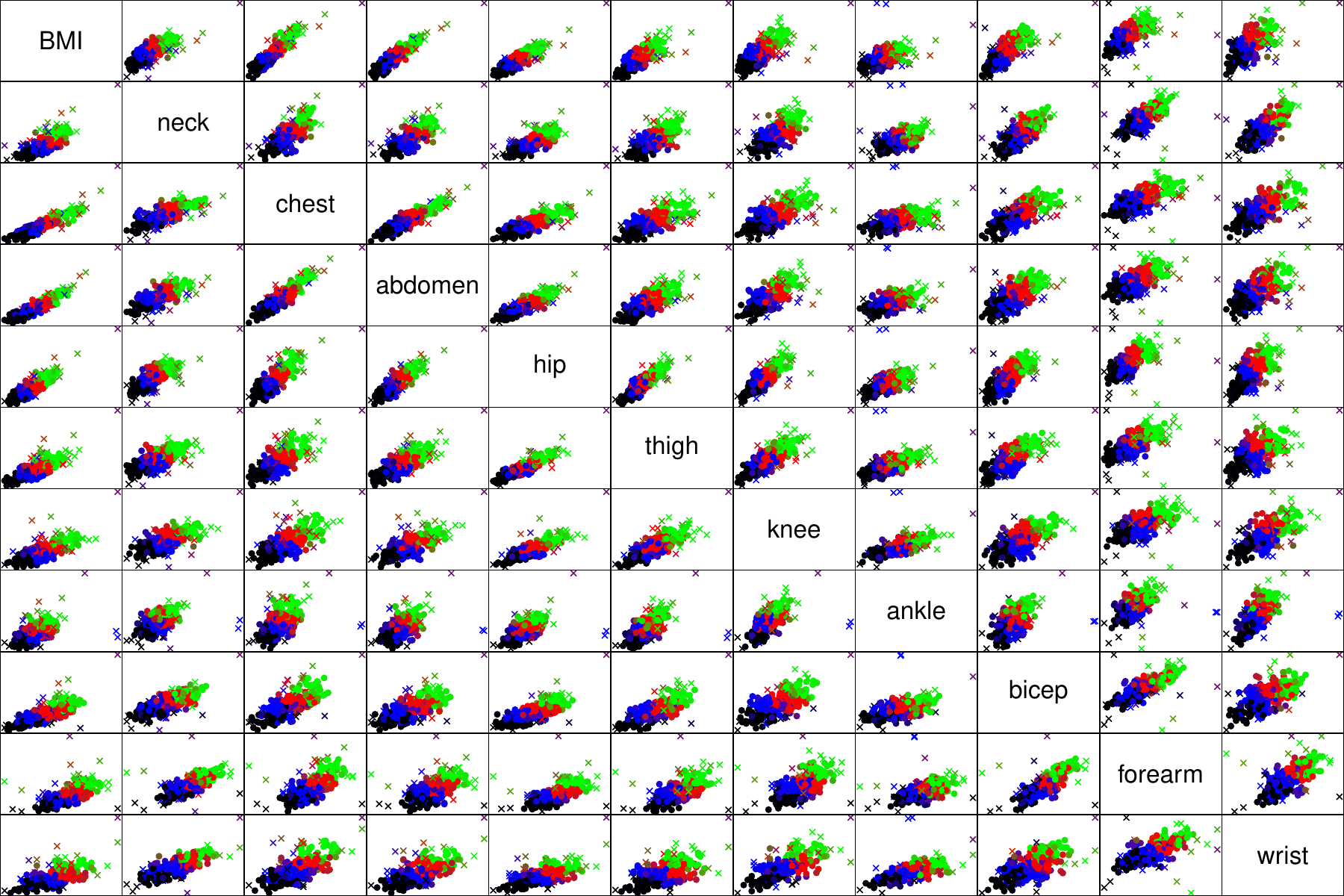} 
\caption{Body fat data: pair plot of the data with clustering results (Cluster 1: black, Cluster 2: blue, Cluster 3: red, Cluster 4: green). Crosses indicate outlying values in at least one of the two variables.}\label{fig_sup: BF_PairLab}
\end{figure}

Additionally, Figure \ref{fig_sup: BF_PairLab} represents the clusters' configuration across variables.

\subsection{OECD regional data set on well-being}\label{sup: oecd}
\begin{figure}[!ht]
    \centering
    \subfloat[][$m = 1.2$ \label{fig_sup: OECD_m12}]{\includegraphics[width=.5\linewidth]{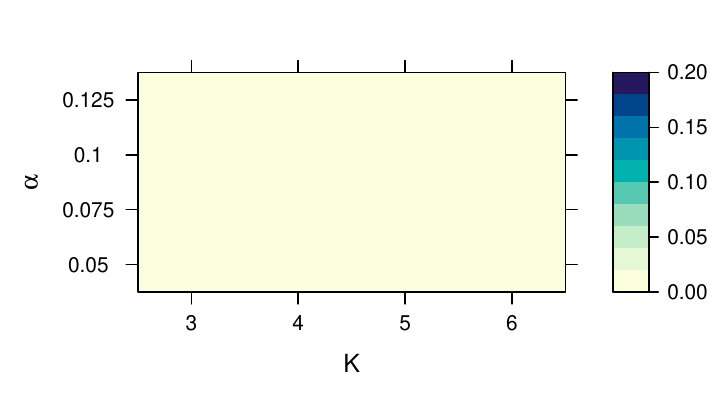}}
    \subfloat[][$m = 1.4$ \label{fig_sup: OECD_m14}]{\includegraphics[width=.5\linewidth]{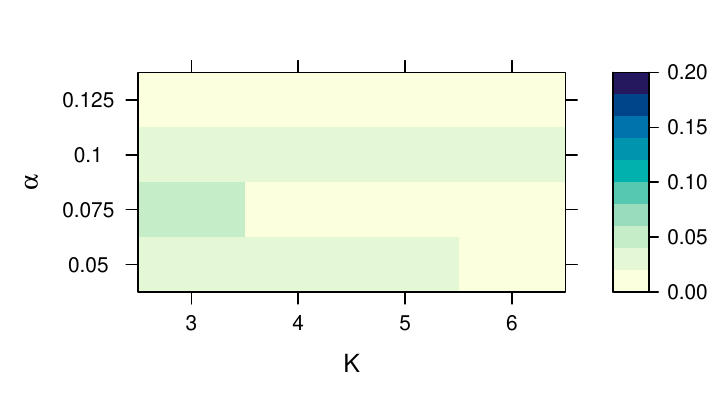}}\\
    \subfloat[][$m = 1.6$ \label{fig_sup: OECD_m16}]{\includegraphics[width=.5\linewidth]{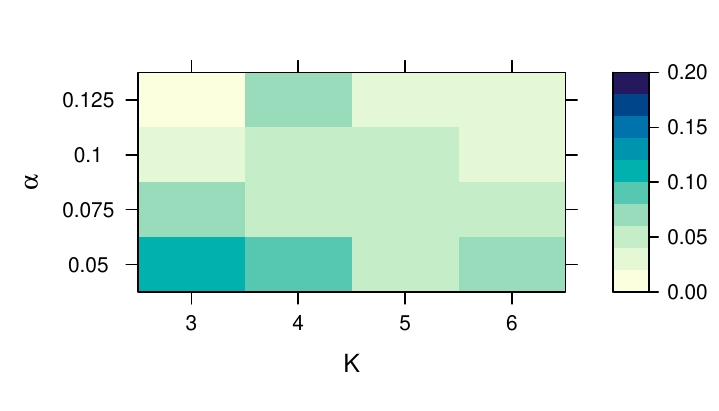}}
    \subfloat[][$m = 1.8$ \label{fig_sup: OECD_m18}]{\includegraphics[width=.5\linewidth]{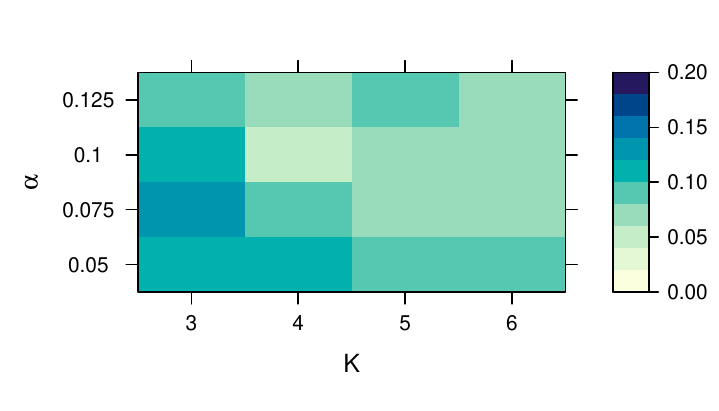}}\\
    \subfloat[][$m = 2$ \label{fig_sup: OECD_m20}]{\includegraphics[width=.5\linewidth]{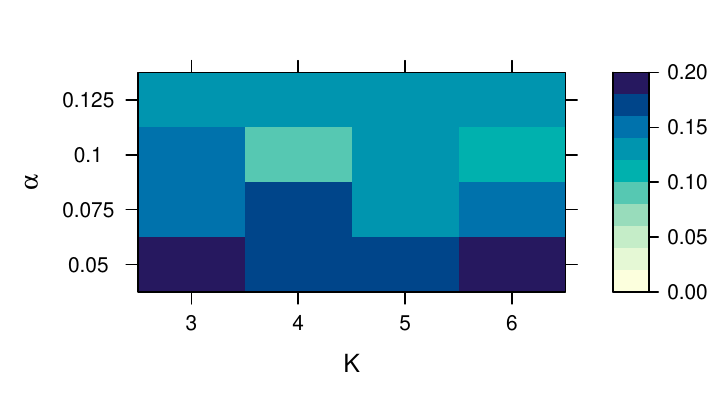}}
    \subfloat[][Hard and weak assignments \label{fig_sup: OECD_m}]{\includegraphics[width=.5\linewidth]{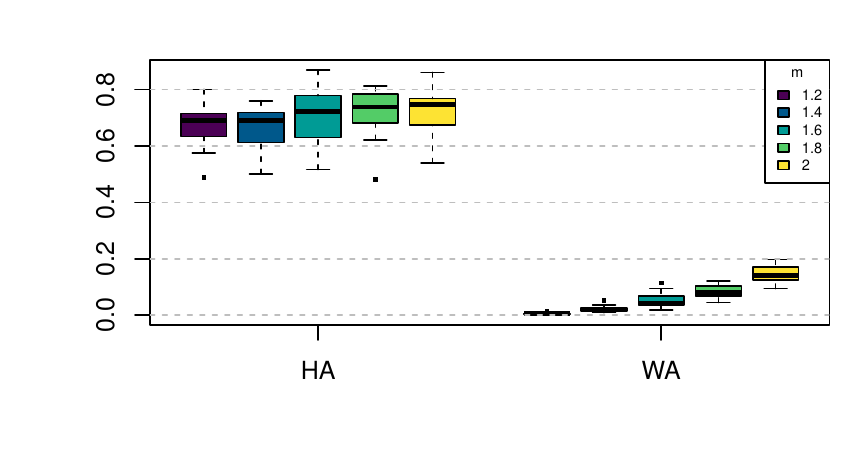}}
    \caption{OECD data: proportion of weak assignments depending on a subset of $K$ and $\alpha$ values, for five different levels of $m$ (\ref{fig_sup: OECD_m12}--\ref{fig_sup: OECD_m20}). Boxplots of the proportions of hard and weak assignments as $m$ varies in \ref{fig_sup: OECD_m}.}
    \label{fig_sup: oecd_m}
\end{figure}

The choice of the parameter $m$ used in Section 6.2 is based on the proportions of weak assignments, which are reported in Figure \ref{fig_sup: oecd_m}. We consider here a reasonable subset for $K$ and $\alpha$, i.e., $\{3, \ldots, 6\}$ and $\{0.05, 0.075, 0.1, 0.125\}$, respectively, with different candidate levels for $m$. The selected value, i.e. $m = 1.8$ (Figure \ref{fig_sup: OECD_m18}), provides a range between $0.05$ and $0.12$ for the proportion of weak assignments, which corresponds to $22$ and $59$ regions, respectively, depending on the other parameters. Additionally, the proportion of hard assignments is relatively stable across different values of $m$ (Figure \ref{fig_sup: OECD_m}). 

\begin{figure}[t]
    \centering
    \includegraphics[width=0.9\linewidth]{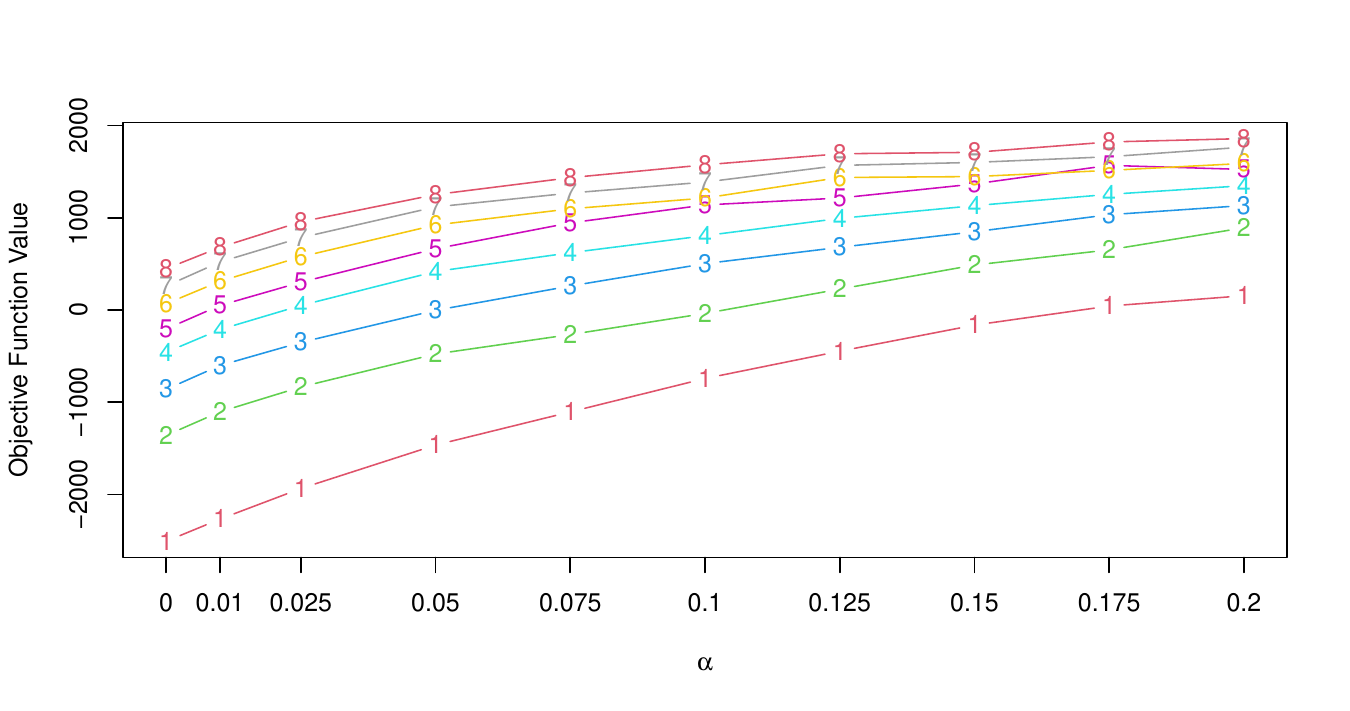}
    \caption{OECD data: objective function curves.}\label{fig_sup: oecd_ctl}
\end{figure}

Looking at the behavior of the objective function curves as $K$ and $\alpha$ vary (Figure \ref{fig_sup: oecd_ctl}), a sensible choice for $K$ results to be $5$. Indeed, the objective function value shows substantial jumps for $K = 1, \ldots, 5$ across the $\alpha$ levels. After $K = 5$, the values become closer to each other when $\alpha$ is sufficiently high, that is, from $0.075$ onward, providing evidence of this level of flagging. In Table \ref{tab: oecd_wa}, we report the complete list of $29$ weakly assigned regions with the respective membership across the five clusters.

\begin{figure}[!htbp]
    \centering
    \includegraphics[width=\linewidth, height=1.1\textwidth]{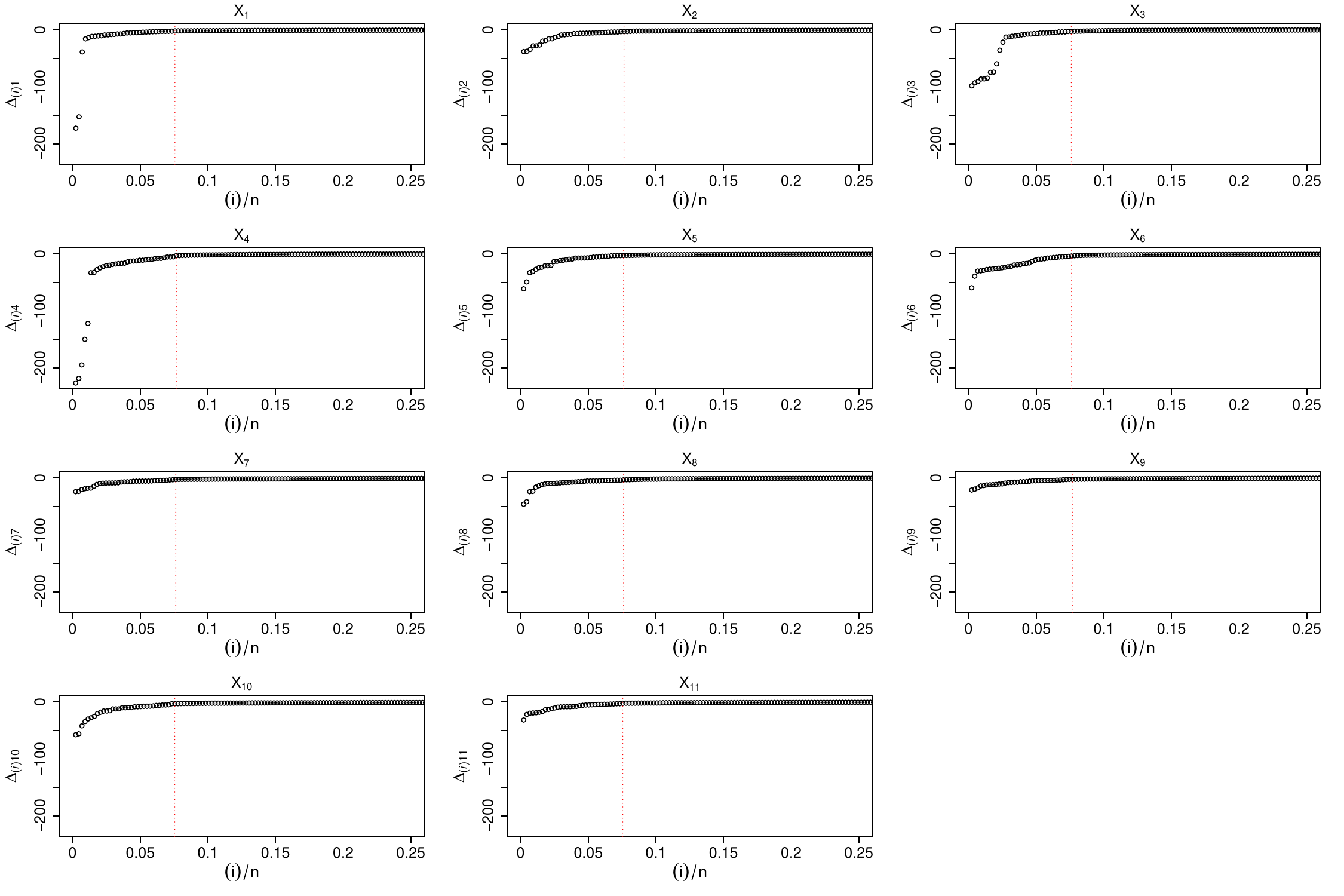}
    \caption{OECD data: $\Delta$ plot for each variable, where units are sorted according to their $\Delta$ values. Vertical, dashed, red line corresponds to $\alpha = 0.075$ when $K = 5$.}
    \label{fig_sup: oecd_delta}
\end{figure}

\begin{figure}[!htbp]
    \centering
    \subfloat[][Cluster 1 \label{fig_sup: OECD_w1}]{\includegraphics[trim={6.5cm 0 7cm 1cm}, clip, width=.23\linewidth]{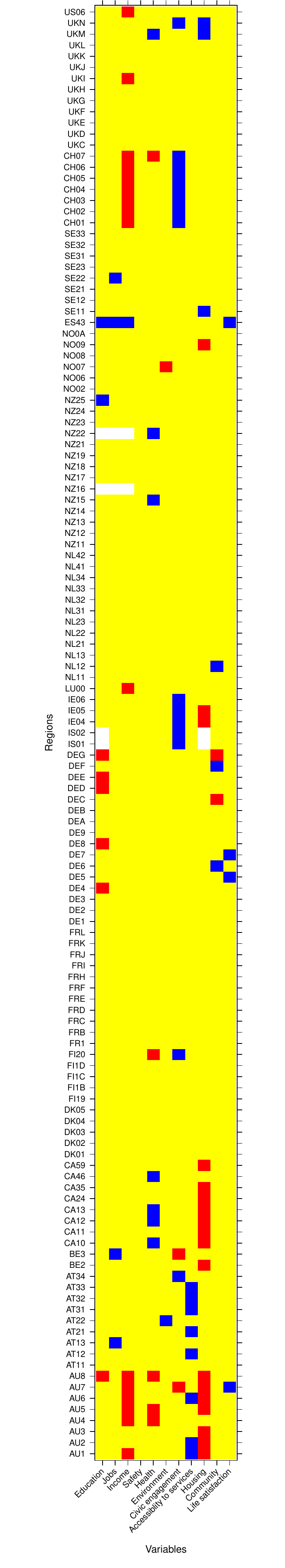}}
    \begin{minipage}[b]{.23\linewidth}
    \subfloat[][Cluster 2 \label{fig_sup: OECD_w2}]{\includegraphics[trim={6.5cm 0 7cm 0}, clip, width=\linewidth]{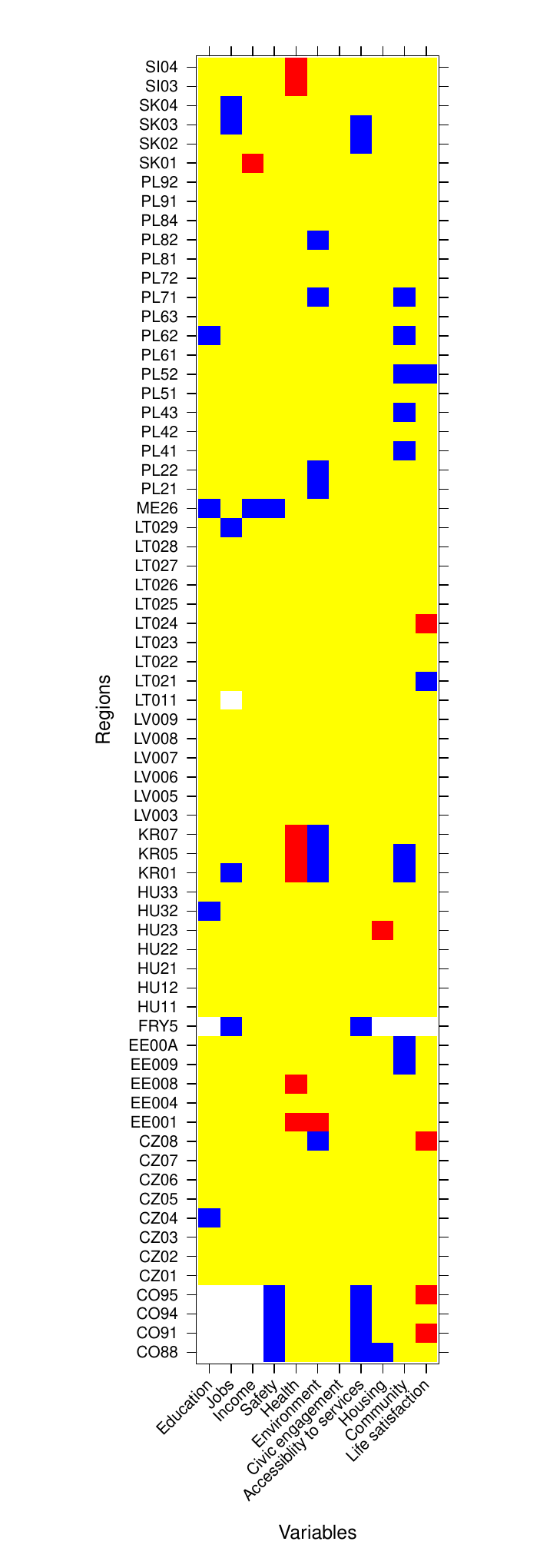}} \\
    \subfloat[][Cluster 3\label{fig_sup: OECD_w3}]{\includegraphics[trim={6.5cm 0 7cm 0}, clip, width=\linewidth]{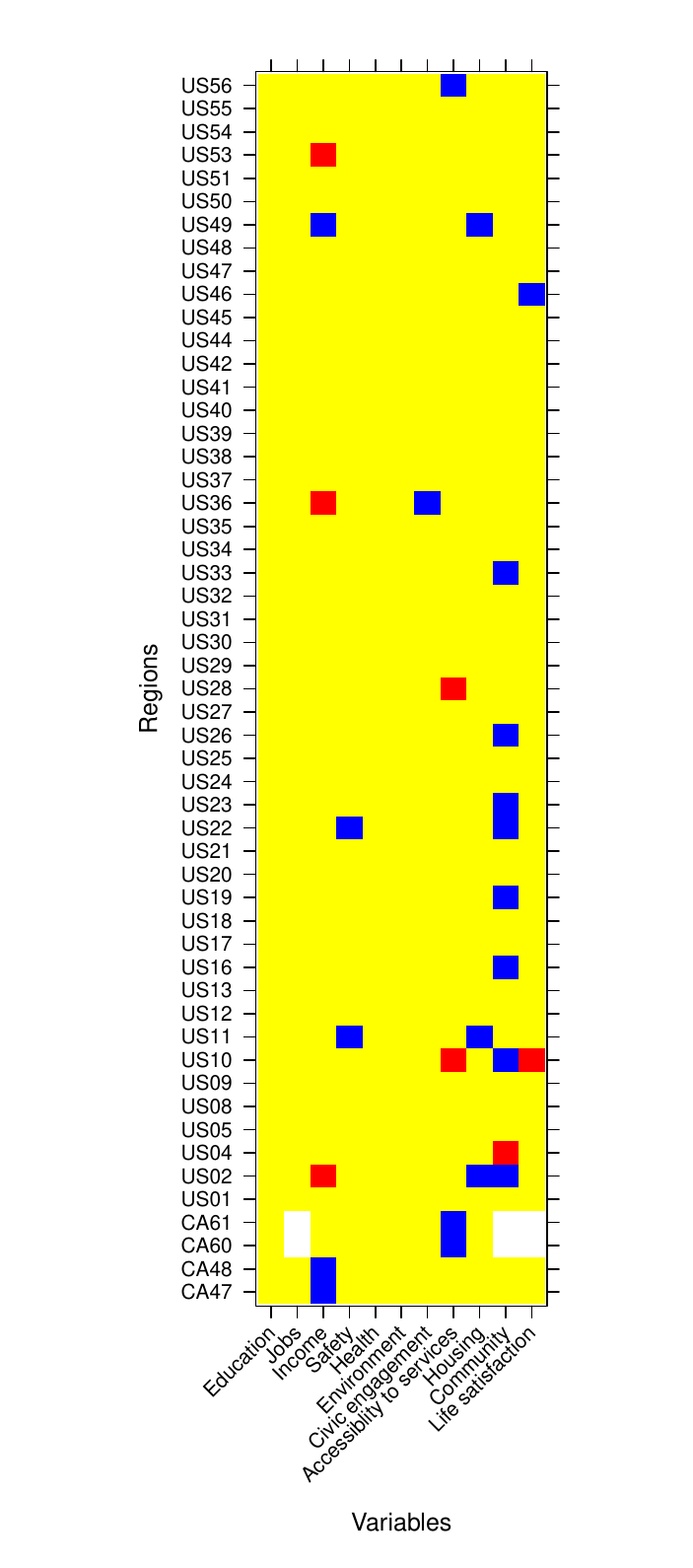}}
    \end{minipage}%
    \subfloat[][Cluster 4\label{fig_sup: OECD_w4}]{\includegraphics[trim={6.5cm 0 7cm 0}, clip, width=.23\linewidth]{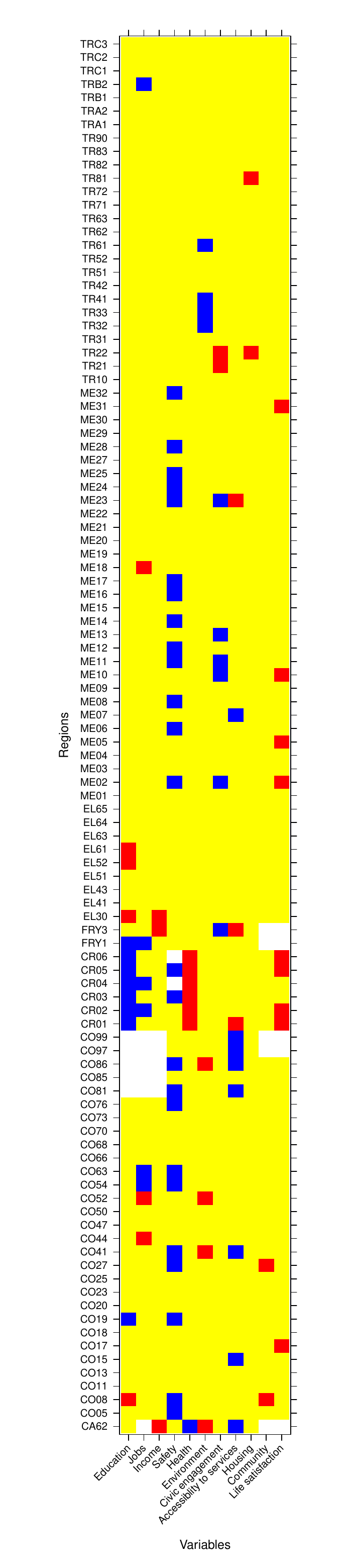}}
    \subfloat[][Cluster 5\label{fig_sup: OECD_w5}]{\includegraphics[trim={6.5cm 0 7cm 0}, clip, width=.23\linewidth]{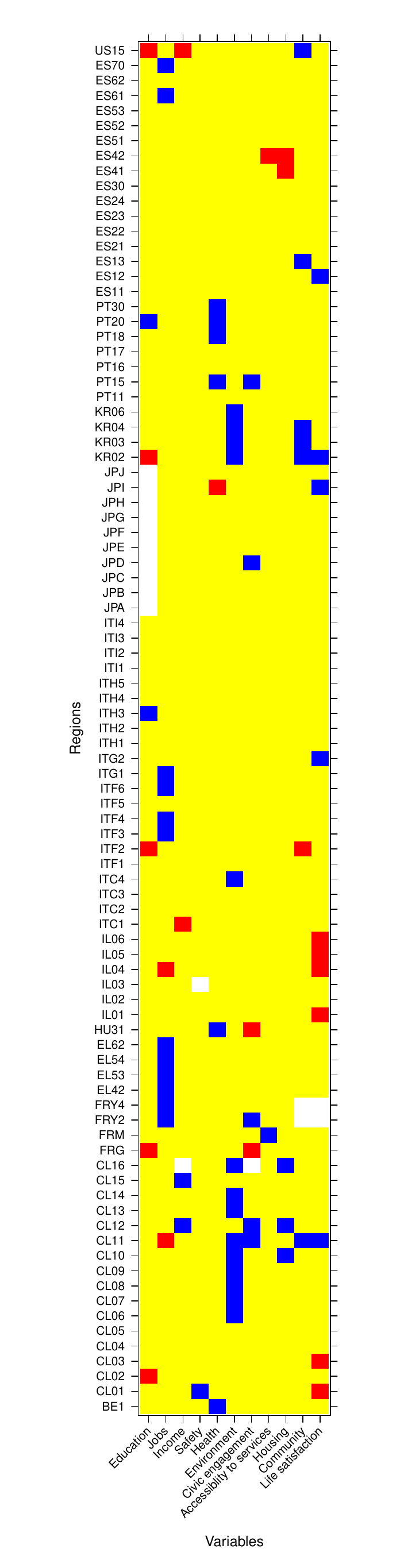}}   
    \caption{OECD data: outlying cells per cluster (yellow: reliable cells; blue: flagged cells imputed with a higher value than the original one; red: flagged cells imputed with a lower value than the original one; white: missing values).}
    \label{fig_sup: oecd_w}
\end{figure}

In Figure \ref{fig_sup: oecd_delta}, it is possible to inspect the $\Delta$ plot for each variable. The level of $\alpha$ proves to be sufficient in every variable supporting the chosen flagging level, which is $0.075$. Furthermore, Figure \ref{fig_sup: oecd_w} shows the unreliable cells for every region grouped by cluster, where the red and blue cells represent lower and higher imputed values than the actual data, respectively, whereas yellow represents reliable cells and white are missing data.

\begin{table}[!htbp]
\caption{Weak assignments for the OECD data.}
\centering
\resizebox{\textwidth}{!}{
\begin{tabular}[t]{l l l r r r r r}
\toprule
& Country & Region & 1 & 2 & 3 & 4 & 5\\
\midrule
\multirow{2}*{Cluster 1}&Canada & Newfoundland and Labrador & 0.86 & 0.01 & 0.03 & 0.01 & 0.09\\
&United States & California & 0.84 & 0.01 & 0.07 & 0.02 & 0.05\\
\midrule
\multirow{6}*{Cluster 2}&Colombia & San Andrés & 0.20 & 0.43 & 0.08 & 0.20 & 0.10\\
&Colombia & Amazonas & 0.02 & 0.62 & 0.01 & 0.31 & 0.04\\
&Colombia & Guainía & 0.02 & 0.47 & 0.01 & 0.45 & 0.06\\
&Colombia & Guaviare & 0.03 & 0.47 & 0.01 & 0.40 & 0.09\\
&France & Mayotte & 0.05 & 0.75 & 0.04 & 0.14 & 0.03\\
&Mexico & Sonora & 0.03 & 0.65 & 0.01 & 0.29 & 0.03\\
\midrule
\multirow{9}*{Cluster 4}&Canada & Nunavut & 0.02 & 0.02 & 0.01 & 0.57 & 0.38\\
&Colombia & Cauca & 0.02 & 0.03 & 0.00 & 0.90 & 0.05\\
&Costa Rica & Central & 0.02 & 0.01 & 0.01 & 0.89 & 0.08\\
&Costa Rica & Central Pacific & 0.01 & 0.02 & 0.00 & 0.54 & 0.43\\
&Costa Rica & Brunca & 0.02 & 0.10 & 0.01 & 0.56 & 0.31\\
&Costa Rica & Huetar Caribbean & 0.01 & 0.03 & 0.00 & 0.81 & 0.15\\
&France & Guadeloupe & 0.14 & 0.26 & 0.04 & 0.33 & 0.24\\
&Greece & North Aegean & 0.03 & 0.03 & 0.01 & 0.78 & 0.15\\
&Greece & Peloponnese & 0.04 & 0.02 & 0.01 & 0.70 & 0.23\\
\midrule
\multirow{12}*{Cluster 5}&Belgium & Brussels Capital Region & 0.09 & 0.01 & 0.02 & 0.03 & 0.86\\
&Chile & Magallanes and Chilean Antarctica & 0.14 & 0.02 & 0.10 & 0.06 & 0.69\\
&Chile & Arica y Parinacota & 0.09 & 0.02 & 0.02 & 0.09 & 0.79\\
&France & Corsica & 0.06 & 0.01 & 0.01 & 0.03 & 0.90\\
&France & Martinique & 0.32 & 0.05 & 0.04 & 0.21 & 0.38\\
&Greece & South Aegean & 0.03 & 0.02 & 0.00 & 0.15 & 0.81\\
&Greece & Western Macedonia & 0.03 & 0.03 & 0.00 & 0.24 & 0.70\\
&Greece & Epirus & 0.01 & 0.01 & 0.00 & 0.39 & 0.58\\
&Greece & Ionian Islands & 0.04 & 0.01 & 0.00 & 0.20 & 0.74\\
&Japan & Shikoku & 0.31 & 0.16 & 0.01 & 0.02 & 0.50\\
&Portugal & Azores (autonomous region) & 0.03 & 0.19 & 0.03 & 0.04 & 0.70\\
&United States & Hawaii & 0.04 & 0.02 & 0.03 & 0.01 & 0.90\\
\bottomrule
\end{tabular}
}
\label{tab: oecd_wa}
\end{table}

\newpage
\bibliographystyle{elsarticle-num}
\bibliography{references}